%% file: 12_5_yr_Scattering_Paper_Draft.tex
\documentclass[twocolumn, times]{aastex63}

\usepackage{amsmath,verbatim, nicefrac,color, soul}
\usepackage{natbib}

\newcommand\Hl[1]{\colorbox{yellow}}

\shorttitle{Monitoring Interstellar Scattering Delays}
\shortauthors{J. E. Turner \lowercase{et al}.}

\begin{document}
\title{The NANOGrav 12.5-Year Data Set: Monitoring Interstellar Scattering Delays}

\input{authors}

\input{abstract.tex}
\keywords{methods: data analysis --
stars: pulsars --
ISM: general -- gravitational waves}

\section{Introduction}\label{intro}
The North American Nanohertz Observatory for Gravitational Waves \citep[NANOGrav;][]{McLaughlin_2013} aims to use a pulsar timing array (PTA) to detect nanohertz frequency gravitational waves. The 12.5-year data set \citep{alam2020nanograv} presents observations of 47 millisecond pulsars (MSPs) with up to sub-$\mu$s precision and finds a strong, common red-noise process consistent with a gravitational-wave background but lacks the quadrupolar correlations necessary to claim a detection \citep{12_stoch}. Pulsar timing precision is largely a result of robust timing models for each MSP, accounting for many phenomena that might affect a pulsar time of arrival (TOA) and potentially mask a gravitational wave signal in our data. 

\par One of the most significant sources of TOA residual uncertainty for PTAs comes from the interaction between a pulsar's radio emission and free electrons in the interstellar medium (ISM). The most significant of these ISM effects is dispersion, in which a frequency-dependent time delay arises from the radio emission propagating through free electrons in the ISM. The delay at a given observing epoch can be related to the product of the integrated column density of free electrons along the line of sight (LOS), known as the dispersion measure (DM), and the inverse square of the observation frequency, $\nu$. Since the Earth, the solar system, the ISM, and the pulsars all have motions that vary the LOS from epoch to epoch, DM is time-dependent. The delay can be corrected by observing a pulsar at multiple frequencies at each observing epoch \citep{handbook,Demorest_2012,NANO_9yr,Jones_2017,keith_2013}.

\par Interstellar scattering also contributes epoch-dependent delays. The phenomenon is the result of a pulsar's radio emission propagating through a nonuniform distribution of free electrons. These delays also vary with time. However, the nature of the propagation for dispersion and scattering results in different frequency dependencies for each phenomenon. As mentioned above, the dispersion delay scales as $\Delta t_{\textrm{DM}}\propto \nu^{-2}$, and, if we assume fluctuations in the ISM can be modeled by a Kolmogorov-like spectrum, it can be shown that time delays from scattering go as $\tau_{\textrm{d}}\propto \nu^{-4.4}$ if inner scale effects are ignored and if the scattering properties of the medium are the same everywhere (or, for a screen, identical across the screen), and if refraction does not modify the scattering \citep{Cordes_1998}. 

\par We see the effects of interstellar scattering in the broadening of pulse profiles and the delaying of pulse arrival times. Scattering also results in interstellar scintillation, which arises from two interrelated phenomenon: diffractive interstellar scintillation (DISS) and refractive interstellar scintillation (RISS) \citep{rickett_1990}. With NANOGrav's observing cadence, DISS is the most observable over a single epoch, primarily because the resulting variability is resolvable over typical observation lengths and bandwidths \citep{11yr_Obs}. More specifically, for gigahertz frequencies and pulsars at DMs $\simeq 50$ pc cm$^{-3}$ the characteristic timescale from DISS is typically on the order of minutes and the characteristic bandwidth from the accompanying pulse broadening is on the order of MHz \citep{Cordes_1998}, although we observe large variations in scintillation parameters for a given DM.

\par \cite{Levin_Scat} examined effects from DISS on pulsars in the NANOGrav 9-year data set \citep{NANO_9yr} and found that, generally, NANOGrav pulsars exhibit scattering delays on the order of 1--100 ns at 1500 MHz. However, even if delays are small compared to TOA errors, if they are correlated over time they could contribute to noise in the data set. Quite a few works, including \cite{Hemberger_2008}, \cite{Coles_2015}, \cite{lentati}, \cite{mckee}, and \cite{Main} have found at least modest evidence that delays are correlated over time. 
Additionally, as NANOGrav's timing precision reaches the sub 100-ns regime for more pulsars, delays from scattering become significant enough to warrant further investigation and possibly mitigation in many pulsars. Since our current timing pipeline does not account for scattering variability, scattering delays may be partially absorbed in DM fits \citep{NANO_9yr}. Because the frequency scaling of these two noise sources is different, this is an additional source of noise in our timing residuals. \cite{Levin_Scat} also showed that many pulsars do not follow  $\nu^{-4.4}$ frequency scaling, instead exhibiting shallower power law behaviors, further motivating the need to separate the effects of dispersion and scattering.

\par In this paper we aim to expand upon the work done in \cite{Levin_Scat} by examining the effects of scattering on TOAs for pulsars in the NANOGrav 12.5-year data set and looking for deviations from the $\nu^{-4.4}$ frequency scaling. We also explore how scattering can give us insight into other information on MSPs and the ISM, including pulsar transverse velocities and the large-scale structure of the ISM in the Milky Way.

\section{Data}

We used observations from the NANOGrav 12.5-year data set \citep{alam2020nanograv}\footnote{\url{http://data.nanograv.org}}. The data was taken and coherently dedispersed with the FPGA-based spectrometers GUPPI (Green Bank Ultimate Pulsar Processing Instrument) and PUPPI (Puerto Rico Ultimate Pulsar Processing Instrument) at the Green Bank Telescope and the Arecibo Observatory, respectively \citep{DuPlain, Ford}. This process was done on 47 pulsars, 11 of which are new to the 12.5-year data set and consequently were not included in the analysis done by \cite{Levin_Scat}.

\par We reused the results from the \cite{Levin_Scat} analysis and augment it by analyzing $\sim$3.6 years of new data not included in the 9-year data set from both telescopes, with the MJD range for most pulsars spanning approximately 56603--57933 (2013 November--2017 June).  Observations at Arecibo were centered near 1380 MHz using bandwidths of 800 MHz with 1 second subintegrations, while observations at Green Bank were centered near 820 and 1500 MHz using 200 and 800 MHz bandwidths, respectively, with 10 second subintegrations. Observations at both telescopes were divided into 1.56 MHz frequency channels, and were $\sim$30 minutes in length. While the NANOGrav 12.5-year data set also includes 327 MHz, 430 MHz, and 2.1 GHz data from Arecibo, the scintles are generally either too narrow to be frequency resolved at our current resolution in the case of 327 and 430 MHz or either too wide or with an insufficient number of scintles to be properly analyzed given our current observation bandwidth in the case of 2.1 GHz. 

\par All observations began with a polarization calibration scan with a 25 Hz noise diode injection for both polarizations. A flux calibrator, QSO J1445$+$099, is also observed once per epoch per frequency. All of the analyses done in this paper used total intensity profiles, which were made by summing the polarizations of the calibrated data. 

\par As mentioned in \cite{alam2020nanograv}, small timing mismatches in both of these backends led to frequency-reversed ``ghost images" of pulses appearing in the data. This can result in large offsets in residuals if uncorrected. This has been accounted for in the 12.5-year data set, and anything left from the subtraction will negligibly affect the information contained in our dynamic spectra. 

\section{Analysis}\label{analysis}
\subsection{Scintillation Parameters}\label{scint_param}
Following a method similar to \cite{Cordes1986} and identical to that of \cite{Levin_Scat}, we created 2D dynamic spectra from each 820 and 1500 MHz observation of all pulsars in our analysis. To create a dynamic spectrum, we calculate the intensity, $S$, of the pulsar's signal at any given observing frequency, $\nu$, and time, $t$, in that observation by the relation
\begin{equation}
\label{dynspec}
S(\nu,t)=\frac{P_{\rm{on}}(\nu,t)-P_{\rm{off}}(\nu,t)}{P_{\rm{bandpass}}(\nu,t)},
\end{equation}
where ${P}_{{\rm{bandpass}}}$ is the total power of the observation as a function of observing frequency and time, and ${P}_{{\rm{on}}}$ and ${P}_{{\rm{off}}}$ are the power in all on- and off-pulse components of the pulse profile, respectively. After smoothing from 2048 pulse profile bins to 64 bins, we define the on-pulse component as the bins in the summed profile that have an intensity $>5\%$ of the maximum within a continuous window.
\par
\begin{figure}[h!]
\includegraphics[scale=.54]{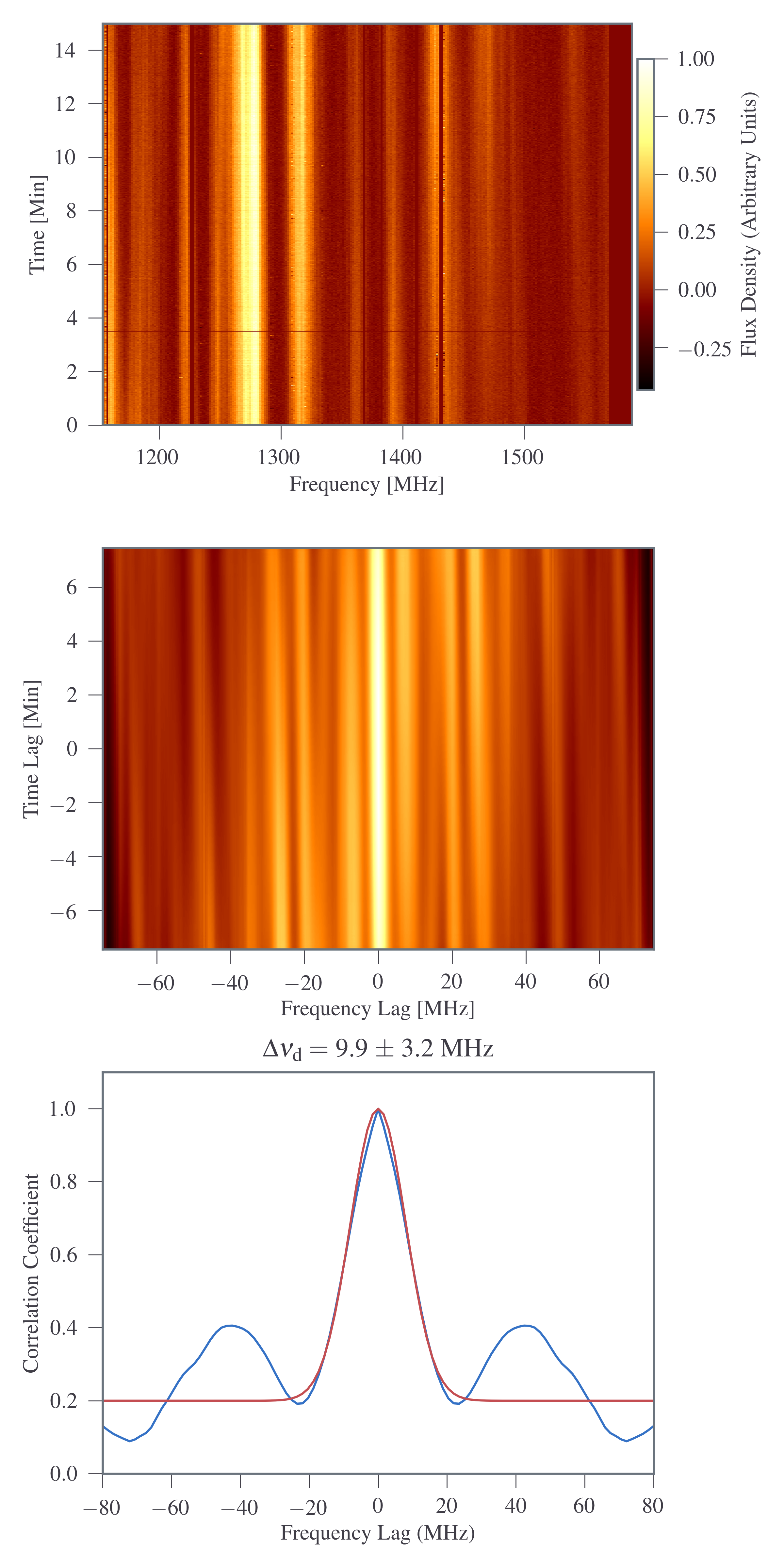}
\centering
\caption{An example stretched dynamic spectra (top), the corresponding 2D ACF (middle), and the resulting values along the time axis of the ACF at lag 0 (bottom, with the 1D ACF in blue and Gaussian fit in red) from  a PSR B1855$+$09 observation  with the Arecibo telescope. The 1$\sigma$ error shown above includes the finite scintle error. Note in this case that the scintles are not fully resolved in time, as is typical in our data because most of our observations are shorter than the scintillation timescales of the pulsars under observation.}
\label{dyn_ex}
\end{figure}

\begin{figure}[h!]
\includegraphics[scale=.54]{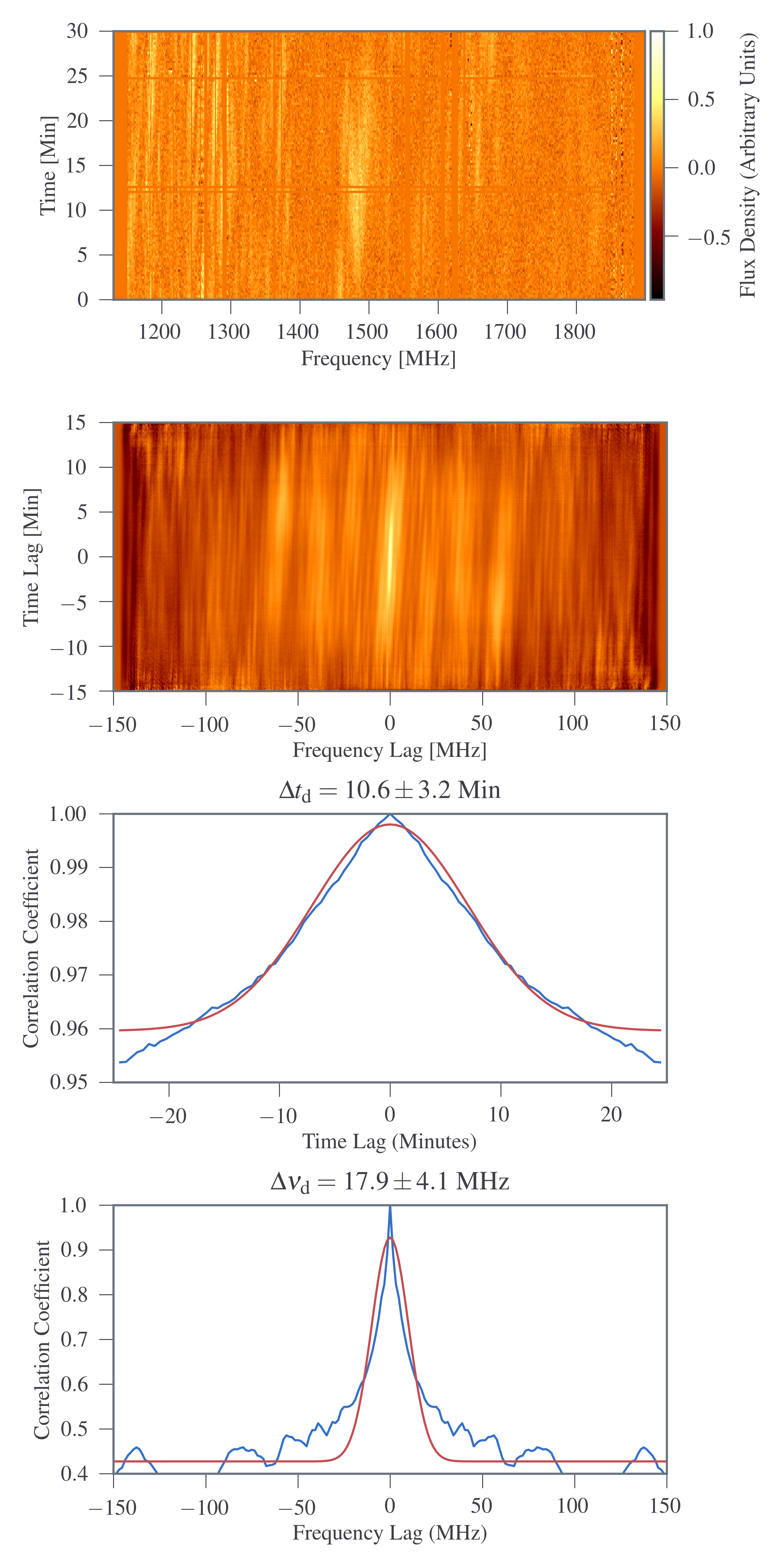}
\centering
\caption{
As in Figure \ref{dyn_ex}, but for an observation of PSR J0636+5128 with the Green Bank telescope. For this pulsar, we are able to measure a scintillation timescale, though it is likely an underestimate due to the short durations of our observations.}
\label{dyn_ex2}
\end{figure}

 These observations were calibrated and excised of radio-frequency interference (RFI) via the median smoothed difference channel zapping algorithm in \textsc{psrchive}'s paz function \citep{2004PASA} and converted into 2D dynamic spectra, such as those seen at the top of Figures \ref{dyn_ex} and \ref{dyn_ex2}. As shown in Figure \ref{dyn_ex}, to determine the scintle sizes at each epoch, we first computed a 2D auto-correlation function (ACF) and summed separately over time and frequency to create a 1D time ACF and a 1D frequency ACF taken at zero time lag and zero frequency lag, respectively. We then fit Gaussian functions to the frequency and time axes at lag 0 of the ACF to obtain estimates for the scintillation bandwidth and timescale, respectively (see Figure \ref{dyn_ex2}).

\par Scattering effects can be estimated based on the size of scintles (maxima) in both time and frequency in a pulsar's dynamic spectra. Here we focus on the scintillation timescale, $\Delta t_\text{d}$, defined as the half-width at $e^{-1}$ of the values along the time axis at ACF lag 0 of the dynamic spectrum's 2D ACF, and the scintillation bandwidth, $\Delta \nu_\text{d}$, defined as the half-width at half-maximum of the values along the frequency ACF at lag 0 of the 2D ACF. The scattering delay, $\tau_\text{d}$, can subsequently be obtained from the scintillation bandwidth via the relation 
\begin{equation}
\label{band_delay}
2\pi \Delta \nu_{\text{d}} \tau_{\text{d}} = C_1,
\end{equation}
where ${C}_{{\rm{1}}}$ is a dimensionless quantity ranging $0.6-1.5$ conditional on the geometry and spectrum of the electron density fluctuations of the medium \citep{Cordes_1998}. In this analysis we assume ${C}_{{\rm{1}}}=1$, as in \cite{Levin_Scat}. We found the results of the 1D and 2D Gaussian fits to the 1D and 2D ACFs, respectively, to be in agreement, and opted to use the 1D ACFs for our analysis since most of the pulsars have scintillation timescales longer than our observation times. If our observations were long enough  to resolve the scintles in both time and frequency, as in \cite{shapiroalbert2019analysis}, we would have used the 2D

\begin{turnpage}
\begin{deluxetable*}{CCCCCCCCCCCCCCCCCC|CC}[ht]
\tablefontsize{\footnotesize}
\tabletypesize{\footnotesize}

\tablecolumns{20}

\tablecaption{Measured Scintillation Parameters \label{table_2}}
\tablehead{ \colhead{} & \multicolumn{12}{C}{\textbf{This work}} & \multicolumn{2}{C}{\textbf{Levin et al. (2016)}} \\ \multicolumn{1}{C}{$\textrm{Pulsar}$} &
\multicolumn{1}{C}{$\overline{\Delta\nu}_{\text{d}}^{1500}$} & \multicolumn{1}{C}{$\overline{\tau}_{\textrm{d}}^{1500}$} & 
\multicolumn{1}{C}{$N^{\textrm{med;1500}}_{\textrm{scint}}$}&
\multicolumn{1}{C}{$N_{\nu}^{1500}$} &
\multicolumn{1}{C}{$\overline{\Delta t}_{\textrm{d}}^{1500}$} & 
\multicolumn{1}{C}{$N_{\textrm{t}}^{1500}$} & 
\multicolumn{1}{C}{$\overline{\Delta\nu}_{\textrm{d}}^{820}$} &
\multicolumn{1}{C}{$\overline{\tau}_{\textrm{d}}^{820}$} &
\multicolumn{1}{C}{$N^{\textrm{med;820}}_{\textrm{scint}}$}&
\multicolumn{1}{C}{$N_{\nu}^{820}$} &
\multicolumn{1}{C}{$\overline{\Delta t}_{\textrm{d}}^{820}$} & 
\multicolumn{1}{C}{$N_{\textrm{t}}^{820}$} & 
\multicolumn{1}{C}{$\overline{\Delta\nu}_{\textrm{d}}^{1500}$} &
\multicolumn{1}{C}{$\overline{\tau}_{\text{d}}^{1500}$} 
 \\  \colhead{} & \multicolumn{1}{C}{\text{(MHz)}} & \multicolumn{1}{C}{\text{(ns)}} & \multicolumn{1}{C}{}& \multicolumn{1}{C}{}  &
\multicolumn{1}{C}{\text{(min)}} & \multicolumn{1}{C}{} & \multicolumn{1}{C}{\text{(MHz)}} & \multicolumn{1}{C}{\text{(ns)}} & \multicolumn{1}{C}{}& \multicolumn{1}{C}{}& \multicolumn{1}{C}{\text{(min)}} & \multicolumn{1}{C}{}& 
\multicolumn{1}{C}{\text{(MHz)}} & \multicolumn{1}{C}{\text{(ns)}} \vspace{-0.15cm}}
\startdata 
\textrm{J}0340{+}4130  & $<$3.3 $\pm$ 1.3  & $>$33 $\pm$ 17 &\textrm{32}&\textrm{10} & \textrm{---} & \textrm{---} & \textrm{---}&\textrm{---} &\textrm{---} & \textrm{---}&\textrm{---} &\textrm{---}   & 9 $\pm$ 3 & 15 $\pm$ 6\\
\textrm{J}0613\text{--}0200   & 4 $\pm$ 3 & 16 $\pm$ 11 &\textrm{21}&\textrm{40} & 10 $\pm$ 4 & \textrm{4} & $<$1.6 $\pm$ 0.2 & $>$78 $\pm$ 17  & \textrm{25}&\textrm{19} & 6 $\pm$ 3 &\textrm{7}   & 11 $\pm$ 4 & 12 $\pm$ 4\\
\textrm{J}0636{+}5128  & 7 $\pm$ 4  & 12 $\pm$ 9 & \textrm{17}&\textrm{24} &9 $\pm$ 3 & \textrm{13} & $<$1.5 $\pm$ 1.0 &$>$31 $\pm$ 37 &\textrm{23}&\textrm{26}&8 $\pm$ 3 & \textrm{17}&\textrm{---} & \textrm{---}\\
\textrm{J}0740{+}6620 & 81 $\pm$ 30 & 1.5 $\pm$ 0.5 & \textrm{2}&\textrm{4} &\textrm{---} & \textrm{---} & 18 $\pm$ 9 & 5 $\pm$ 3 &\textrm{3}&\textrm{16}&\textrm{---} & \textrm{---}&\textrm{---} & \textrm{---}\\ 
\textrm{J}0931\text{$--$}1902  & 20 $\pm$ 4  & 7 $\pm$ 2 & \textrm{9}&\textrm{4} &$>30$ & \textrm{---} &\textrm{---} & \textrm{---} &\textrm{---} & \textrm{---} &\textrm{---} &\textrm{---} & \textrm{50}  & \textrm{3}\\
\textrm{J}1012{+}5307  & 40 $\pm$ 18  & 4 $\pm$ 2 & \textrm{6}&\textrm{1} &\textrm{---} & \textrm{---}&\textrm{---} &\textrm{---} & \textrm{---} & \textrm{---} &\textrm{---} &\textrm{---}  & 66 $\pm$ 6 & 2.5 $\pm$ 0.1 \\
\textrm{J}1024\text{$--$}0719  & 36 $\pm$ 3 & 4.4 $\pm$ 0.3 &\textrm{5}&\textrm{3} &$>30$ & \textrm{---} & 10 $\pm$ 3 & 11 $\pm$ 4 & \textrm{4}&\textrm{6}  & 17 $\pm$ 4 &\textrm{1} & 47 $\pm$ 18 & 3 $\pm$ 1 \\
\textrm{J}1125{+}7819  & 50 $\pm$ 20  & 1.9 $\pm$ 0.9 & \textrm{3} & \textrm{25} & \textrm{---} & \textrm{---} & 7 $\pm$ 3 & 17 $\pm$ 8  & \textrm{6} & \textrm{37}  & 12 $\pm$ 4 & \textrm{23} & \textrm{---} & \textrm{---}\\
\textrm{J}1455\text{$--$}3330   & 43 $\pm$ 12 & 3.2 $\pm$ 0.9 &\textrm{5}&\textrm{5}  &\textrm{---} & \textrm{---} & 5 $\pm$ 1 & 27 $\pm$ 7 & \textrm{8}&\textrm{12}  &\textrm{---} &\textrm{---}  & 70 $\pm$ 18 & 4 $\pm$ 1\\
\textrm{J}1614\text{$--$}2230   & 6 $\pm$ 3 & 16 $\pm$ 6 &\textrm{25}&\textrm{16}& 9 $\pm$ 3 & \textrm{1}  & $<$2.4 $\pm$ 0.6 & $>$33 $\pm$ 14 &\textrm{17}&\textrm{3} &5 $\pm$ 1 & \textrm{1} & 9 $\pm$ 3 &16 $\pm$ 5 \\
\textrm{J}1640{+}2224   & 50 $\pm$ 26 & 1.8 $\pm$ 0.9 &\textrm{4}&\textrm{17}& $>30$ & \textrm{---} & \textrm{---} & \textrm{---} &\textrm{---}& \textrm{---} &\textrm{---} & \textrm{---}& 56 $\pm$ 15 & 3 $\pm$ 1  \\
\textrm{J}1713{+}0747   & 23 $\pm$ 15 & 3 $\pm$ 2 &\textrm{6}&\textrm{72}& $>30$ & \textrm{---} & \textrm{---} & \textrm{---} &\textrm{---}&\textrm{---} &\textrm{---} & \textrm{---}& 21 $\pm$ 9 & 7 $\pm$ 2  \\
\textrm{J}1738{+}0333   & 19 $\pm$ 8 & 6 $\pm$ 3 & \textrm{9}&\textrm{18}&$>30$ & \textrm{---} &\textrm{---} & \textrm{---} &\textrm{---}& \textrm{---}&\textrm{---} &\textrm{---} & 17 $\pm$ 8 & 9 $\pm$ 2 \\
\textrm{J}1744\text{$--$}1134   & 33 $\pm$ 13 & 3.3 $\pm$ 1.6 & \textrm{5}&\textrm{18}& >30 &\textrm{---} & 10 $\pm$ 3 & 12 $\pm$ 4 &\textrm{5}& \textrm{21}& \textrm{---} & \textrm{---}  & 42 $\pm$ 9 & 4 $\pm$ 1\\
\textrm{J}1853{+}1303   & 11 $\pm$ 6 & 8 $\pm$ 4 &\textrm{14} &\textrm{6} & $>30$  &\textrm{---} & \textrm{---} & \textrm{---} &\textrm{---}& \textrm{---}&\textrm{---} & \textrm{---} & 13 $\pm$ 5 & 12 $\pm$ 5
\\
\textrm{B}1855{+}09  & 10 $\pm$ 5 & 8 $\pm$ 4  &\textrm{11}&\textrm{25} & $>30$ &\textrm{---}& 9 $\pm$ 3 & 18 $\pm$ 6 & 5 &\textrm{1}& \textrm{---} & \textrm{---} & 5 $\pm$ 2 & 21 $\pm$ 10
\\
\textrm{J}1909\text{$--$}3744   & 28 $\pm$ 13 & 4 $\pm$ 2 &\textrm{6}& \textrm{89} & \textrm{---} & \textrm{---} & <4 $\pm$ 1 & >21 $\pm$ 12 &\textrm{10}&\textrm{18}& 6 $\pm$ 2 & \textrm{5} & 39 $\pm$ 15 & 5 $\pm$ 2
\\
\textrm{J}1910{+}1256   & 2.3 $\pm$ 0.8  & 47 $\pm$ 23 & \textrm{71}&\textrm{17} &$>30$ &\textrm{---} & $<$3.4 $\pm$ 2.6 &$>$90 $\pm$ 44 & \textrm{26}&\textrm{3}& $>30$ & \textrm{---}& 2.3 $\pm$ 0.9 &58 $\pm$ 17 
\\
\textrm{J}1918\text{$--$}0642  & 9 $\pm$ 3 & 13 $\pm$ 5 &\textrm{16}&\textrm{28} & $>30$ & \textrm{---} & \textrm{---} & \textrm{---} &\textrm{---}&\textrm{---}&\textrm{---} & \textrm{---} & 15 $\pm$ 5 & 10 $\pm$ 3
\\
\textrm{J}1923{+}2515   & 18 $\pm$ 4 & 2.4 $\pm$ 1.6 &\textrm{5}&\textrm{8}& \textrm{---} &\textrm{---}& \textrm{---} &\textrm{---}& \textrm{---} & \textrm{---}&\textrm{---} & \textrm{---} & 22 $\pm$ 10 & 6 $\pm$ 1
\\
\textrm{B}1937{+}21   & $<$1.5 $\pm$ 0.8 & $>$ 76 $\pm$ 43 &\textrm{30}&\textrm{43}& 9 $\pm$ 3 &\textrm{23} & \textrm{---} &\textrm{---}& \textrm{---} &\textrm{---}&\textrm{---} &\textrm{---} & 2.8 $\pm$ 1.3 & 44 $\pm$ 21
\\
\textrm{J}1944{+}0907  & 8 $\pm$ 6 & 7 $\pm$ 6 & 17 & 28 & \text{---} & \text{---}  & 3 $\pm$ 3 & 5 $\pm$ 7 & 14 & 3 & \text{---} & \text{---}  & 11 $\pm$ 5 & 10 $\pm$ 6 \\  
\textrm{J}2010\text{$--$}1323   & 8 $\pm$ 3 & 13 $\pm$ 6 & \textrm{18} &\textrm{29} & 7 $\pm$ 6 &\textrm{2} &\textrm{---} & \textrm{---} &\textrm{---}&  \textrm{---}&\textrm{---} &\textrm{---} & 7 $\pm$ 2 & 19 $\pm$ 6
\\
\textrm{J}2043{+}1711   & 56 $\pm$ 27 & 1.8 $\pm$ 0.9 &\textrm{4}&\textrm{2}& $>30$ &\textrm{---} & \textrm{---} & \textrm{---}& \textrm{---}&\textrm{---}&\textrm{---} & \textrm{---} & \textrm{86}&\textrm{2}\\
\textrm{J}2145\text{$--$}0750   & 43 $\pm$ 11 & 3 $\pm$ 1 &\textrm{4}&\textrm{7} & $>30$ & \textrm{---} & 7 $\pm$ 4 & 9 $\pm$ 5 &\textrm{5}&\textrm{19}&$>30$ &\textrm{---} & 48 $\pm$ 13 & 2.8 $\pm$ 0.7
\\
\textrm{J}2229{+}2643  & 46 $\pm$ 15 & 2.9 $\pm$ 0.8 & \textrm{4} & \textrm{6} & \textrm{---} & \textrm{---} & \textrm{---} & \textrm{---} & \textrm{---} & \textrm{---} & \textrm{---} & \textrm{---} & \textrm{---} & \textrm{---}\\
\textrm{J}2302{+}4442   & 8 $\pm$ 7 & 16 $\pm$ 8 & \textrm{20}&\textrm{15} & \textrm{---} & \textrm{---} & $<$ 1.5 $\pm$ 0.2 &$>$ 37 $\pm$ 15 &\textrm{26} &\textrm{20}&\textrm{---} & \textrm{---} & 10 $\pm$ 2 & 14 $\pm$ 3 \\
\textrm{J}2317{+}1439  & 46 $\pm$ 14  & 2.8 $\pm$ 0.8 & \textrm{5}&\textrm{13} &$>30$ & \textrm{---} & 12 $\pm$ 6 &13 $\pm$ 5 &\textrm{5}&\textrm{1}&\textrm{---} & \textrm{---} & \textrm 42 $\pm$ 12 &3 $\pm$ 1
\enddata
\tablecomments{All parameters with bars ($\overline{\tau}_{\rm d}$, $\overline{\Delta\nu}_{\rm d}$, etc.) represent ensemble weighted averages of the individual measurements, with 1$\sigma$ errors shown. $N^{\textrm{med}}_{\textrm{scint;}}$ represents the median number of scintles and $N_{\textrm{t}}$ and $N_{\nu}$ indicate the number of estimates made for that quantity. $\tau_{\textrm{d}}$ values represent the scattering delays, while $\Delta \nu_{\textrm{d}}$ values represent scintillation bandwidths. All measurements and errors have been rounded to the last significant digit shown. Values with only one measurement use their measured uncertainties as opposed to weighted errors. Due to our short observation lengths, it is likely that all of our $\overline{\Delta t_{\rm d}}$ values are biased lower than their true averages.}
\end{deluxetable*}
\end{turnpage}
\clearpage
 
\begin{deluxetable*}{CCC|CCC|CCC}

\tablewidth{0pt}
\tablecolumns{9}

\tablecaption{NE2001 Electron Density Model Predicted Scintillation Parameters \label{table_1}}
\tablehead{ \colhead{Pulsar} & \colhead{Period} & \colhead{DM} & \colhead{$\tau_{\text{d;1500}}^{\text{NE2001}}$} &  \colhead{$\Delta\nu^{\text{NE2001}}_{\text{d;1500}}$} &  \colhead{$\Delta\text{t}^{\text{NE2001}}_{\text{d;1500}}$} & \colhead{$\tau_{\text{d;820}}^{\text{NE2001}}$} &  \colhead{$\Delta\nu^{\text{NE2001}}_{\text{d;820}}$} & 
\colhead{$\Delta\text{t}^{\text{NE2001}}_{\text{d;820}}$} \\ \colhead{} & \colhead{\text{(ms)}} & \colhead{\text{(pc cm$^{-3}$)}} & \colhead{\text{(ns)}} & \colhead{\text{(MHz)}} & \colhead{\text{(min)}} & \colhead{\text{(ns)}} & \colhead{\text{(MHz)}} &  \colhead{(min)}   \vspace{0.05cm}}
\startdata
\text{J}0023{$+$}0923 &  3.05 &  14.3 &  2.5 &  42.5 &  17.4 &  40 &  9.1 &  13.2 \\
 \text{J}0030{$+$}0451 &  {4.87} &  {4.3} &  {0.06} &  {1900} &  {710} &  {0.8} &  {405} &  {540} \\
 \text{J}0340{$+$}4130  &   {3.29}  &   {49.6}  &   {50}  &   {2.1}  &   {17.1}  &  {700} &  {0.5} &  {12.9}\\
 \text{J}0613\text{$--$}0200  &   {3.06}  &   {38.8}  &   {20}  &   {6.4}  &   {16.4}  &  {230} &  {1.4} &  {12.5}\\
 \text{J}0636{$+$}5128 &  {2.87} &  {11.1} &  {1.9} &  {55.1} &  {95} &  {30} &  {11.8} &  {72} \\
 \text{J}0645{$+$}5158 &  {8.85} &  {18.2} &  {6.1} &  {17.2} &  {20.3} &  {90} &  {3.7} &  {15.4} \\
 \text{J}0740{$+$}6620 &  {2.89} &  {15.0} &  {3.2} &  {33.0} &  {12.6} &  {50} &  {7.1} &  {9.5} \\
 \text{J}0931\text{$--$}1902  &   {4.64}  &   {41.5}  &   {20}  &   {4.3}  &   {30.5}  &  {350} &  {0.9} &  {23.2}\\
 \text{J}1012{$+$}5307  &   {5.26}  &   {9.0}  &   {1.2}  &   {88}  &   {16.6}  &  {20} &  {18.8} &  {12.6}\\
 \text{J}1024\text{$--$}0719  &   {5.16}  &   {6.5}  &   {0.2}  &   {610}  &   {16.8}  &  {2.4} &  {130} &  {12.7}\\
 \text{J}1125{$+$}7819 &  {4.2} &  {12.0} &  {1.5} &  {70.1} &  {21.9} &  {20} &  {15} &  {16.7} \\
 \text{J}1453{$+$}1902 &  {5.79} &  {14.1} &  {3.1} &  {34.1} &  {26.4} &  {40} &  {7.3} &  {20.1} \\   
 \text{J}1455\text{$--$}3330  &   {7.99}  &   {13.6}  &   {1.0}  &   {110}  &   {103}  &  {10} &  {23.3} &  {78}\\
 \text{J}1600\text{$--$}3053 &  {3.60} &  {52.3} &  {90} &  {1.1} &  {5.1} &  {1,300} &  {0.24} &  {3.9} \\
 \text{J}1614\text{$--$}2230  &   {3.15}  &   {34.5}  &   {30}  &   {3.6}  &   {5.4}  &  {420} &  {0.8} &  {4.1}\\
 \text{J}1640{$+$}2224  &   {3.16}  &   {18.4}  &   {5.8}  &   {18.1}  &   {15.3}  &  {80} &  {3.9} &  {11.7}\\ 
 \text{J}1643\text{$--$}1224 &  {4.62} &  {62.3} &  {90} &  {1.2} &  {5.9} &  {1,300} &  {0.25} & \textrm{4.4} \\
 \text{J}1713{$+$}0747  &   {4.57}  &   {16.0}  &   {4.1}  &   {25.6}  &   {37.7}  &  {60} &  {5.5} &  {28.6}\\ 
 \text{J}1738{$+$}0333 &  {5.85} &  {33.8} &  {20} &  {5.00} &  {5.5} &  {300} &  {1.1} &  {4.1} \\
 \text{J}1741{$+$}1351 &  {3.75} &  {24.2} &  {0.7} &  {160} &  {48.8} &  {9.3} &  {35} &  {37.1} \\
 \text{J}1744\text{$--$}1134  &   {4.08}  &   {3.1}  &   {0.02}  &   {5,700}  &   {335}  &  {0.3} &  {1,200} &  {253}\\ 
 \text{J}1747\text{$--$}4036 &  {1.65} &  {153.0} &  {2,400} &  {0.04} &  {2.4} &  {30,000} &  {0.01} &  {1.8} \\
 \text{J}1832\text{$--$}0836 &  {2.72} &  {28.2} &  {20} &  {5.9} &  {2.9} &  {250} &  {1.3} &  {2.2}\\
 \text{J}1853{$+$}1303  &   {4.09}  &   {30.6}  &   {6.2}  &   {52.2}  &   {16.3}  &  {90} &  {3.6} &  {39.7}\\
 \text{B}1855{$+$}09  &   {5.36}  &   {13.3}  &   {2.2}  &   {4.9}  &   {53.9}  &  {30} &  {10.4} &  {41.1}\\
 \text{J}1903{$+$}0327 &  {2.15} &  {297.5} &   {240,000} &  {0.0004} &  {0.5} &  {3,400,000} &  {0.00009} &  {0.5} \\
 \text{J}1909\text{$--$}3744  &   {2.95}  &   {10.4}  &   {1.5}  &   {68}  &   {7.9}  &  {20} &  {14.6} &  {6.0}\\
 \text{J}1910{$+$}1256  &   {4.98}  &   {38.1}  &   {8.4}  &   {12.5}  &   {19.0}  &  {120} &  {2.7} &  {14.5}\\ 
 \text{J}1911{$+$}1347  &   {4.63}  &   {31.0}  &   {5.0}    &   {20.8}   &    {21.7} &  {70}   &   {4.5}   &  {16.6} \\ 
 \text{J}1918\text{$--$}0642  &   {7.65}  &   {26.6}  &   {10}  &   {10.0}  &   {21.3}  &  {150} &  {2.1} &  {16.1}\\ 
 \text{J}1923{$+$}2515  &   {3.88}  &   {18.9}  &   {1.7}  &   {61}  &   {22.7}  &  {20} &  {13} &  {17.2}\\ 
 \text{B}1937{$+$}21  &   {1.56}  &   {71.0}  &   {130}  &   {0.8}  &   {62.0}  &  {190} &  {0.2} &  {46.1}\\
 \text{J}1944{$+$}0907 &  {5.19} &  {24.3} &  {2.9} &  {36} &  {9.8} &  {40} &  {7.7} &  {7.5}\\
 \text{J}1946{$+$}3417 &  {3.17} &  {110.2} &  {210} &  {0.5} &  {2.0} &  {3,100} &  {0.1} &  {1.5} \\
 \text{B}1953{$+$}29 &  {6.13} &  {104.5} &  {240} &  {0.43} &  {4.3} & {3,500} &  {0.09} &  {3.3} \\
 \text{J}2010\text{$--$}1323 &   {5.22}  &   {22.2}  &   {6.7}  &   {15.7}  &   {16.7}  &  {100} &  {3.4} &  {12.8}\\
 \text{J}2017{$+$}0603 &  {2.90} &  {23.9} &  {3.8} &  {27.9} &  {52.7} &  {50} &  {6.0} &  {40.1} \\
 \text{J}2033{$+$}1734 &  {5.95} &  {25.1} &  {3.0} &  {35.4} &  {53.7} &  {40} &  {7.6} &  {40.9} \\
 \text{J}2043{$+$}1711  &   {2.38}  &   {20.7}  &   {2.0}  &   {51}  &   {30.8}  &  {30} &  {11} &  {23.8}\\
 \text{J}2145\text{$--$}0750  &   {16.05}  &   {9.0}  &   {0.5}  &   {200}  &   {60.5}  &  {7.5} &  {42.3} &  {45.9}\\
 \text{J}2214{$+$}3000 &  {3.12} &  {22.5} &  {3.1} &  {33.8} &  {41.2} &  {40} &  {7.2} &  {31.2} \\
 \text{J}2229{$+$}2643 &  {2.98} &  {22.7} &  {4.2} &  {25.2} &  {18.8} &  {60} &  {5.4} &  {14.3} \\
 \text{J}2234{$+$}0611 &  {3.58} &  {10.8} &  {0.8} &  {136} &  {14.9} &  {10} &  {29.1} &  {11.3} \\
 \text{J}2234{$+$}0944 &  {3.63} &  {17.8} &  {3.3} & {31.5} &  {9.9} &  {0.5} &  {6.7} &  {7.5}\\
 \text{J}2302{$+$}4442  &   {5.19}  &   {13.8}  &   {0.9}  &   {120}  &   {103}  &  {10} &  {26.4} &  {79}\\
 \text{J}2317{$+$}1439  &   {3.45}  &   {21.9}  &   {1.9}  &   {54}  &   {57.4}  &  {30} &  {11.6} &  {43.5}\\
 \text{J}2322{$+$}2057 &  {4.81} &  {13.4} &  {1.0} &  {104} &  {25.3} &  {10} &  {22.3} &  {19.2}
\enddata
\tablecomments{Predictions of scattering delays, scintillation bandwidths, and scintillation timescales made by the NE2001 electron density model \citep{NE2001}. We calculated $\Delta t_{\textrm{d}}$ values using transverse velocities derived from proper motions rather than the 100 km/s transverse velocity that NE2001 assumes. DM distances were used for calculating transverse velocities if current parallax measurements were negative or if errors on parallax measurements were larger than around 25\%.}
\vspace{-1.05cm}
\end{deluxetable*}

\noindent ACFs since there would have been a sufficient number of scintles within the observing time.
\par Uncertainties in our scattering delay measurements are an addition in quadrature of the finite scintle error, which can be approximated as
\begin{equation}
\label{finite_scintle}
\begin{split}
\epsilon & \approx \tau_{\text{d}} N_{\rm{scint}}^{\nicefrac{-1}{2}} \\
& \approx \tau_{\text{d}}[(1+\eta_{\text{t}}T/\Delta t_\text{d})(1+\eta_\nu B/\Delta \nu_\text{d})]^{\nicefrac{-1}{2}},
\end{split}
\end{equation}
where ${N}_{{\rm{scint}}}$ is the number of scintles, $T$ and $B$ are total integration time and total bandwidth, respectively, and ${\eta }_{{\rm{t}}}$ and ${\eta}_{\nu}$ are filling factors ranging from $0.1-0.3$ depending on the definitions of characteristic timescale and scintillation bandwidth, and in our case both set to 0.2 \citep{Finite_Scint}.

Since all of our observations are at most 30 minutes in length, generally, $T< \Delta t_\text{d}$, and as a result $1+\eta_{\text{t}}T/\Delta t_\text{d} \approx 1$. This allows us to rely exclusively on the observing and scintillation bandwidths when calculating $\epsilon$. It should be noted that this is a conservative approach that can only overestimate our reported uncertainties: if $T \gtrsim \Delta t_d$, then Equation \ref{finite_scintle} shows that we will underestimate $N_{\rm scint}$ and overestimate $\epsilon$  \citep{Levin_Scat}. 

\par The limited frequency resolution also introduces selection effects into our data. We are unable to reliably measure scintillation bandwidths smaller than our 1.5-MHz wide channel widths. As a result, some of the average scattering delays quoted for the most highly scattered pulsars are lower limits. Due to this bias, we treat an individual scattering measurement $\gtrsim 30$ ns (about three channel widths) as a lower limit for a given epoch.

\par To account for the wide bandwidth of our observations, we assumed a Kolmogorov medium to stretch each observation's dynamic spectrum by
 by $\nu^{4.4}$, with the frequency axis being re-scaled to reference frequencies of 820 and 1500 MHz for the respective observing bands as in \cite{Levin_Scat}. If the scaling index used for the stretching is correct, then all scintles in a given dynamic spectrum should be roughly equal in size. Understretching a spectrum (i.e., the epoch has a true index steeper than 4.4) would result in an overestimation of the true scattering delay at a given epoch, and vice-versa for overstretching. In some cases, this stretching can result in scintles at the lower end of the band appearing wider than the width of an individual channel despite physically being narrower and vice-versa. This means we can then derive scintillation bandwidths smaller than our channel widths. However, we still interpret our measurements as averages over entire observing bands. As a result, our upper limits will still be determined based on the unstretched channel width at the center of each band. 
 
We have carried out simulations to determine the errors due to the assumption that $-4.4$ is the proper index to use for stretching by placing discrete scintles with some characteristic frequency scaling, stretching using the $-4.4$ scaling, and then measuring the resultant frequency scaling. For index values between $-1$ and $-5$, the average fractional error due to stretching by an incorrect index is roughly 10\%. Furthermore, the index values measured will always be biased high, i.e. a flatter scaling than $-4.4$.
 
\begin{deluxetable*}{CCCCC|CCCCC}[ht]

\centering

\tablecolumns{10}

\tablecaption{Comparison with Previously Published Scintillation Parameters \label{table_compare}}
\tablehead{ \colhead{} & \multicolumn{4}{C}{\textbf{This work}} & \multicolumn{5}{C}{\textbf{Previously Published Values}} \\ \multicolumn{1}{C}{\textrm{Pulsar}} & \multicolumn{1}{C}{$\overline{\tau}$_{\text{d}}} & 
\multicolumn{1}{C}{$\overline{\Delta \nu}$_{\text{d}}} &
\multicolumn{1}{C}{$\overline{\Delta t}$_{\text{d}}} &
\multicolumn{1}{C}{\nu} & 
\multicolumn{1}{C}{$\tau$_{\text{d, scaled}}} & 
\multicolumn{1}{C}{$\Delta \nu$_{\text{d, scaled}}} &
\multicolumn{1}{C}{$\Delta t$_{\text{d, scaled}}} &
\multicolumn{1}{C}{$\nu_{\textrm{original}}$} &
\multicolumn{1}{C}{\textrm{Reference}}
\\  \colhead{} & \multicolumn{1}{C}{\text{(ns)}} & \multicolumn{1}{C}{\text{(MHz)}} & \multicolumn{1}{C}{\text{(min)}} & \multicolumn{1}{C}{\text{(MHz)}} & \multicolumn{1}{C}{\text{(ns)}}& \multicolumn{1}{C}{\text{(MHz)}} & \multicolumn{1}{C}{\text{(min)}} & \multicolumn{1}{C}{\text{(MHz)}}& \multicolumn{1}{C}{} \vspace{0.15cm}}
\startdata 
\text{J}0340{+}4130 & $>$33 $\pm$ 17 & $<$3.3 $\pm$ 1.3  & \textrm{---} & 1500 & 43 $\pm$ 2 & 3.7 $\pm$ 0.2 & 16 $\pm$ 1 & 1500 & \text{\cite{shapiroalbert2019analysis}} \\ 
\text{J}0613\text{$--$}0200 & 16 $\pm$ 11 & 4 $\pm$ 3 & 10 $\pm$ 4 & 1500 & 61 & 3 & 26 & 1369 & \text{\cite{coles_2010}} \\
\rotatebox{90}{$\,=$} & \rotatebox{90}{$\,=$} & \rotatebox{90}{$\,=$} & \rotatebox{90}{$\,=$} & \rotatebox{90}{$\,=$} &  97$^{*}$ & 2$^{*}$ & 75$^{*}$ & 1500$^{*}$ & \text{\cite{keith_2013}} \\
\rotatebox{90}{$\,=$} & \rotatebox{90}{$\,=$} & \rotatebox{90}{$\,=$} & \rotatebox{90}{$\,=$} & \rotatebox{90}{$\,=$} &  21 $\pm$ 1 & 7.7 $\pm$ 0.5 & 11 $\pm$ 1 & 1500 & \text{\cite{shapiroalbert2019analysis}} \\
\rotatebox{90}{$\,=$} & \rotatebox{90}{$\,=$} & \rotatebox{90}{$\,=$} & \rotatebox{90}{$\,=$} & \rotatebox{90}{$\,=$} &  50\text{$--$}200$^{\dagger}$ & 0.8\text{$--$}3.2 $^{\dagger}$ & \textrm{---} & 1350$^{\dagger}$ & \text{\cite{Main}}$^{\dagger}$ \\
\text{J}1024\text{$--$}0717 & 10 $\pm$ 3 & 11 $\pm$ 4 & 17 $\pm$ 4 & 820 & 5 & 33 & 56 & 685 & \text{\cite{coles_2010}} \\
\rotatebox{90}{$\,=$}  & 4.4 $\pm$ 0.3 & 36 $\pm$ 3 & $>$30 & 1500 & 0.59$^{*}$ & 268$^{*}$ & 70$^{*}$ & 1500$^{*}$ & \text{\cite{keith_2013}} \\
\text{J}1614\text{$--$}2230 & 16 $\pm$ 6 & 6$\pm$ 3 & 9 $\pm$ 3 & 1500 & 29 $\pm$ 2 & 5.5 $\pm$ 0.4 & 12 $\pm$ 1 & 1500 & \text{\cite{shapiroalbert2019analysis}} \\ 
\text{J}1713{+}0747 & 3 $\pm$ 2 & 23 $\pm$ 15 & $>$30 & 1500 & 7$^{*}$ & 24$^{*}$ & 48$^{*}$ & 1500$^{*}$ & \text{\cite{keith_2013}} \\
\text{J}1744\text{$--$}1144 & 12 $\pm$ 4 & 10 $\pm$ 3 & \textrm{---} & 820 & 27 & 6.0 & 26 & 660 & \text{\cite{Johnston_98}} \\
\rotatebox{90}{$\,=$} &  \rotatebox{90}{$\,=$} & \rotatebox{90}{$\,=$} & \rotatebox{90}{$\,=$} &  \rotatebox{90}{$\,=$} & 6 & 28 & 58 &  685 & \text{\cite{coles_2010}} \\
\rotatebox{90}{$\,=$} & 3.3 $\pm$ 1.6 & 33 $\pm$ 13 & >30 & 1500 & 3$^{*}$ & 59$^{*}$ & 35$^{*}$ & 1500$^{*}$ & \text{\cite{keith_2013}} \\
\text{B}1855{+}09 & 18 $\pm$ 6 & 9 $\pm$ 3 & \textrm{---} & 820 & 16 & 10 & 21 & 685 & \text{\cite{coles_2010}} \\
\rotatebox{90}{$\,=$} & 8 $\pm$ 4 & 10 $\pm$ 5 & $>$ 30 &  1500 & 13 & 12 & 37 & 1369 & \text{\cite{coles_2010}} \\
\rotatebox{90}{$\,=$} & \rotatebox{90}{$\,=$} & \rotatebox{90}{$\,=$} & \rotatebox{90}{$\,=$}  & \rotatebox{90}{$\,=$} & 29$^{*}$ & 6$^{*}$ & 24 & 1500$^{*}$ & \text{\cite{keith_2013}} \\
\text{J}1909\text{$--$}3744 & >21 $\pm$ 12 & <4 $\pm$ 2 & 6 $\pm$ 2 & 820 & 10 & 17 & 41 &  685 & \text{\cite{coles_2010}} \\
\rotatebox{90}{$\,=$} & 4 $\pm$ 2 & 28 $\pm$ 13   & \textrm{---} & 1500  & 2 $\pm$ 0.8 & 81 $\pm$ 31 & > 82 & 1500 & \text{\cite{shapiroalbert2019analysis}}\\
\rotatebox{90}{$\,=$} & \rotatebox{90}{$\,=$} & \rotatebox{90}{$\,=$} & \rotatebox{90}{$\,=$} & \rotatebox{90}{$\,=$} & 4$^{*}$ & 37$^{*}$ & 38$^{*}$ & 1500$^{*}$ & \text{\cite{keith_2013}} \\
\text{B}1937{+}21 & $>$76$\pm$43 & $<$1.5 $\pm$ 0.8 & 9 $\pm$ 3 & 1500 & 48 & 3 & 7 & 1369 & \text{\cite{coles_2010}}\\
\rotatebox{90}{$\,=$}  &  \rotatebox{90}{$\,=$}  & \rotatebox{90}{$\,=$} & \rotatebox{90}{$\,=$} & \rotatebox{90}{$\,=$} & 127 & 1 & 8 & 1400 & \text{\cite{cordes_1937}}\\
\rotatebox{90}{$\,=$}  &  \rotatebox{90}{$\,=$}  & \rotatebox{90}{$\,=$} & \rotatebox{90}{$\,=$} & \rotatebox{90}{$\,=$} & 130$^{*}$ & 1$^{*}$ & 6$^{*}$ & 1500 & \text{\cite{keith_2013}} \\
\text{J}2145\text{$--$}0750 & 9 $\pm$ 5 & 7 $\pm$ 4 & $>$30 & 820 & 6 & 25 & 58 & 685 & \text{\cite{coles_2010}}\\
\rotatebox{90}{$\,=$} & 3.3 $\pm$ 0.8 & 43 $\pm$ 11 & $>$30 & 1500 & 0.82$^{*}$ & 194$^{*}$ & 57$^{*}$ & 1500$^{*}$ & \text{\cite{keith_2013}}
\enddata
\tablecomments{Published values were reported at observing frequency $\nu_{\textrm{original}}$ and converted to the values at the frequency closest to the one used in our paper using a scaling index of $\xi = -4.4$. For consistency, we only examined scintillation measurements taken at comparable frequencies. Additionally, as is discussed later, we do not find consistent scaling behavior along the LOSs to different pulsars, and as a result of this variability we felt that attempting to scale scintillation measurements taken at largely disjointed frequencies would not make for a sound comparison. Our scintillation timescale averages are lower than many of the previously measured values at similar frequencies, further providing evidence for the possibility of our timescale averages being biased low as a result of our short observation lengths. \\ $^{*}$Only values that were already scaled were reported in the original publication. \\ $^{\dagger}$No average value was quoted, but scattering delays were found within this range.}
\end{deluxetable*}

\subsection{Scaling Behavior}\label{scale_beh}
\par
\begin{figure}[!ht]
\includegraphics[scale=.52]{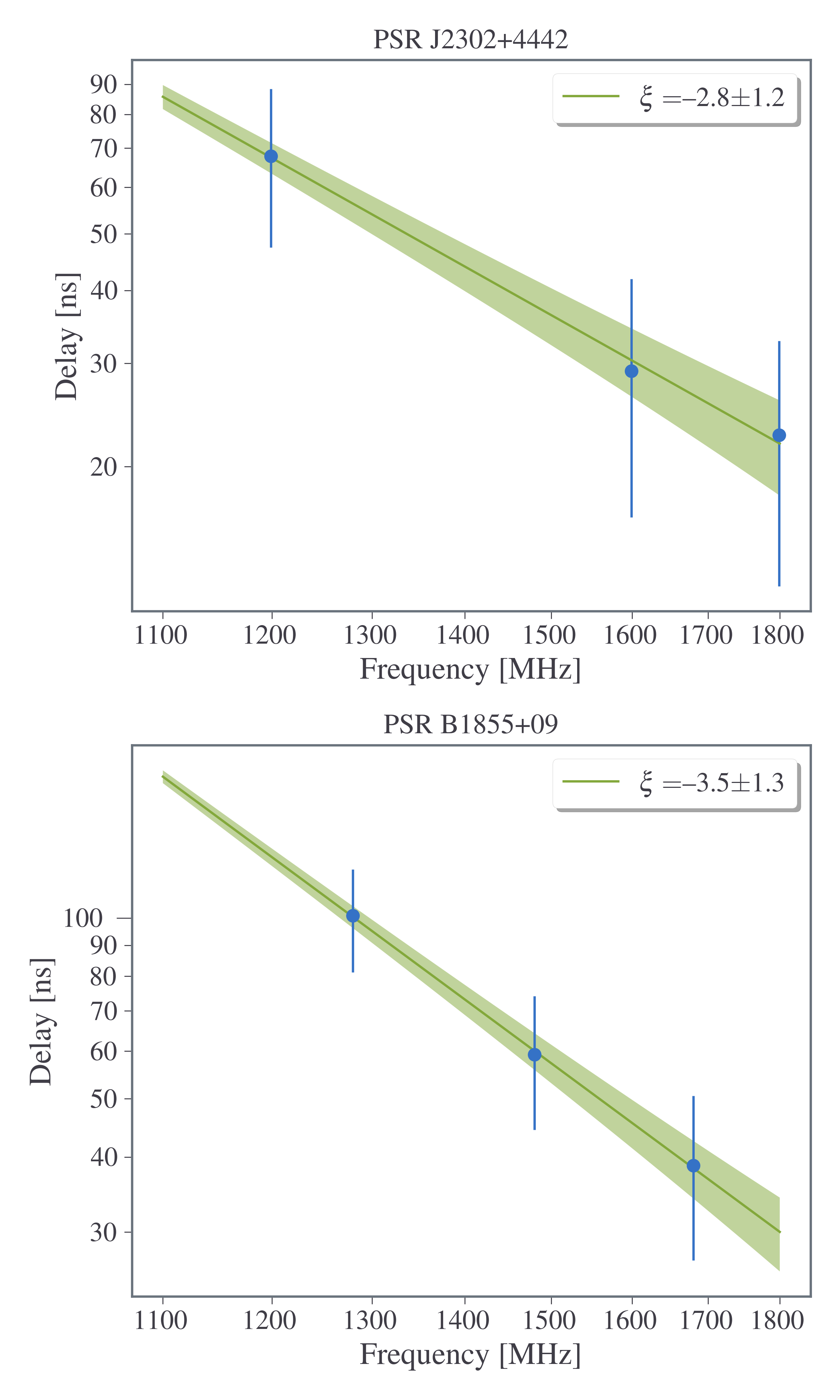}
\centering
\caption{Top: A subband fit for PSR J2302$+$4442 at MJD 57921. Bottom: A subband fit for PSR B1855$+$09 at MJD 57608.}
\label{scale_fit}
\end{figure}

 With a large enough observation bandwidth, it is possible to place constraints on the scaling behavior of scattering delays as a function of frequency. \cite{Levin_Scat} were able to break up a few unstretched wideband observations at 1500 MHz into four equal subbands of 200 MHz each, determine $\Delta \nu_{\textrm{d}} \textrm{ and } \tau_{\textrm{d}}$ in each unstretched subband using the ACF method described in Section \ref{analysis}, and perform a weighted linear fit for $\tau_{\textrm{d}}$ in semi-log space of the form $\nu^{\xi}$ to estimate the scaling index $\xi$ for a given epoch. Some examples of these fits can be seen in Figure \ref{scale_fit}.

\begin{figure}[!ht]
\includegraphics[scale=.52]{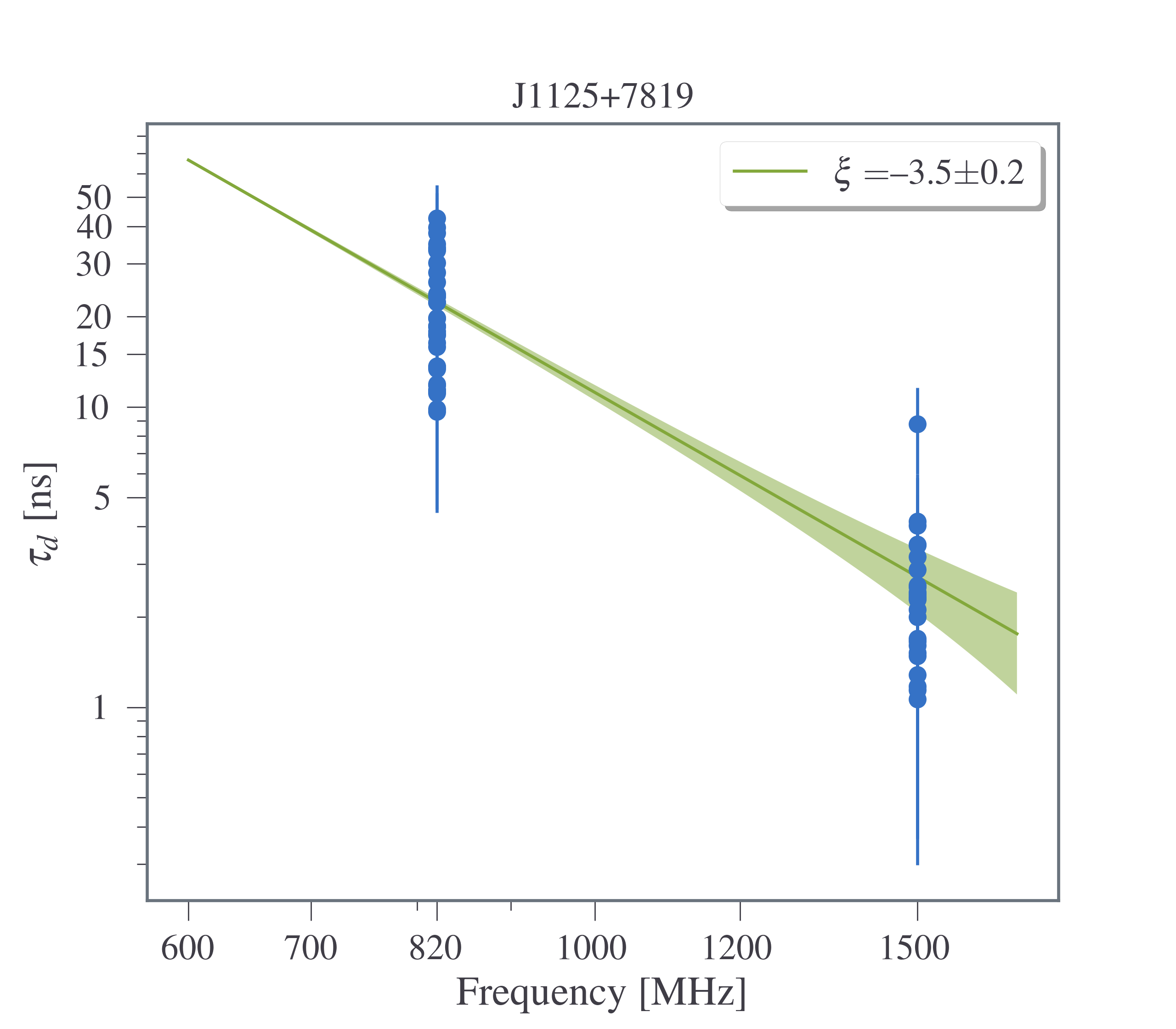}
\centering
\caption{An example fit in semi-log space of the scaling index $\xi$ over the 820 and 1500 MHz bands for PSR J1125$+$7819. Each point indicates a measured epoch of scattering delay in a given frequency band with its corresponding 1$\sigma$ error.}
\label{ex_fit}
\end{figure}
\par We applied this method to four of the pulsars and, as discussed in more detail in Section \ref{scale_dis}, found similar results to \cite{Levin_Scat}. We also use this method to look at the time dependence of this scaling index. We found it feasible to perform this method only in the 1500 MHz band, as the 800 MHz band has only 200 MHz of bandwidth and we would have an insufficient number of scintles per subband to effectively utilize this approach.

\par For many of the other pulsars in our data set, we took advantage of our dual frequency measurements to examine scaling indices across a wider frequency range. In this multiband method, we took the weighted averages of scattering delays at 820 and 1500 MHz for pulsars with measurements at both frequencies and performed the same fit described above. An example of one of these fits is shown in Figure \ref{ex_fit}. While this multiband method examine scaling indices differently than \cite{Levin_Scat}, our ability to utilize multiple frequency bands augments \cite{Levin_Scat}, which only used 1500 MHz-band data. Examining time variability was not possible using the multiband method, since we rarely, if ever, had epochs (observations within a span of about a week) in which we had detectable measurements for two frequencies.
\par In addition to our multiband method, we were able to utilize the original scaling analysis from \cite{Levin_Scat} on four pulsars to determine the variation in scaling index over time, as well as PSRs B1855$+$09 and J2302$+$4442 to compare the results of the two analyses.

\par It is important to note that these weighted averages from the multiband analysis are determined using  measurements from dynamic spectra that have already been stretched by $\nu^{4.4}$ to account for the wide bandwidth. This will result in some errors on the calculated scaling index. Note that, however, a true $\nu^{-4.4}$ scaling would still yield $\xi = -4.4$.

\par In the future, new developments such as wideband receivers should allow us to achieve the S/N and frequency range necessary to determine the scaling index of an epoch by looking at unstretched spectra and maximizing the S/N of that epoch's frequency ACF \citep{lam_opt}.

\subsection{Transverse Velocities}
\label{t_velo}
\par After recovering interstellar scattering parameters, we estimated transverse velocities for pulsars with measured scintillation timescales. Transverse velocities for many NANOGrav pulsars have already been inferred from proper motions, defined as 
\begin{equation}
\label{prop_mot}
V_{\textrm{pm}}=4.74\mu D_{\textrm{kpc}},
\end{equation}
where $V_{\textrm{pm}}$ is in units of km s$^{-1}$, $\mu$ is proper motion in units of mas yr$^{-1}$, and $D_{\textrm{kpc}}$ is the distance to the pulsar in kpc, but they can also be estimated from scintillation behavior, assuming the surrounding ISM can be interpreted as stationary relative to the pulsar in question. 
\par Merging the expressions for transverse velocity from \cite{gupta_velo} and \cite{Cordes_1998} for greater generality, we have 
\begin{equation}
\label{iss_velo}
V_{\textrm{ISS}}=A_{\textrm{ISS}}\frac{\sqrt{\Delta \nu_{\textrm{d,MHz}}D_{\textrm{kpc}}x}}{\nu_{\textrm{GHz}}\Delta t_{\textrm{d,s}}},
\end{equation}
where $\Delta \nu_{\textrm{d,MHz}}$ is the scintillation bandwidth in MHz, $\Delta t_{\textrm{d,s}}$ is the scintillation timescale in seconds, $D_{\textrm{kpc}}$ is the distance to the pulsar in kpc, $\nu_{\textrm{GHz}}$ is the observation frequency in GHz, and $A_{\textrm{ISS}}$ is a factor dependent on assumptions  regarding the geometry and uniformity of the medium. 

\par In this analysis, we have assumed a thin screen and a Kolmogorov medium as in \cite{Cordes_1998}, and so $A_{\textrm{ISS}} = 2.53 \times 10^4 \ \textrm{km s}^{-1}$. As used in \cite{gupta_velo}, we define $x=D_{\textrm{o}}/D_{\textrm{p}}$, where $D_{\textrm{o}}$ is the distance from the observer to the screen and $D_{\textrm{p}}$ is the distance from the screen to the pulsar. For the calculation of $V_{\textrm{ISS}}$, we assume the screen is halfway between us and the observed pulsars, and so $D_{\textrm{o}}=D_{\textrm{p}}$, meaning $x=1$. Additionally, we ignore  orbital velocities (for  binary pulsars) and the Earth's velocity, and assume the screen is isotropic. We encourage our readers to look at \cite{Rickett_2014}, \cite{Reardon}, \cite{Reardon_2020}, and \cite{Main} for examples of significant orbital and annual variations of scintillation timescales, non-zero screen velocities, and non-isotropic scattering. For the calculation of $V_{\textrm{pm}}$, we used distances determined by parallax measurements if  $\sigma_D/D<0.25$, otherwise we used the DM distance determined by the NE2001 electron density model.

\par The ability to independently determine transverse velocities from different sets of physical quantities also helps us determine whether the ISM behaves as Kolmogorov with a scattering screen at the halfway point. We expect transverse velocities derived from proper motions to be more accurate, as proper motions are generally measured with much greater precision and with fewer selection effects than scintillation parameters. Consequently, comparisons of those results serve as a strong indicator of the accuracy of ISS-derived transverse velocities. In addition, there are a number of pulsars, in particular, non-recycled, for which we are unable to measure high quality timing-derived proper motions, so it is useful to have alternative ways of measuring transverse velocities. 

\section{Results}

\subsection{Scintillation Parameters and Variations}
\label{parm_var}
Our measurements of interstellar scattering delays, scintillation bandwidths, and scintillation timescales are given in Table \ref{table_2}, with barred parameters ($\overline{\tau}_{\rm d}$, $\overline{\Delta\nu}_{\rm d}$, etc.) representing the ensemble weighted averages of the individual observations. We determined values for $\overline{\tau}_{\rm d}$ by calculating $\tau_{\rm d}$ values for individual epochs and then averaging them, rather than directly converting $\overline{\Delta\nu}_{\rm d}$. For comparison, in Tables \ref{table_1} and \ref{table_compare}, we also list the predicted scintillation parameters from the NE2001 Galactic electron density model \citep{NE2001} and the results from previous studies of these pulsars, respectively. Due to our short observation lengths, our $\overline{\Delta t_{\rm d}}$ values should probably be taken as lower limits in most cases since there were many epochs where scintles were not resolvable in time. An clear exception to this rule is PSR B1937$+$21, for which all scintles were smaller than our observation length. Other likely exceptions to this rule likely include but are not necessarily limited to PSR J1125$+$7819 at 820 MHz and PSR J0636$+$5128. There are a few pulsars where we quote scintillation bandwidths but neither scintillation timescales nor timescale upper limits; since many epochs from these pulsars contain the beginnings and ends of many scintles but never complete scintles, these pulsars all likely have scintillation timescales within 5$-$10 minutes of our observation lengths of 30 minutes. 
\par For Table \ref{table_1}, the predicted $\Delta t_{\textrm{d}}$ values were calculated using transverse velocities derived from proper motions. Additionally, DM distances were used for calculating transverse

\begin{figure*}[!ht]
\centering
\includegraphics[width=.48\textwidth]{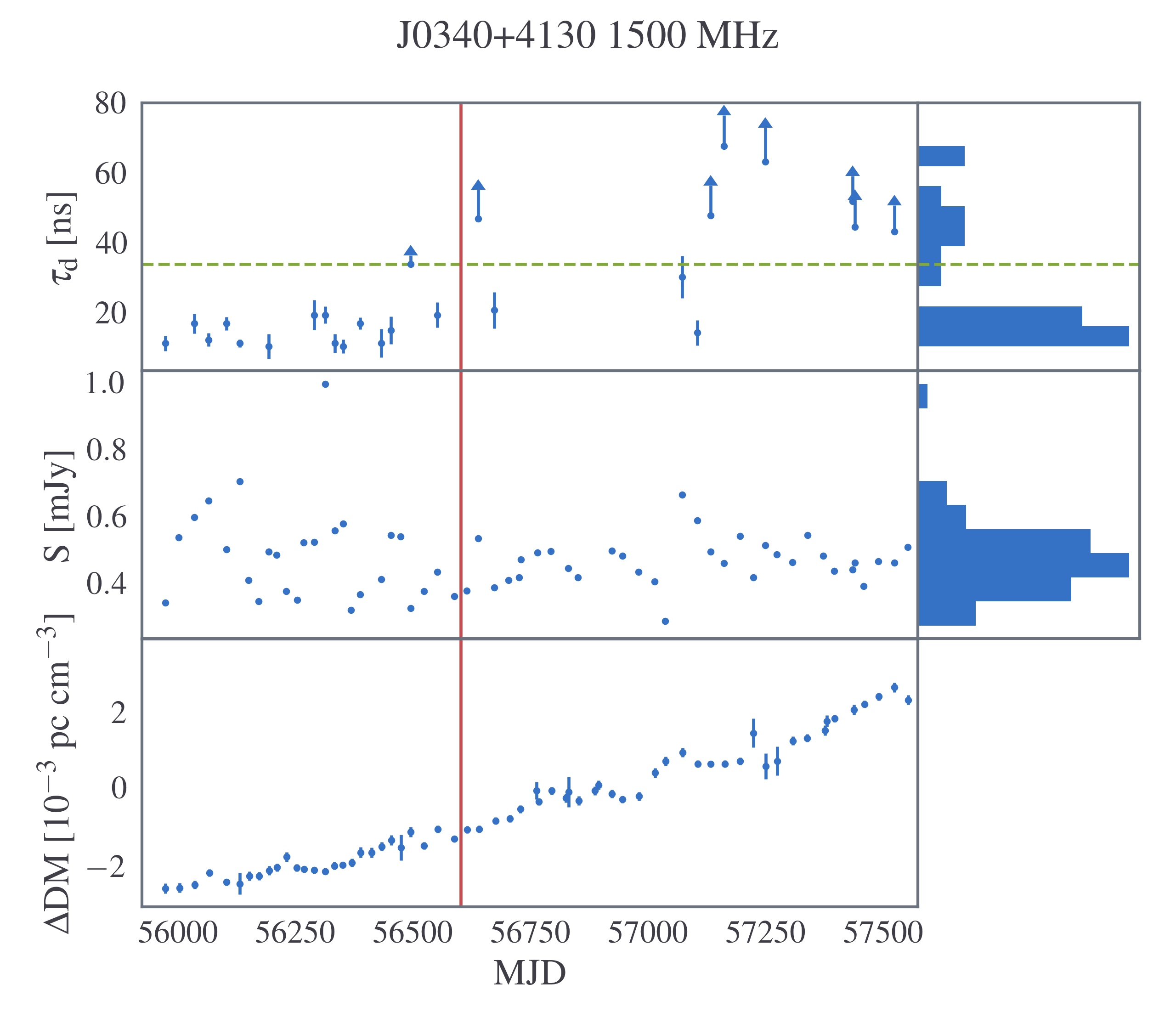}\qquad
\includegraphics[width=.48\textwidth]{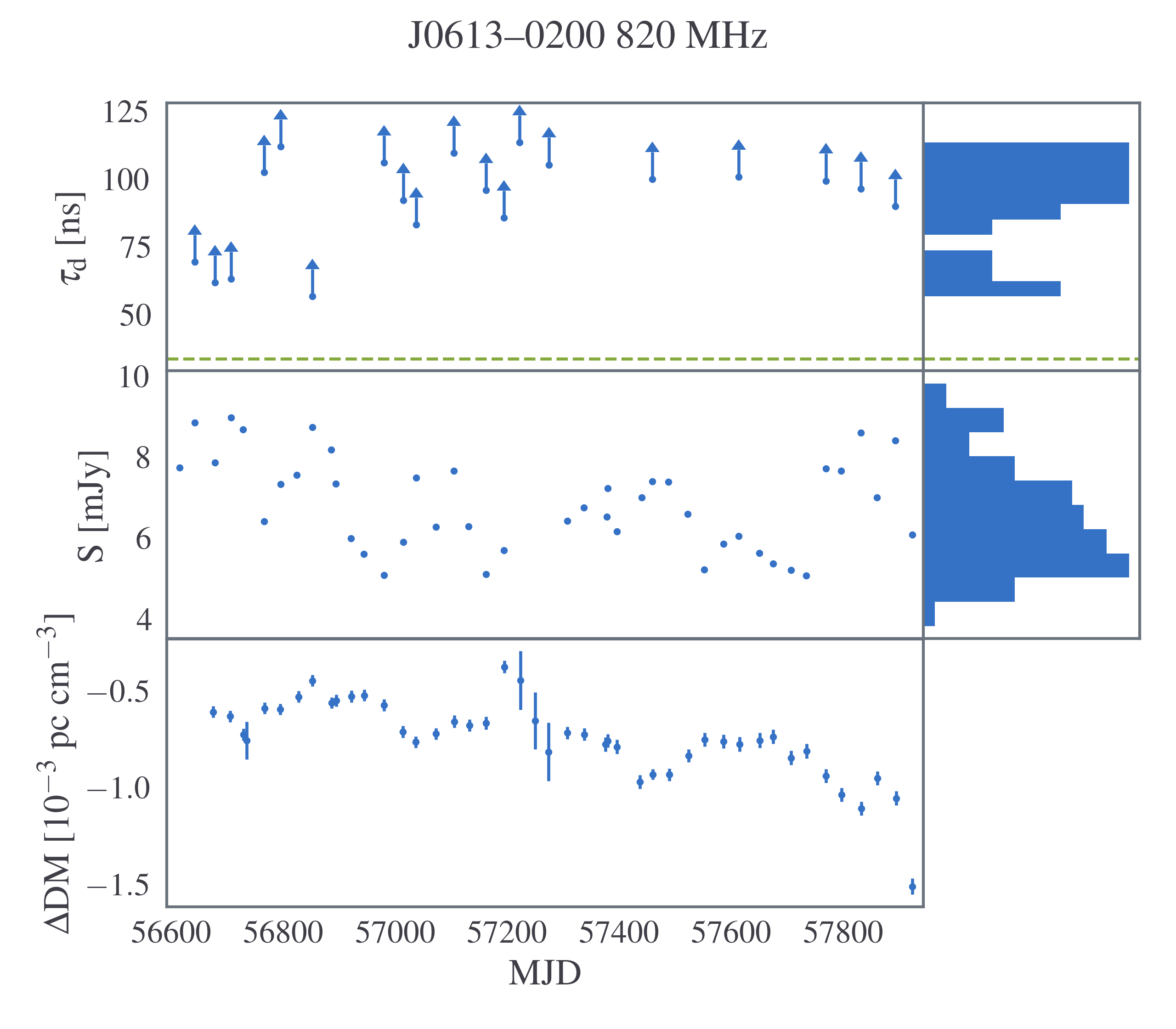}\\
\includegraphics[width=.48\textwidth]{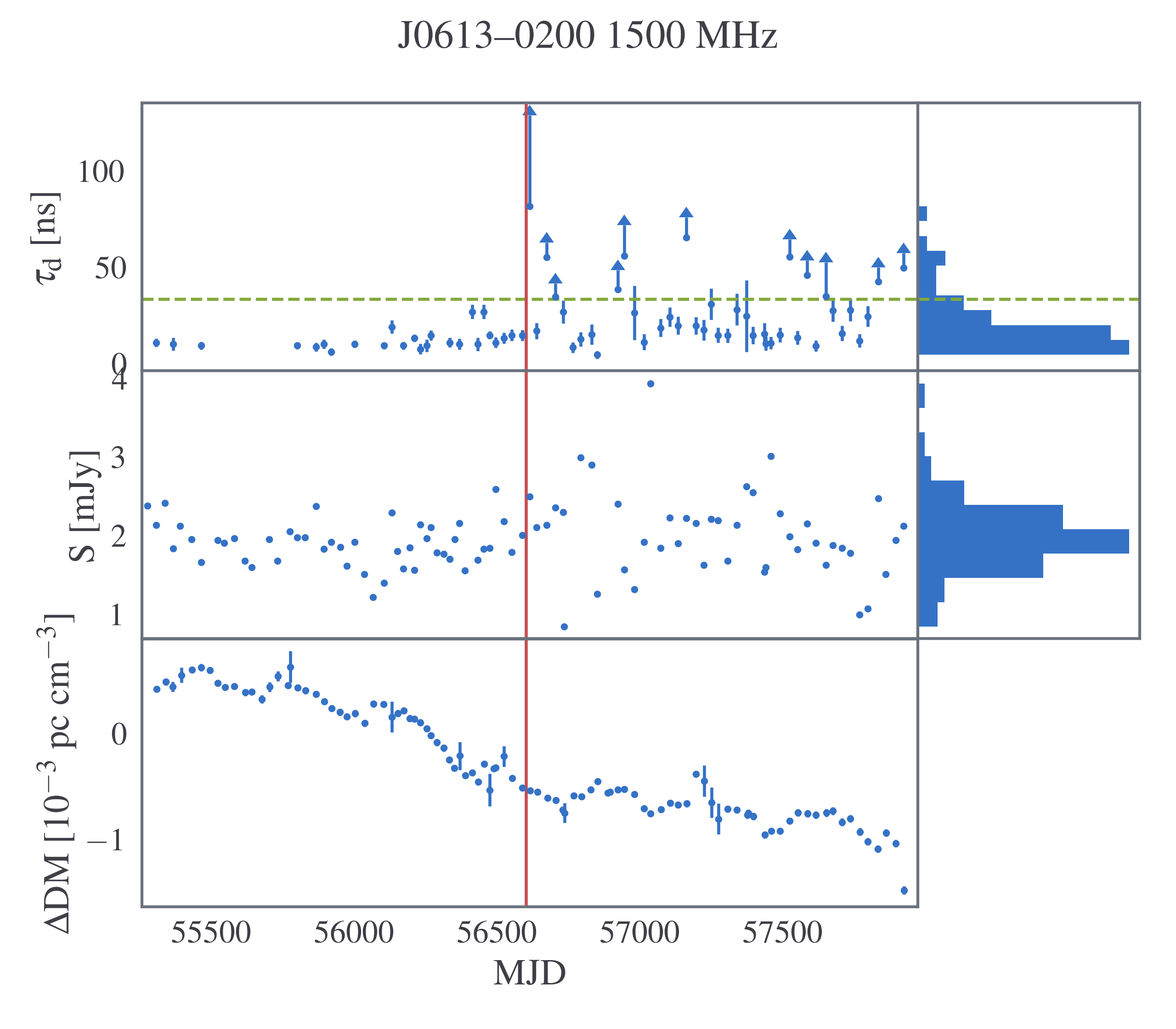}\qquad
\includegraphics[width=.48\textwidth]{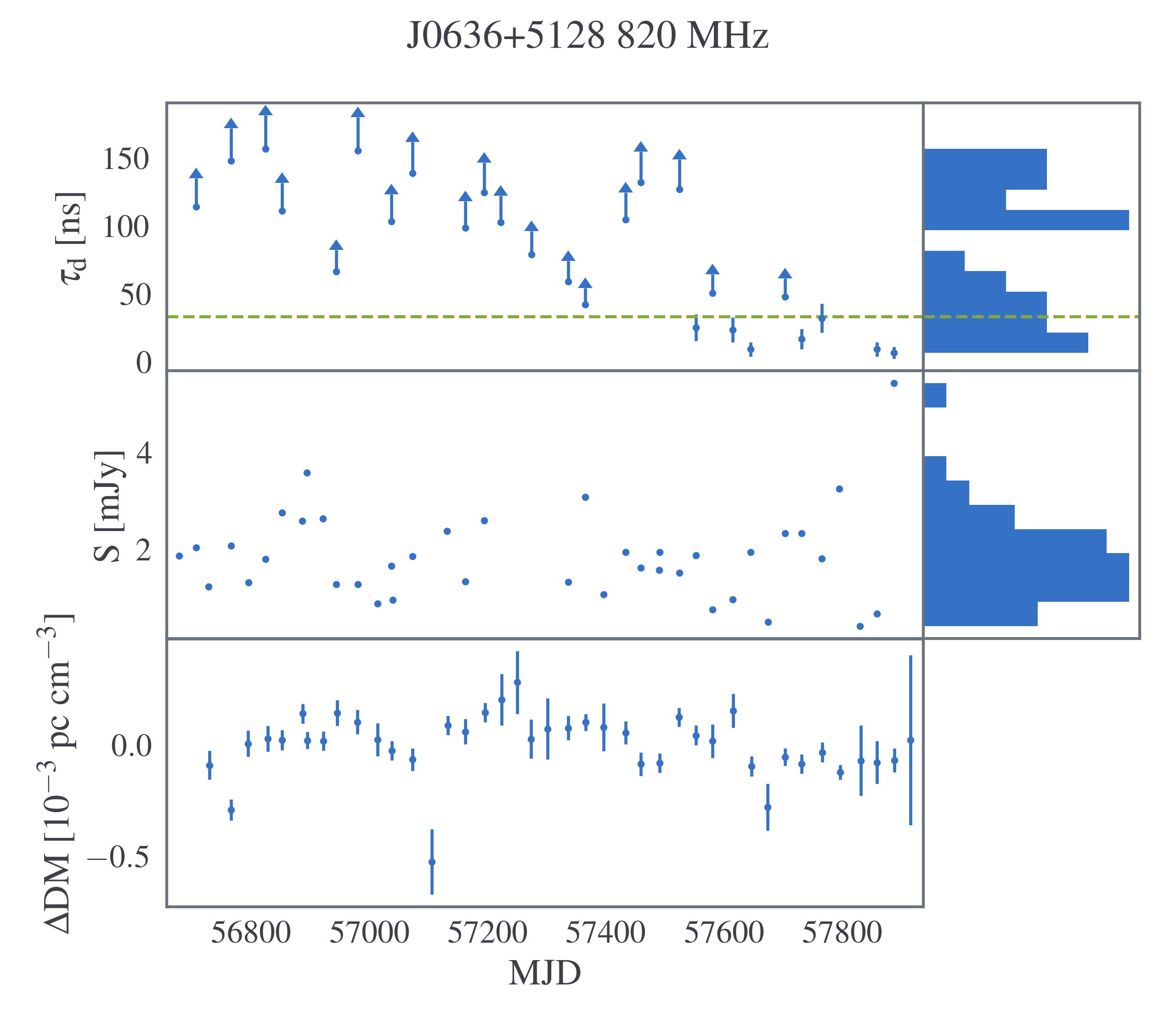}\\
\includegraphics[width=.48\textwidth]{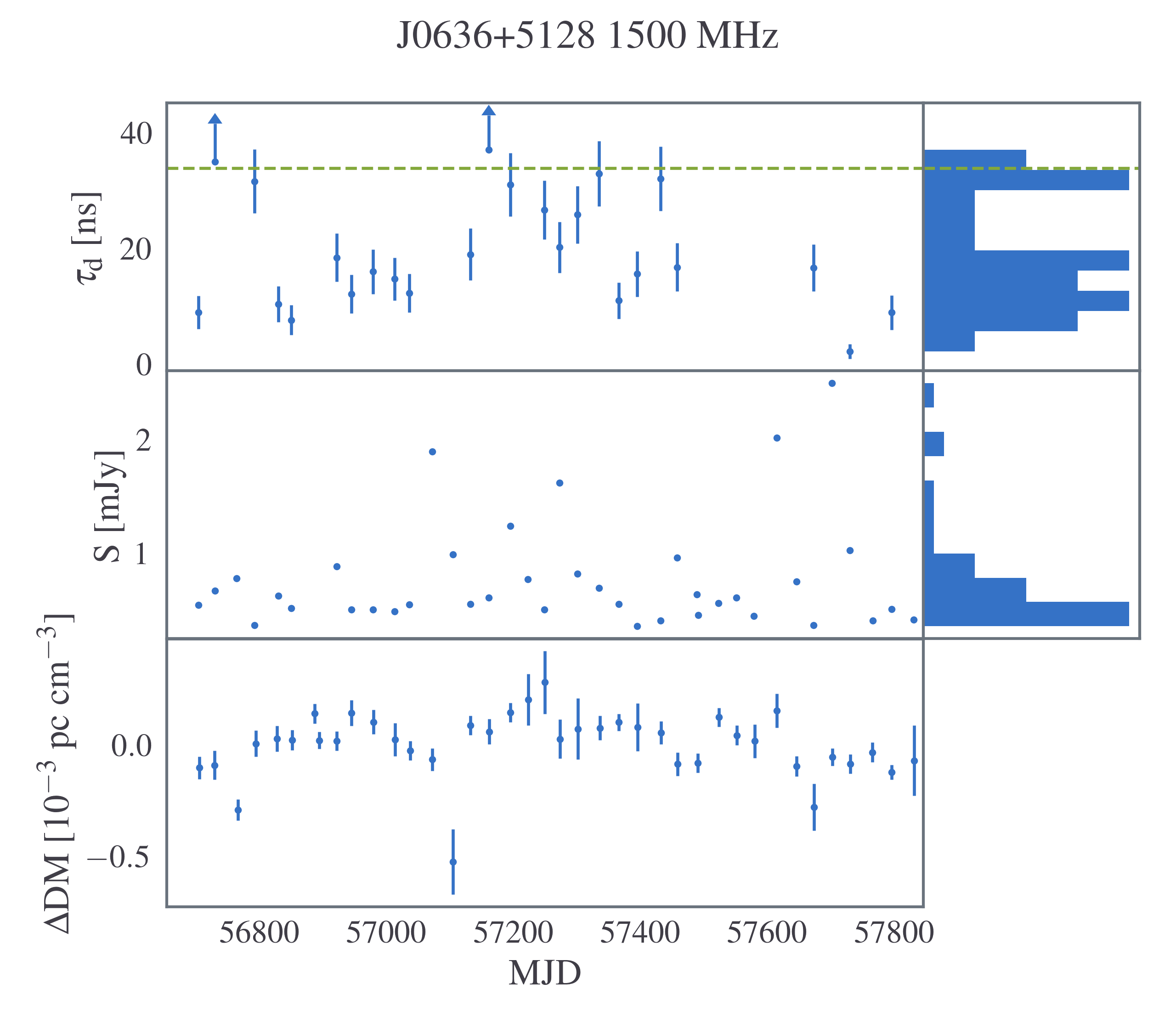}\qquad
\includegraphics[width=.48\textwidth]{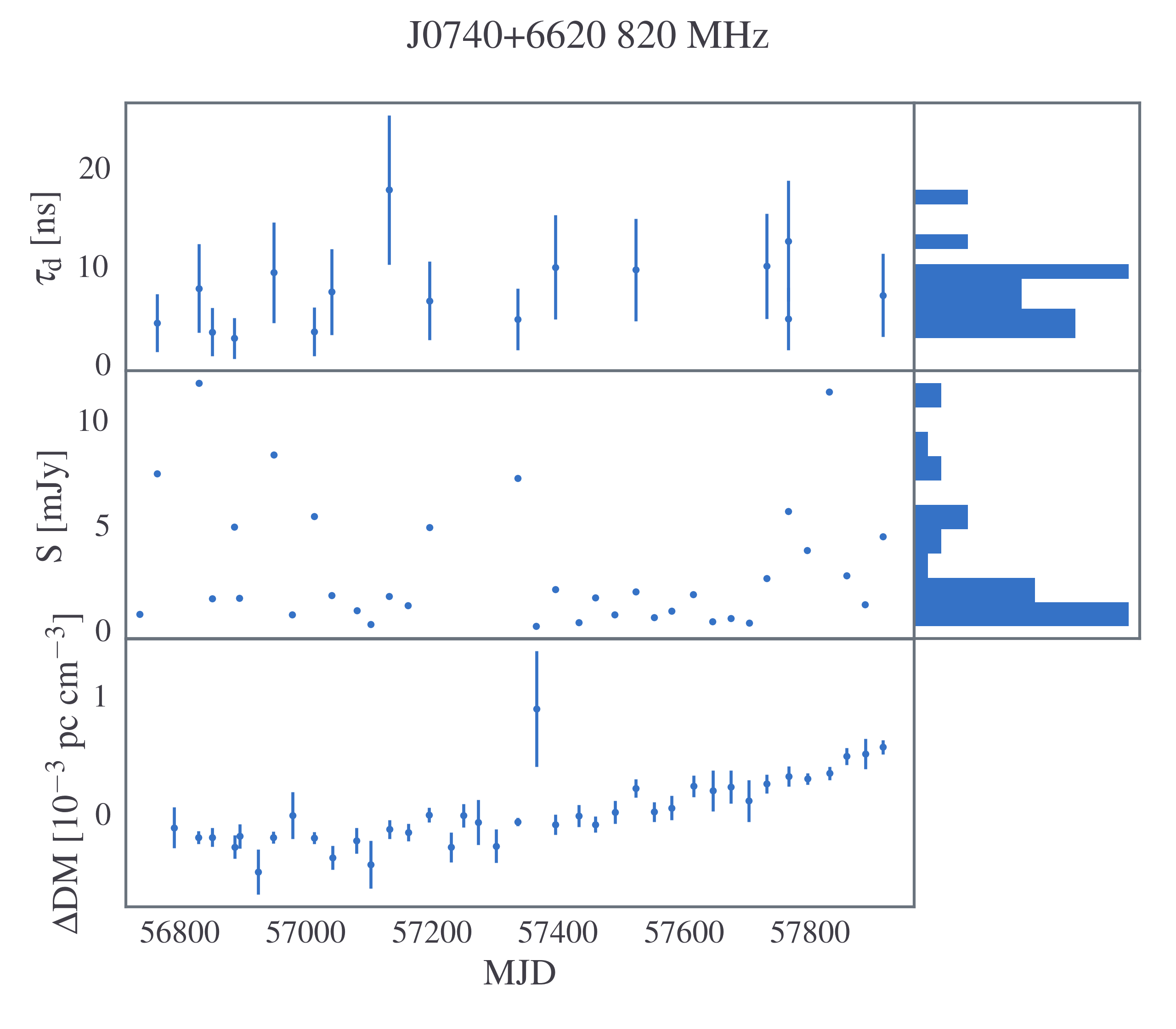}

\label{dmx_plots_1}
\end{figure*}

\begin{figure*}[!ht]
\centering
\includegraphics[width=.48\textwidth]{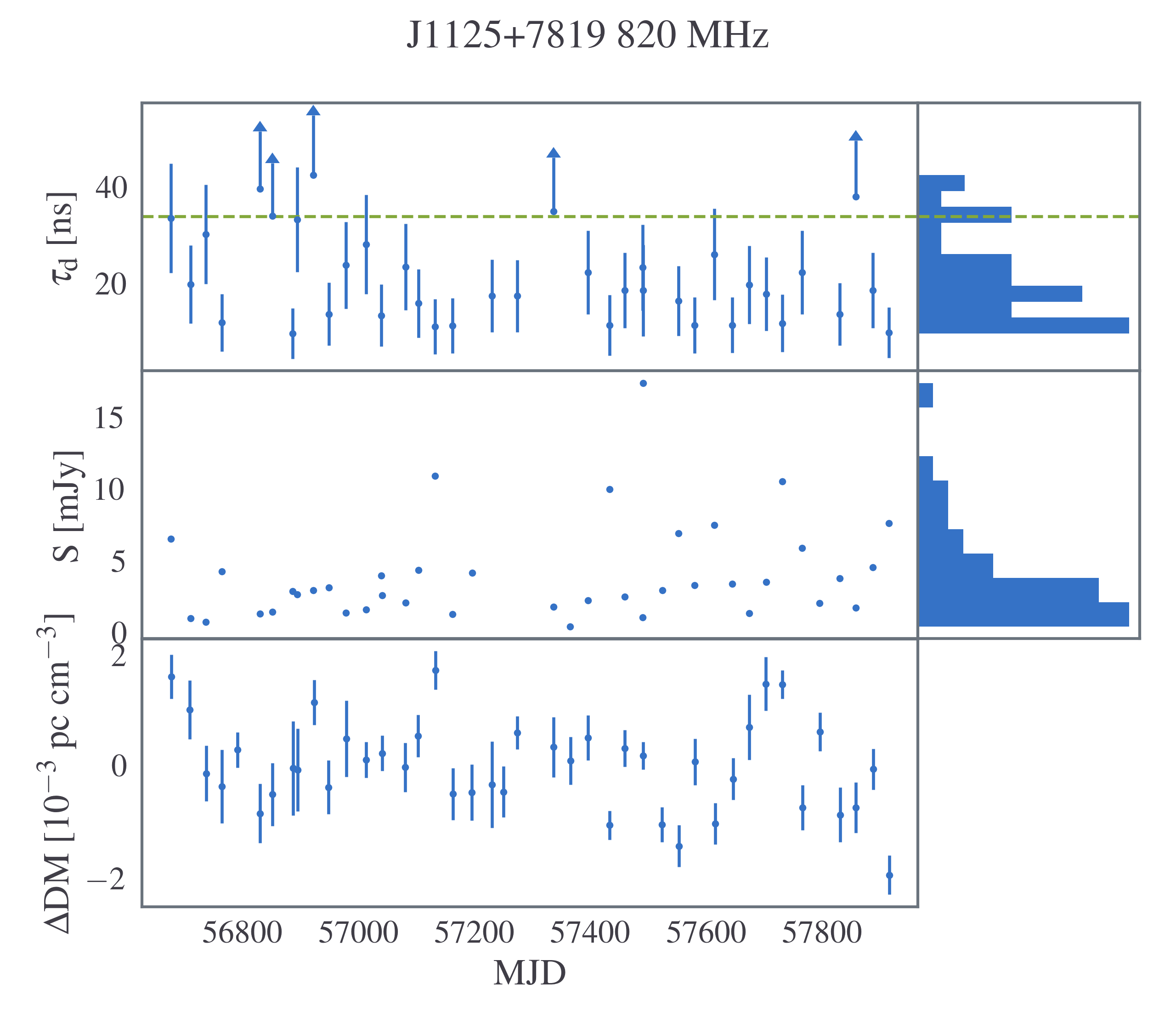}\qquad
\includegraphics[width=.48\textwidth]{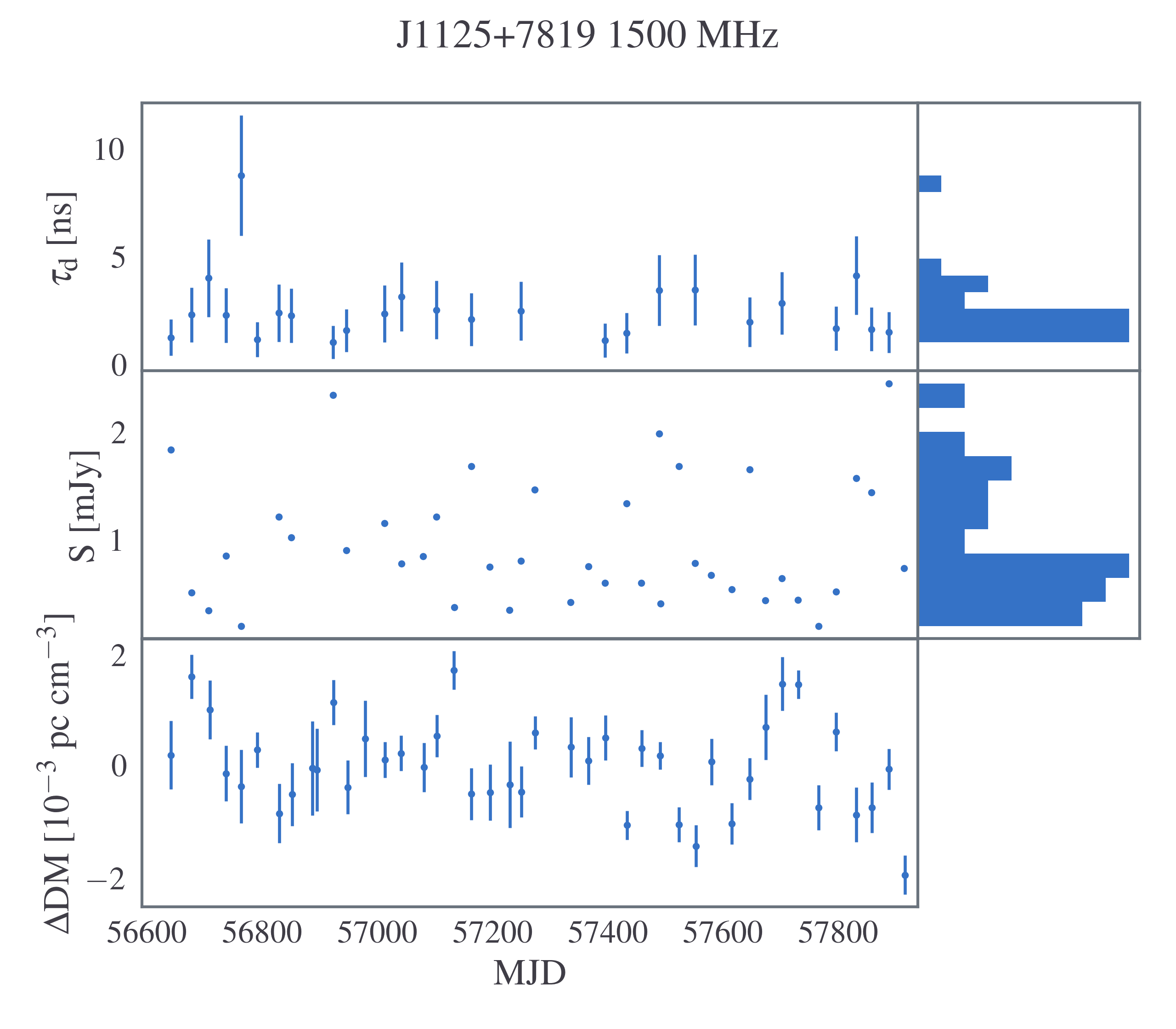}\\
\includegraphics[width=.48\textwidth]{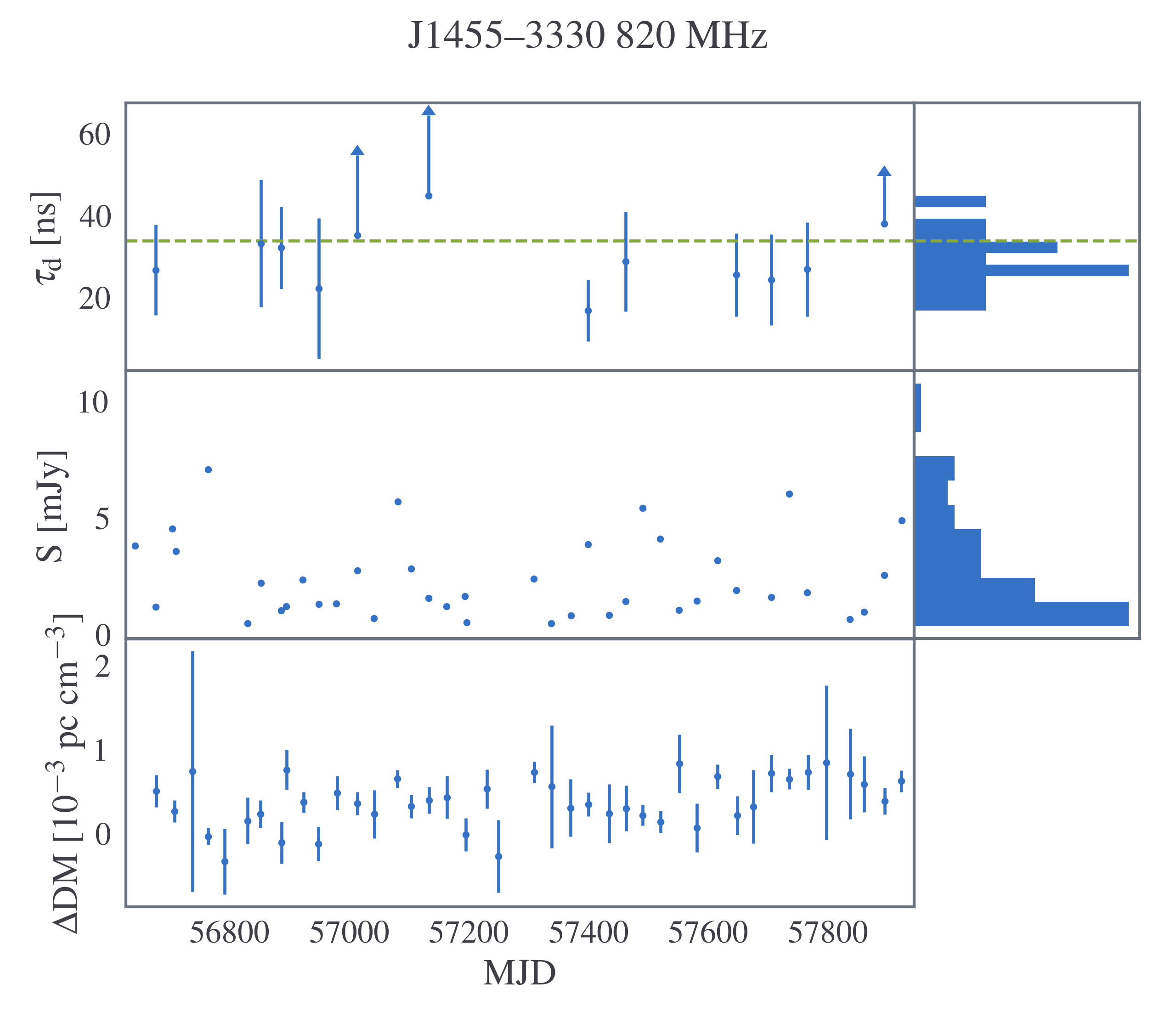}\qquad
\includegraphics[width=.48\textwidth]{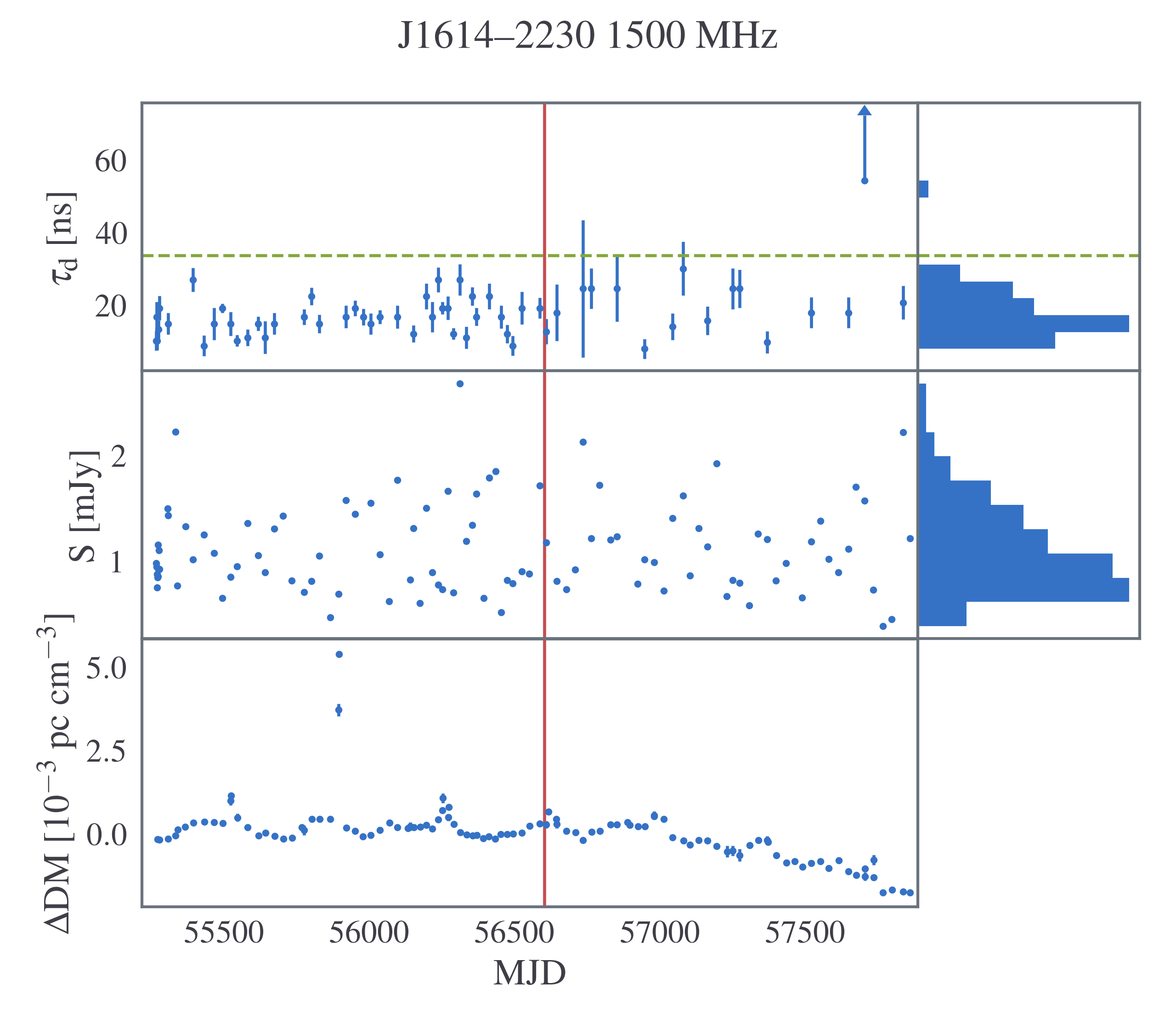}\\
\includegraphics[width=.48\textwidth]{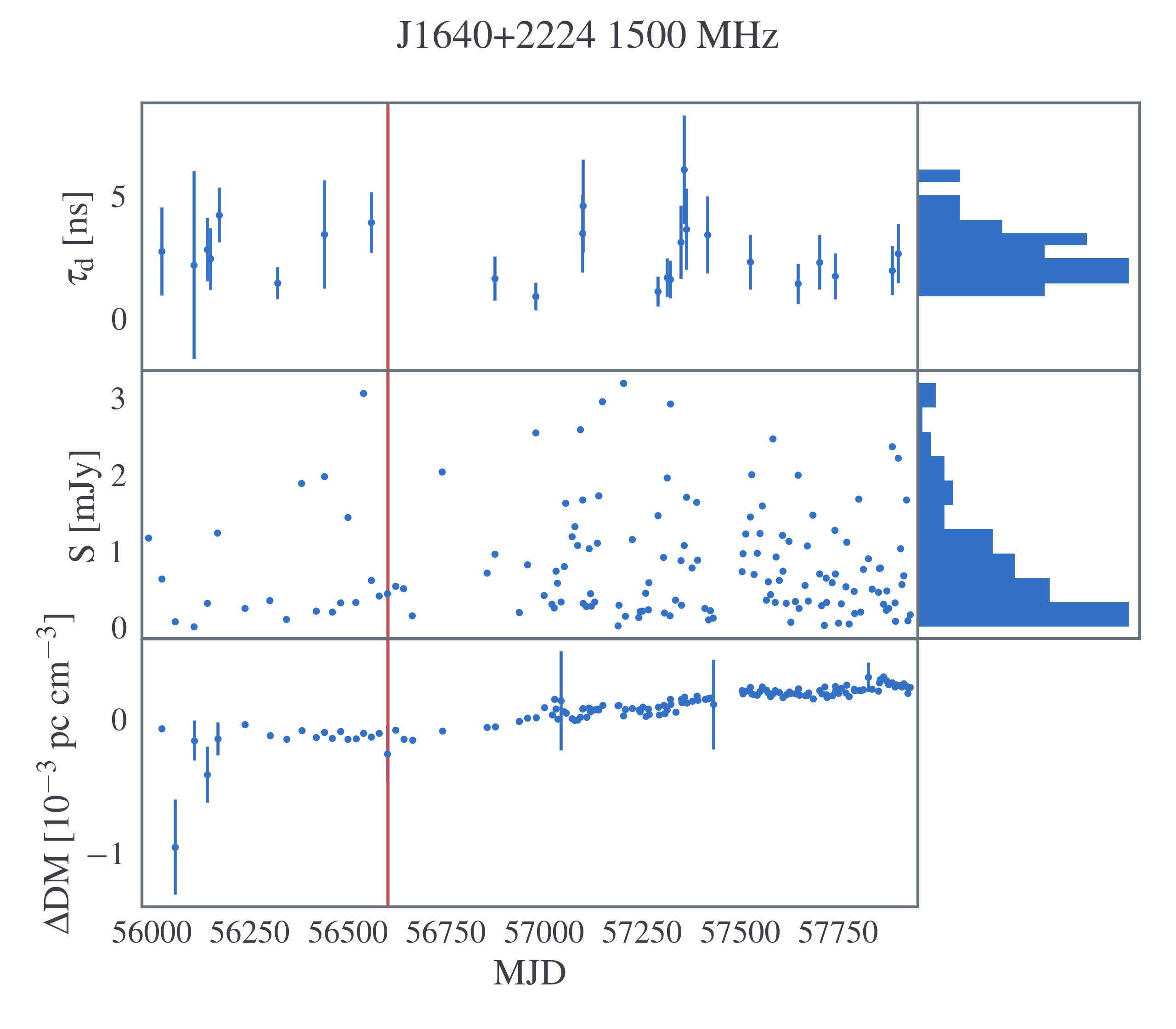}\qquad
\includegraphics[width=.48\textwidth]{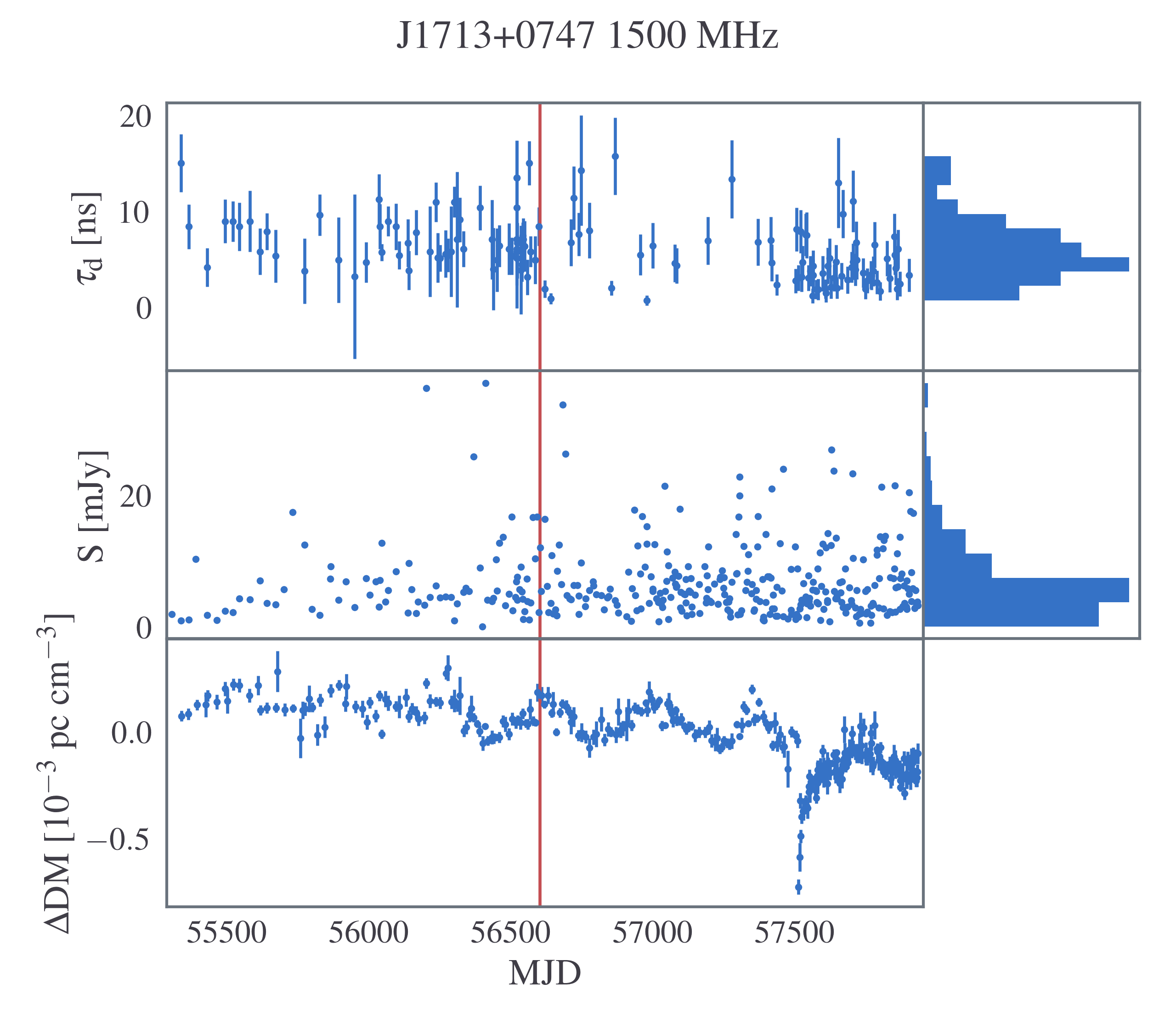}
\label{dmx_plots_2}
\end{figure*}

\begin{figure*}[!ht]

\centering
\includegraphics[width=.48\textwidth]{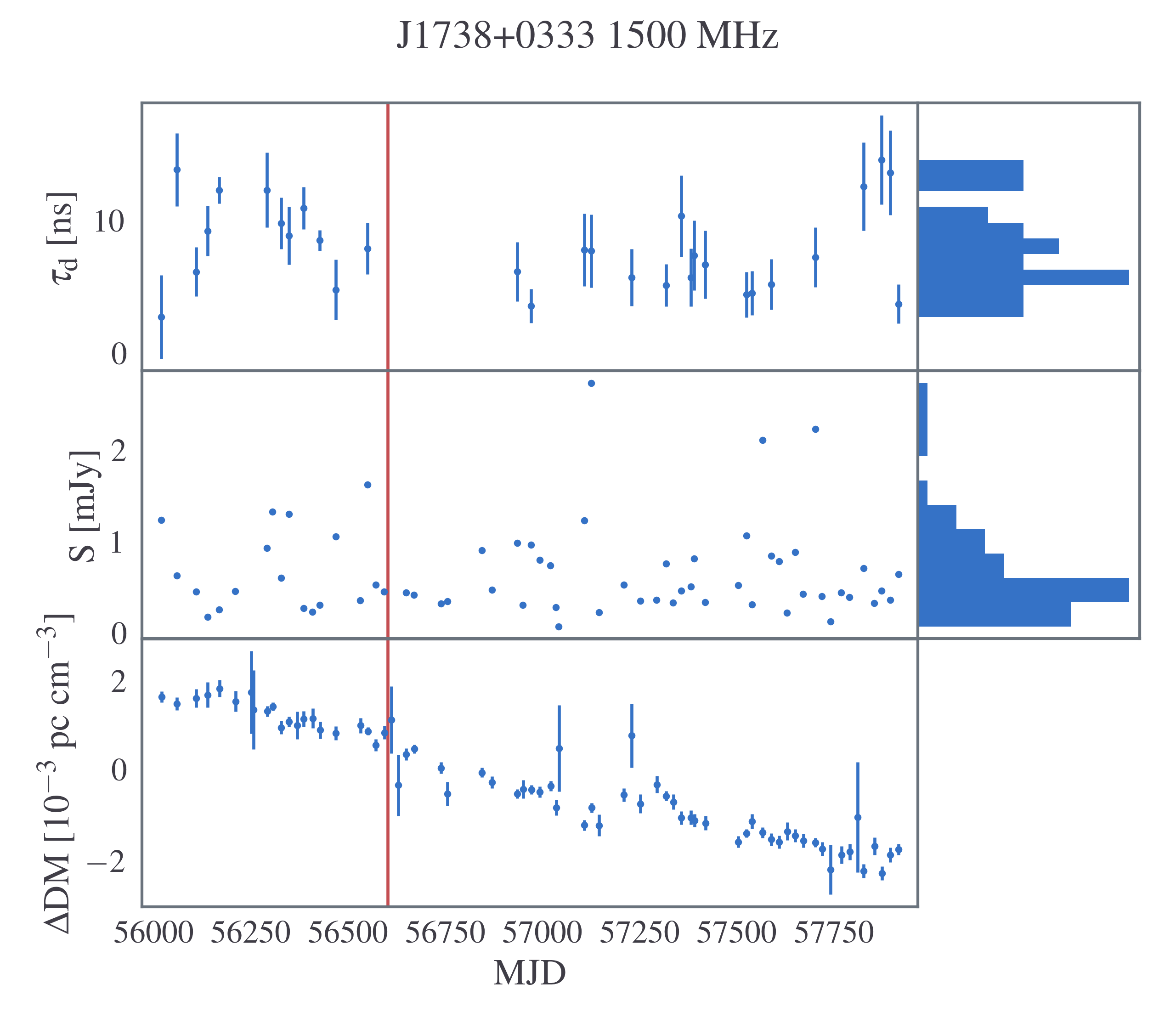}\qquad
\includegraphics[width=.48\textwidth]{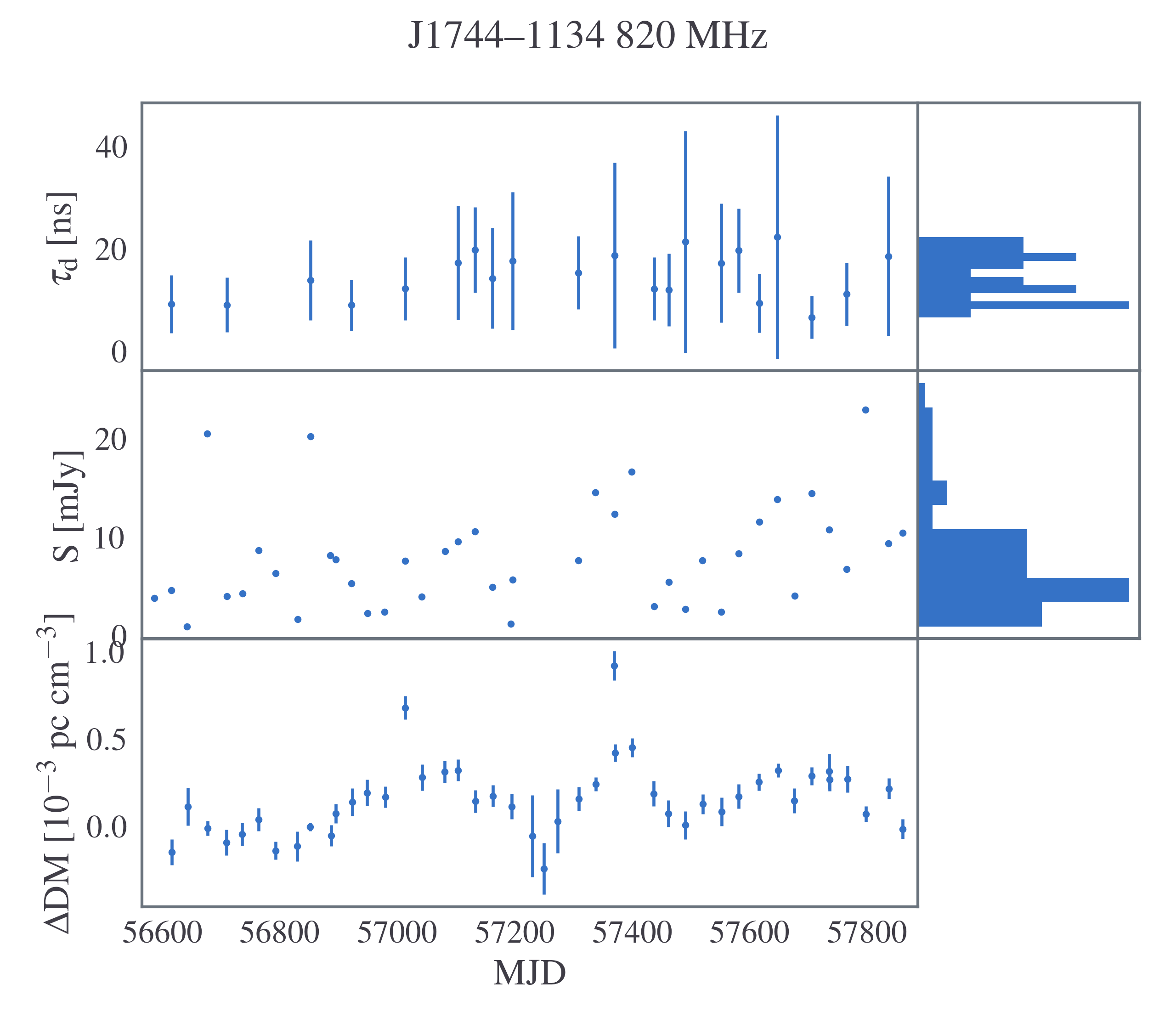}\\
\includegraphics[width=.48\textwidth]{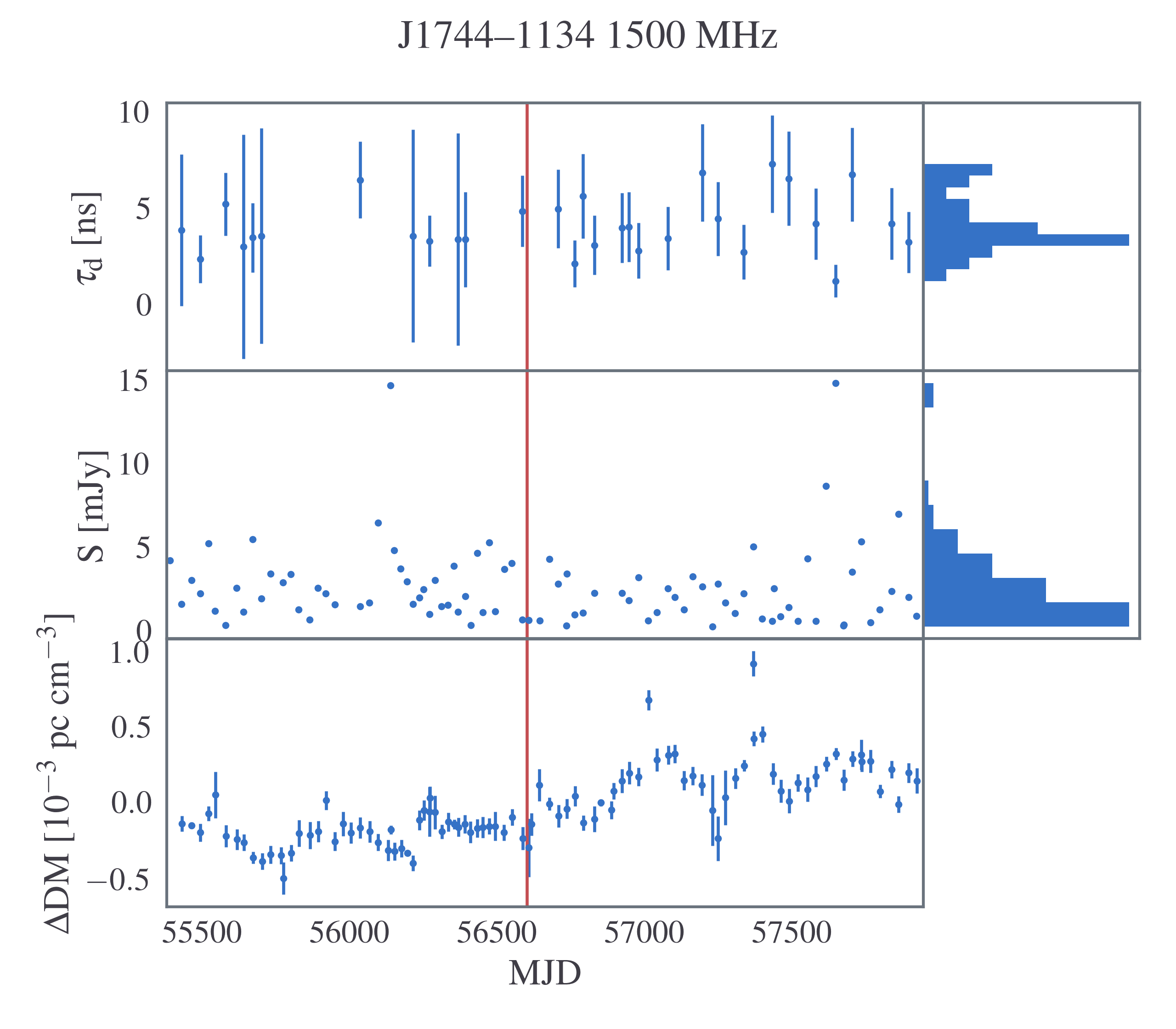}\qquad
\includegraphics[width=.48\textwidth]{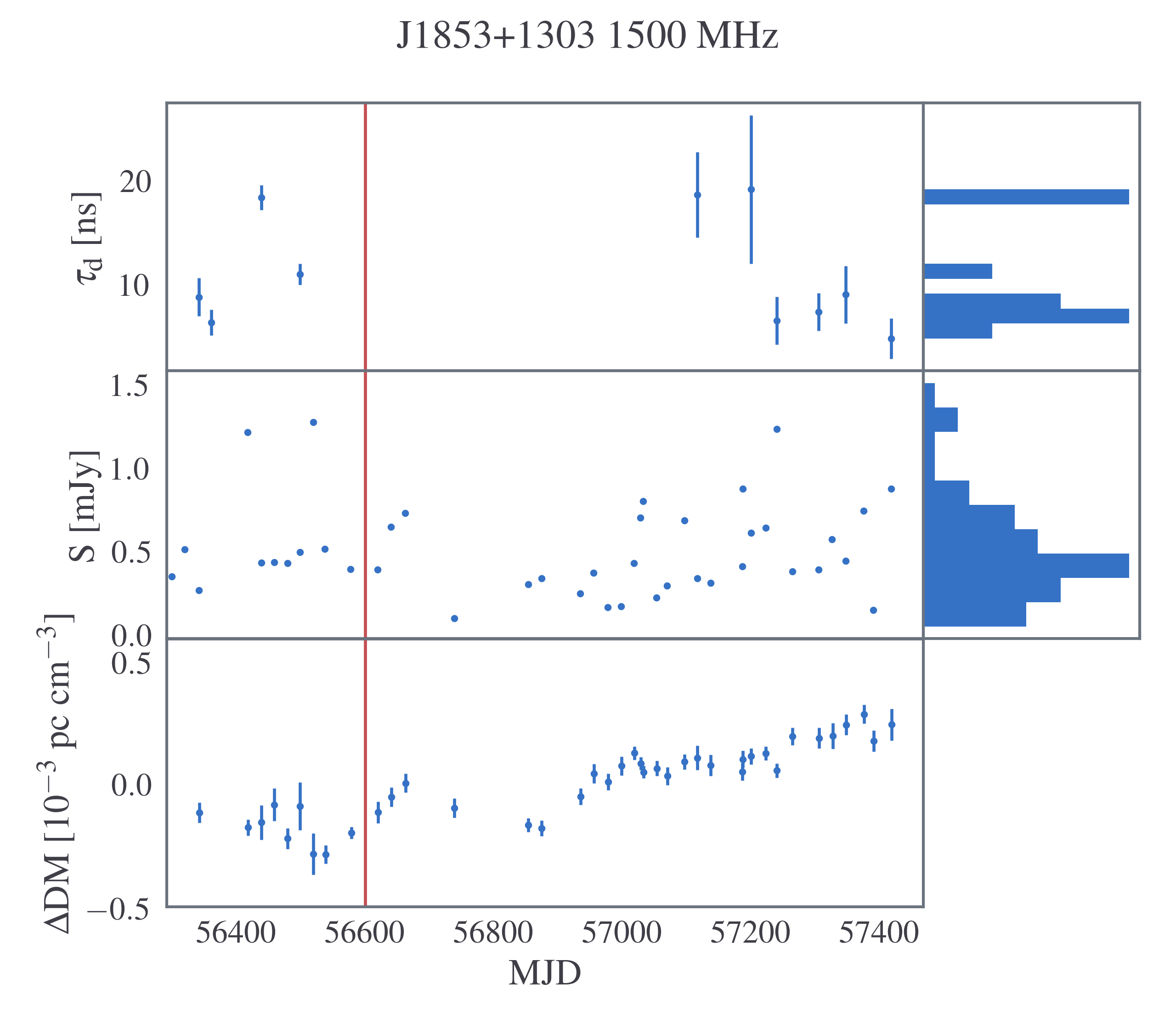}\\
\includegraphics[width=.48\textwidth]{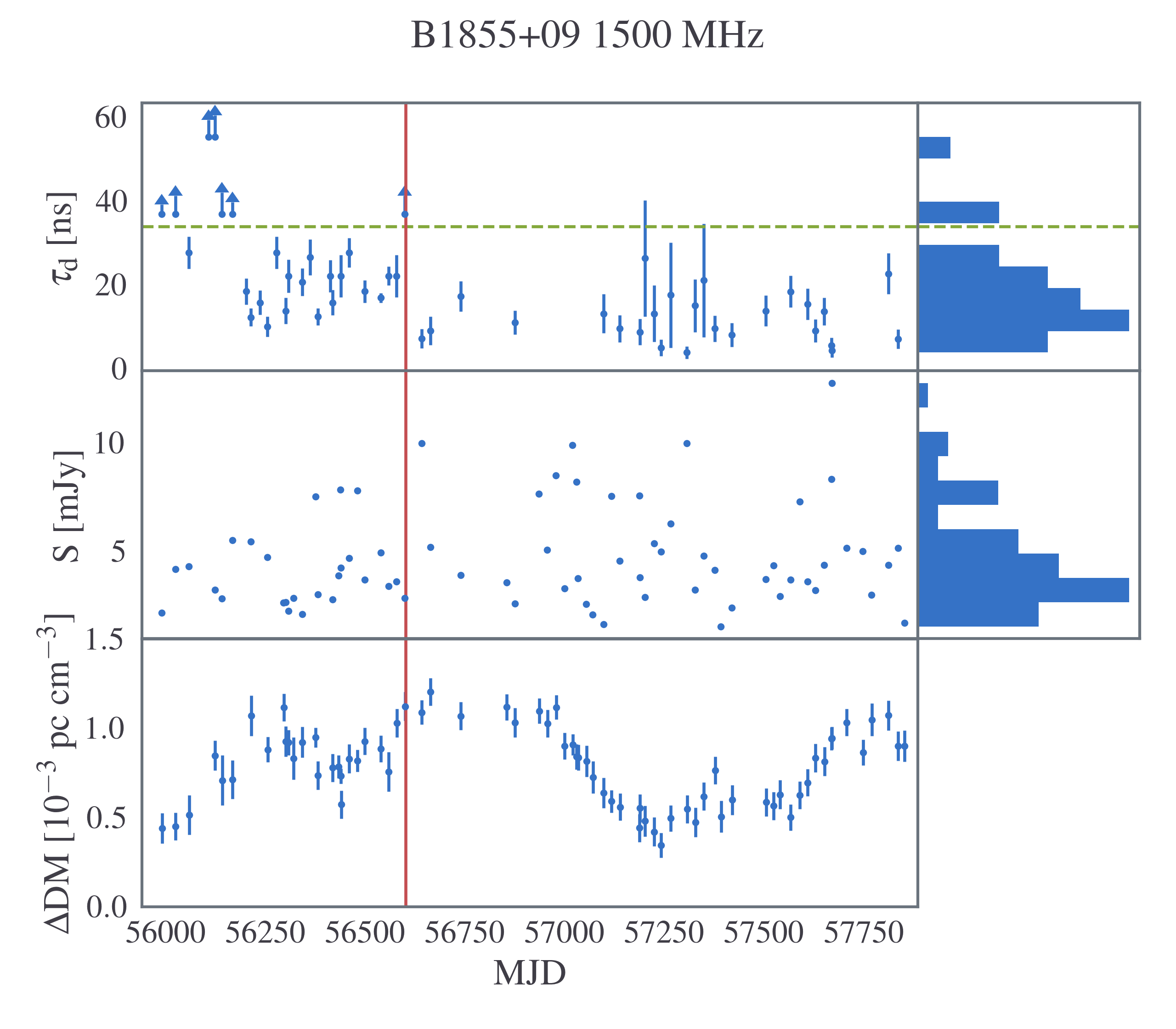}\qquad
\includegraphics[width=.48\textwidth]{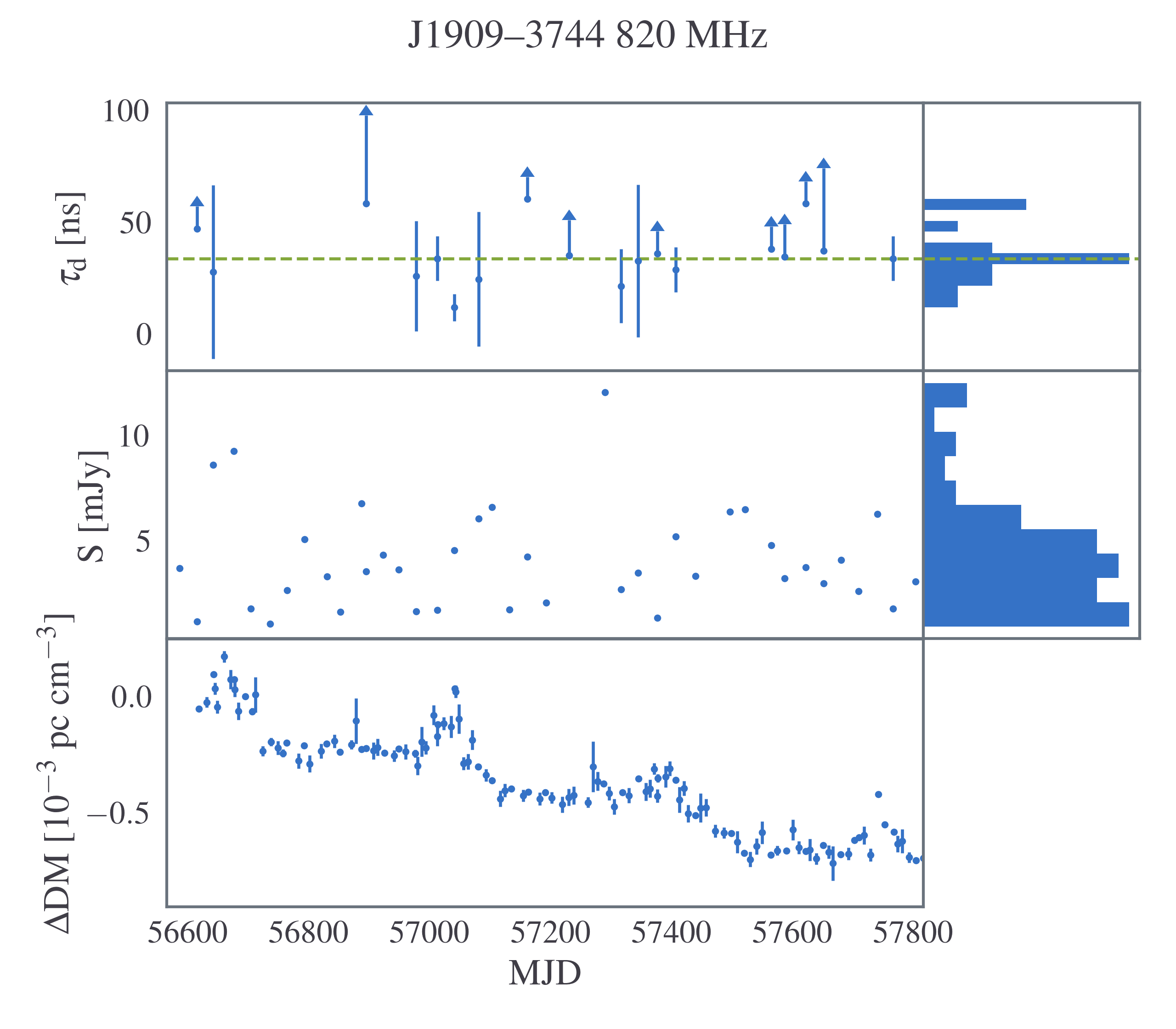}
\label{dmx_plots_4}
\end{figure*}
\begin{figure*}[!ht]
\centering
\includegraphics[width=.48\textwidth]{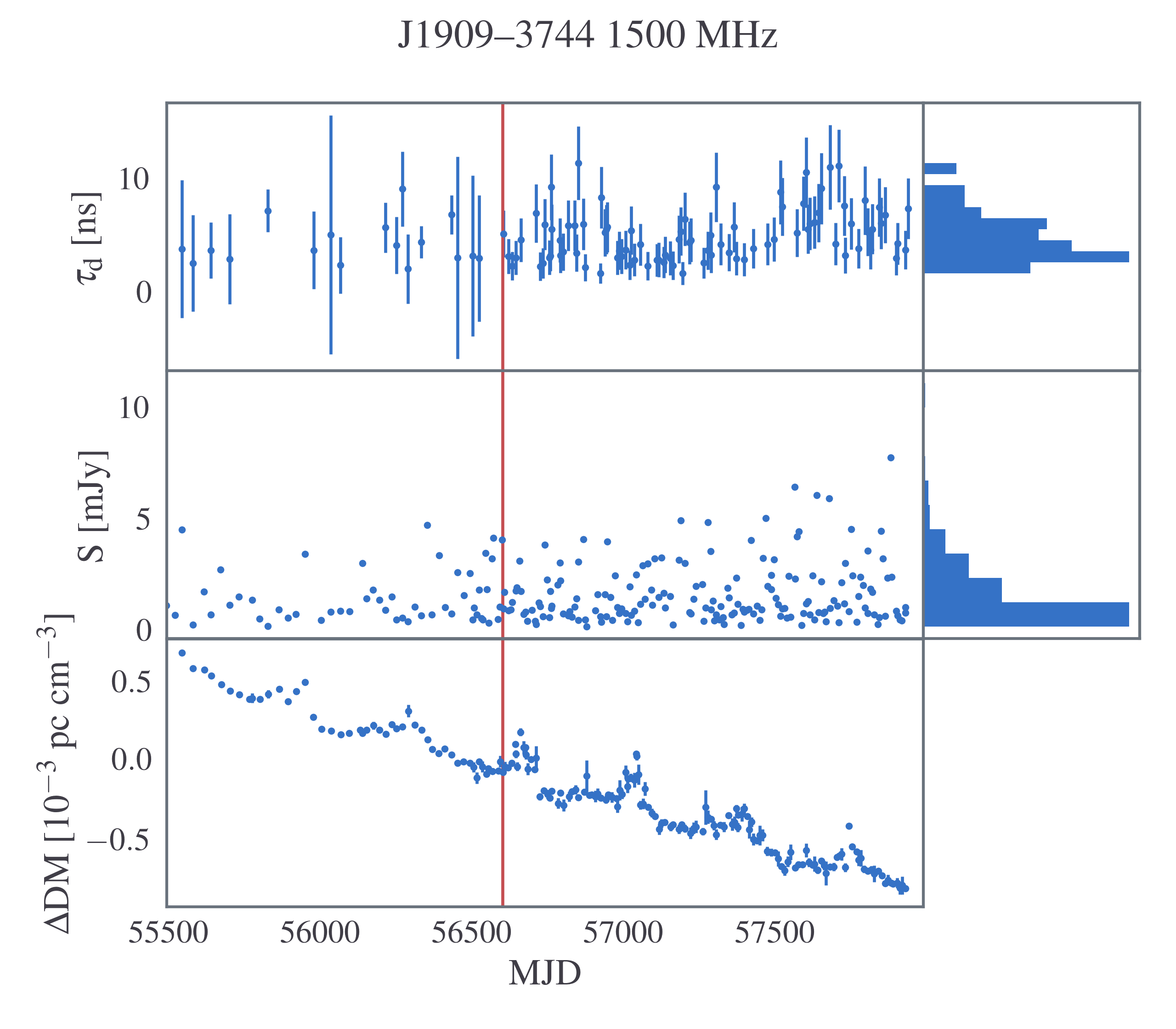}\qquad
\includegraphics[width=.48\textwidth]{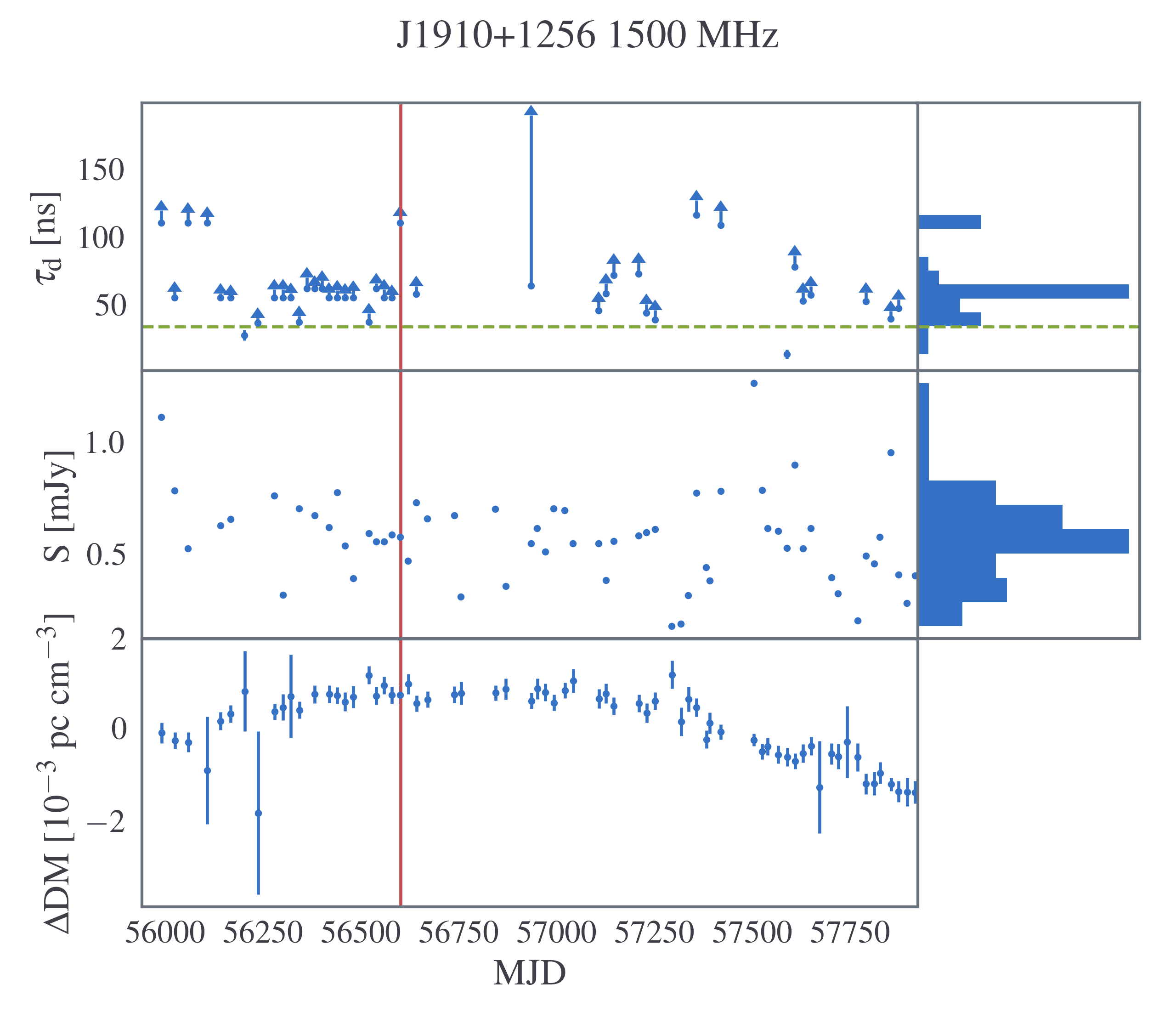}\\
\includegraphics[width=.48\textwidth]{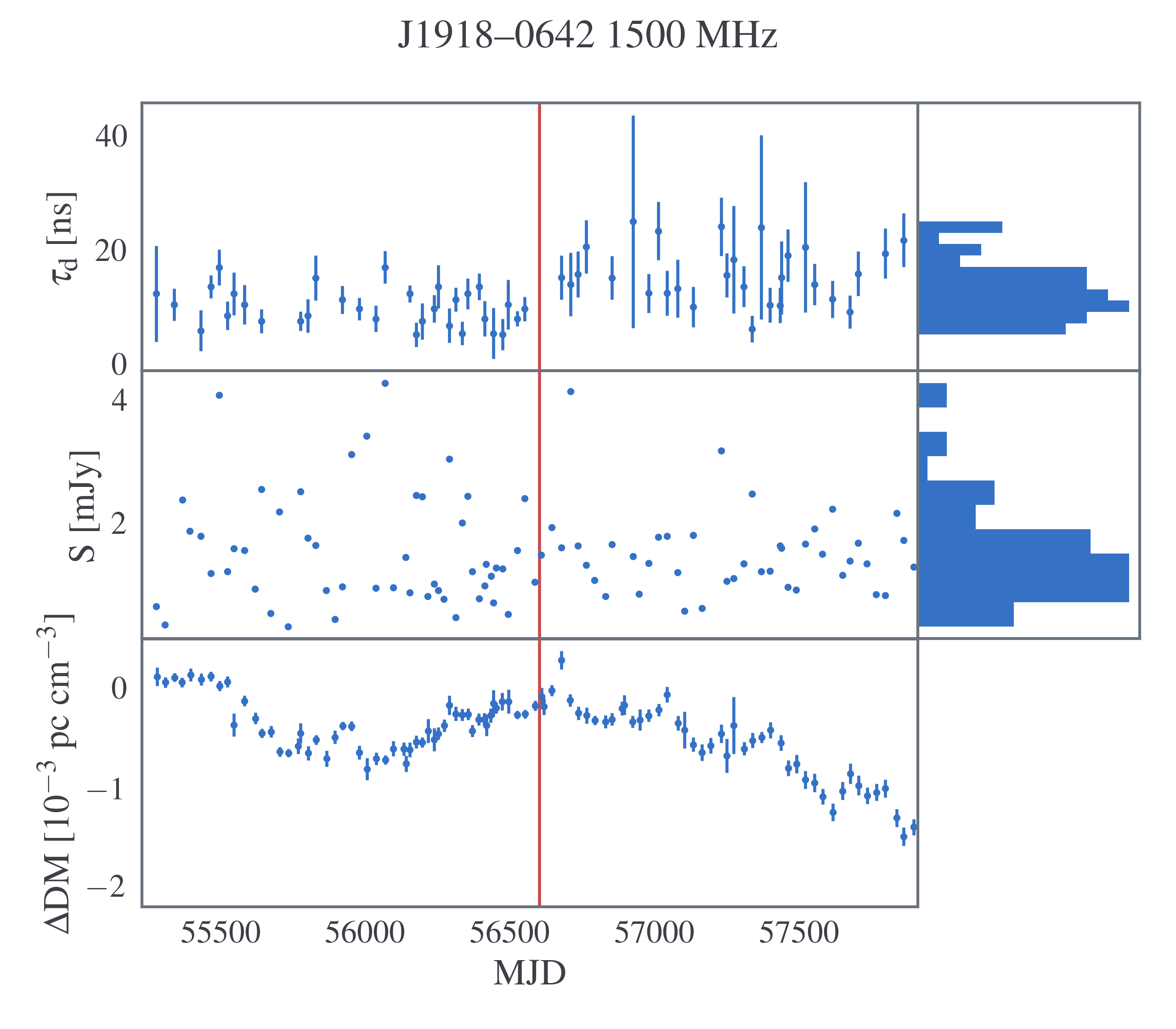}\qquad
\includegraphics[width=.48\textwidth]{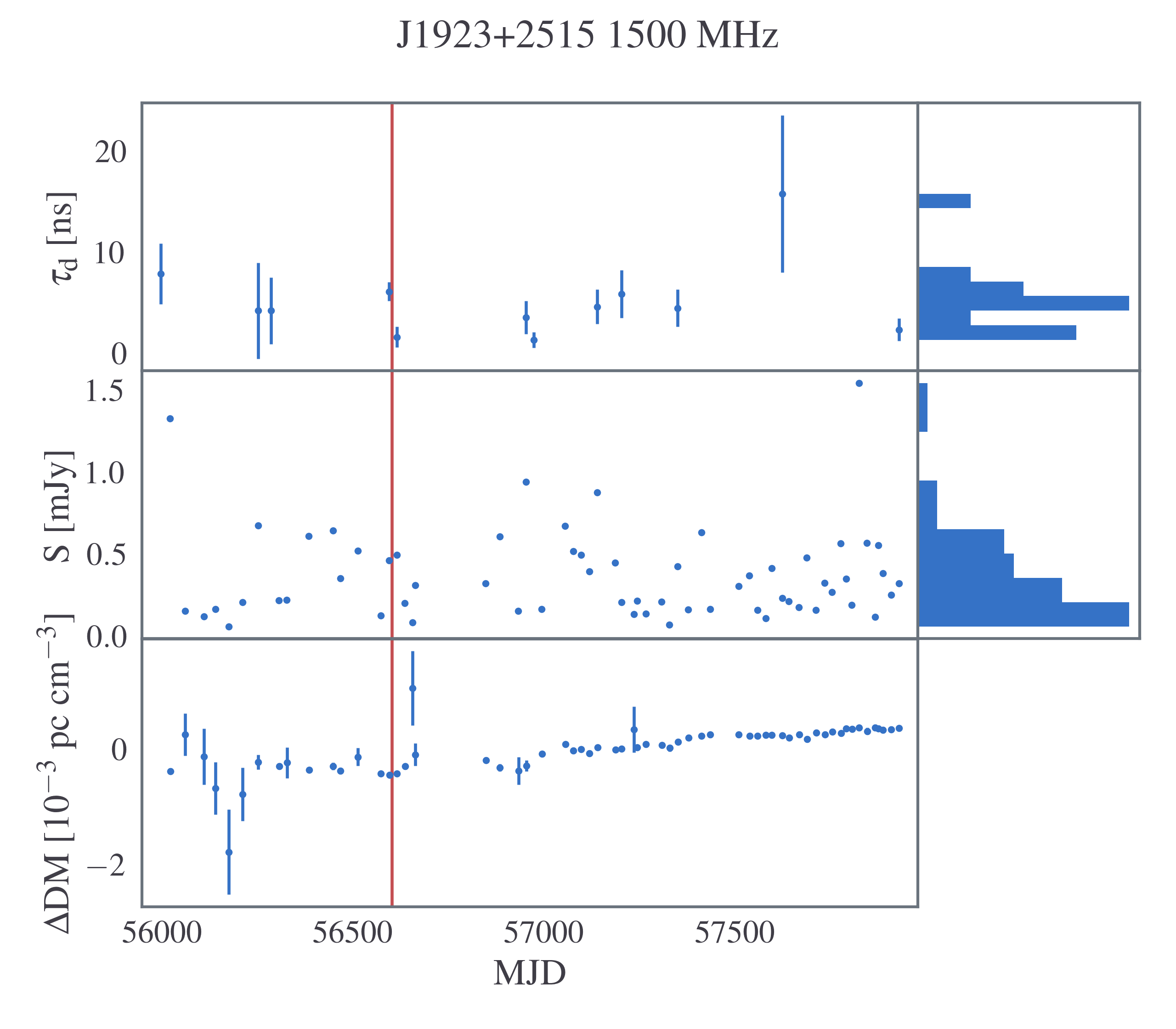}\\
\includegraphics[width=.48\textwidth]{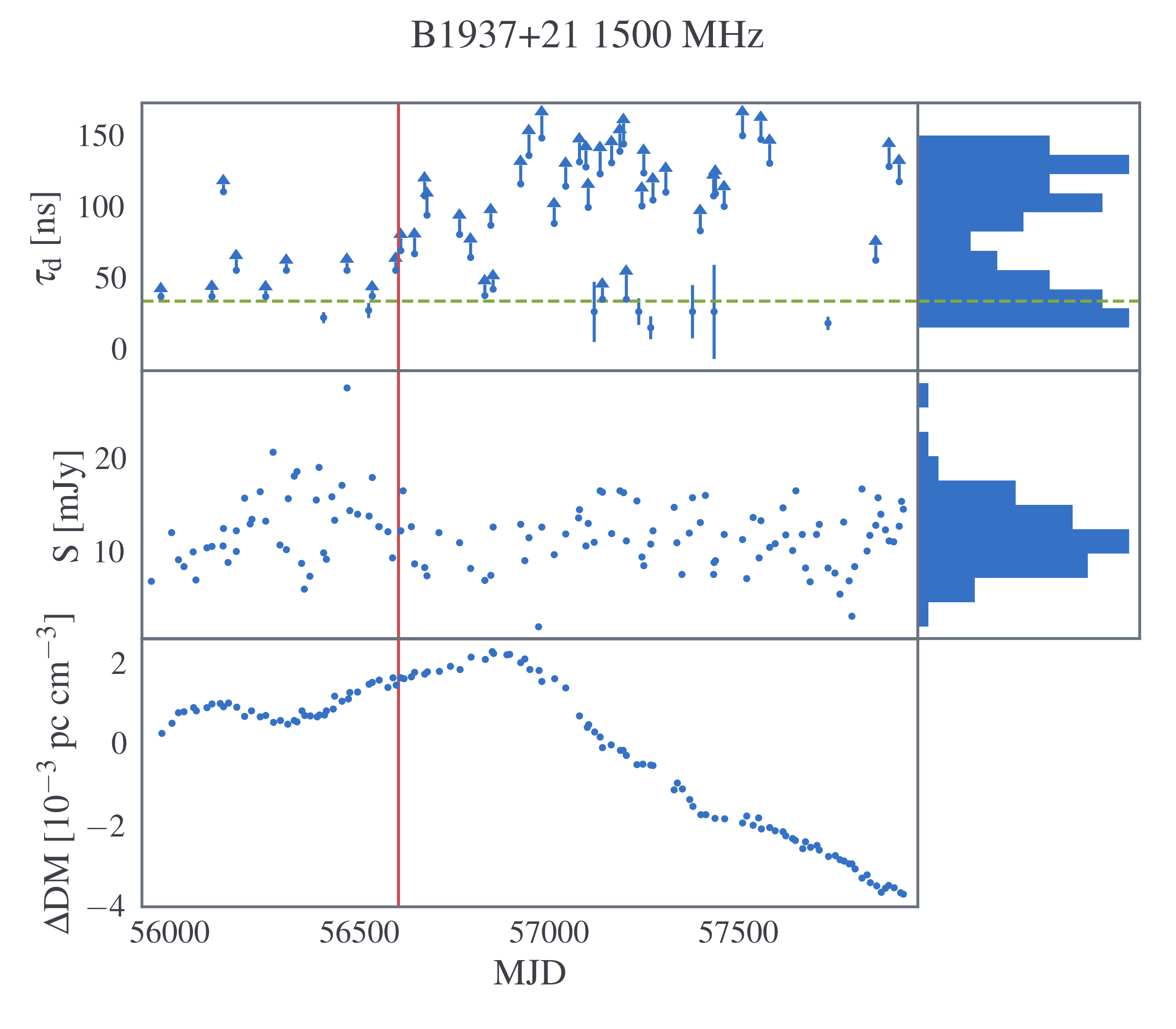}\qquad
\includegraphics[width=.48\textwidth]{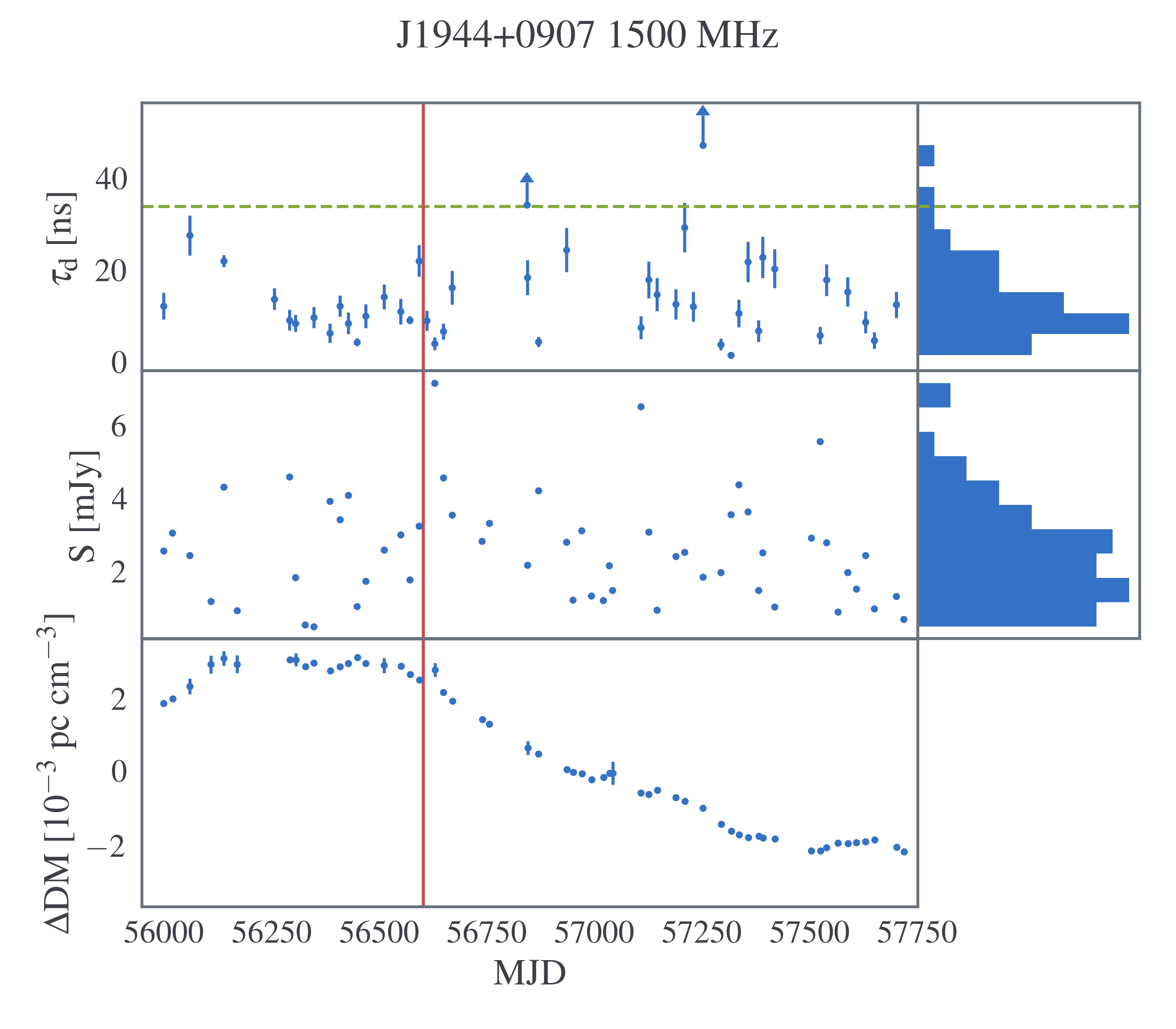}

\label{dmx_plots_5}
\end{figure*}
\begin{figure*}[!ht]
\includegraphics[width=.47\textwidth]{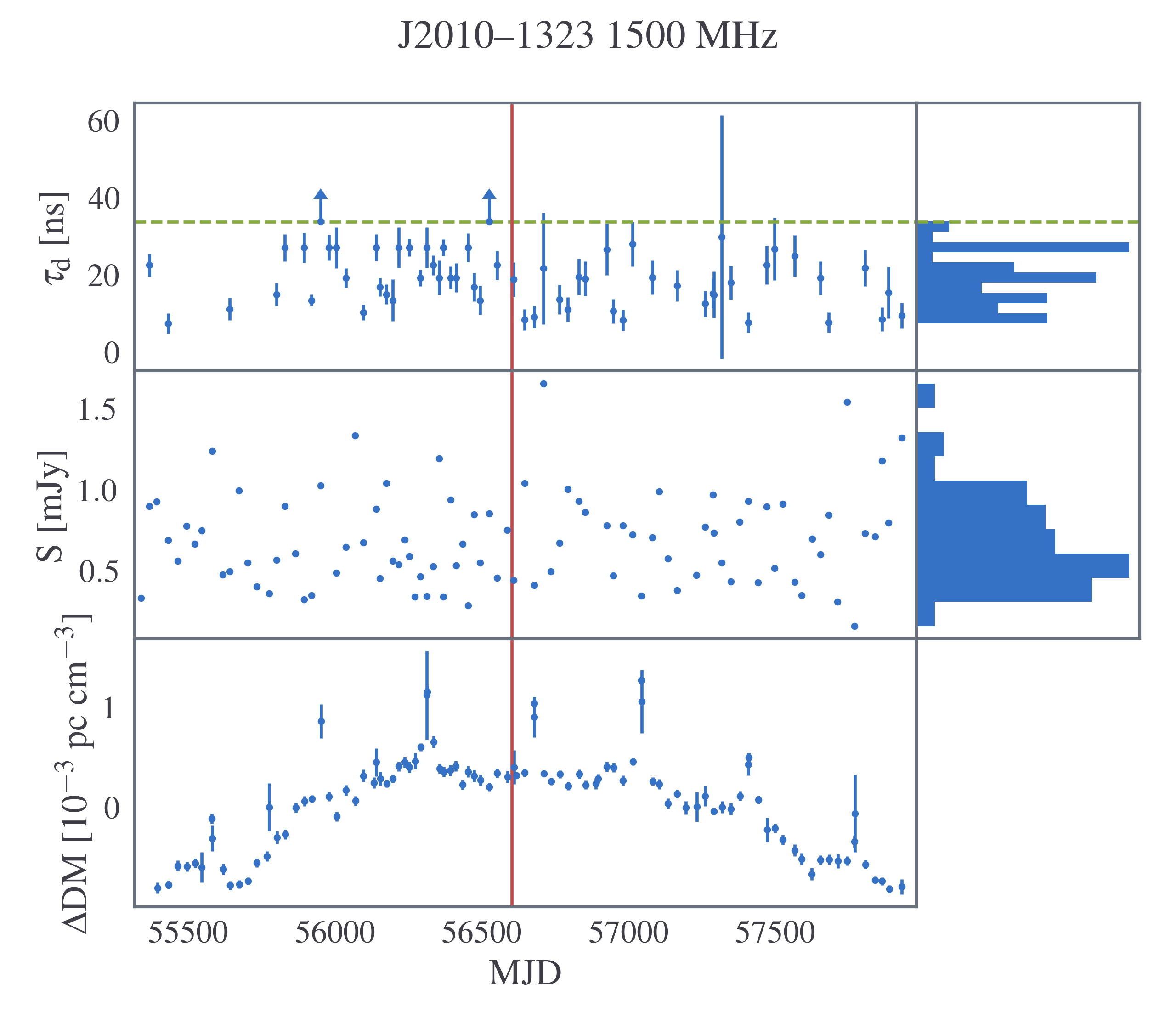}\qquad
\includegraphics[width=.47\textwidth]{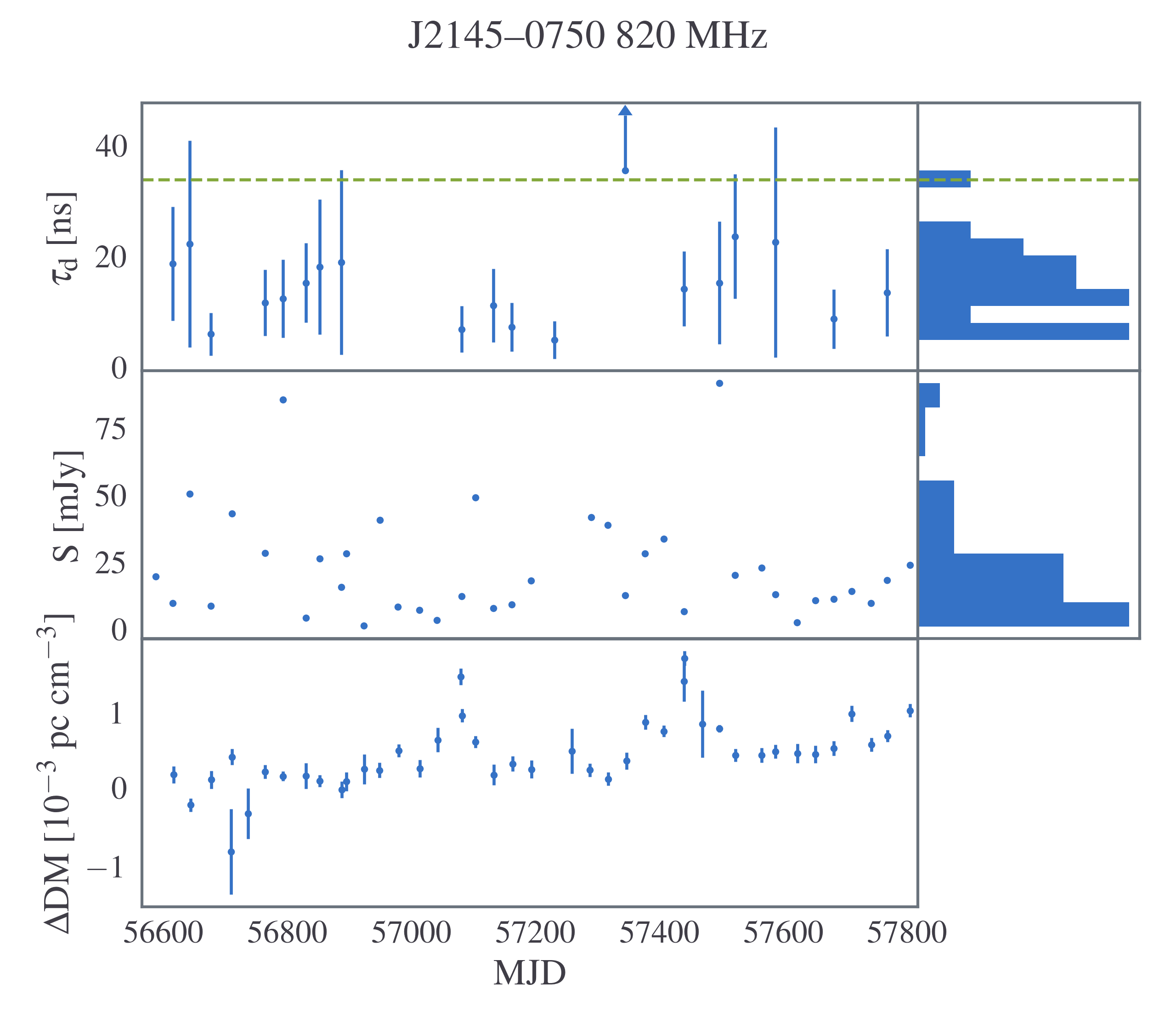}\\
\includegraphics[width=.47\textwidth]{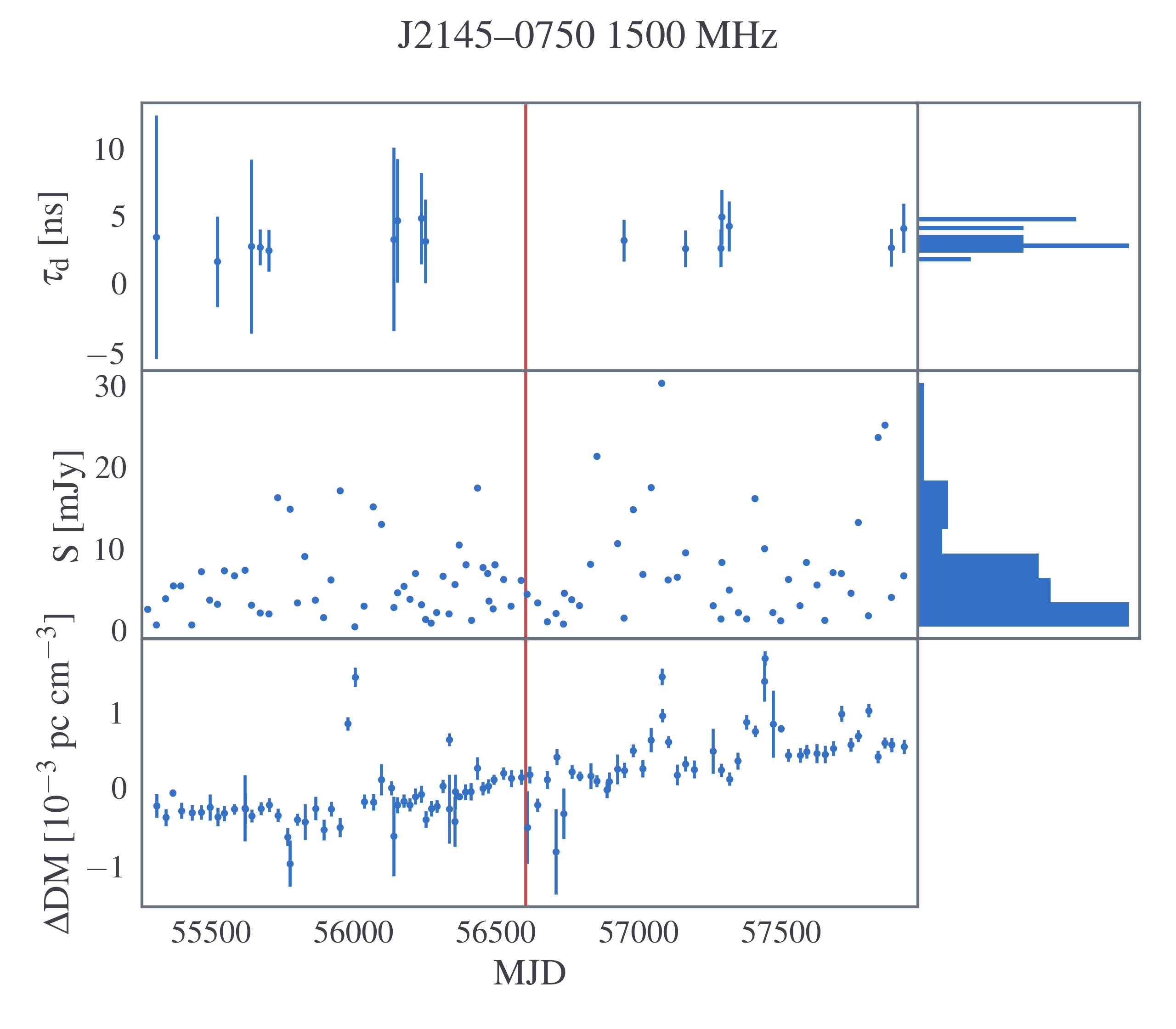}\qquad
\includegraphics[width=.47\textwidth]{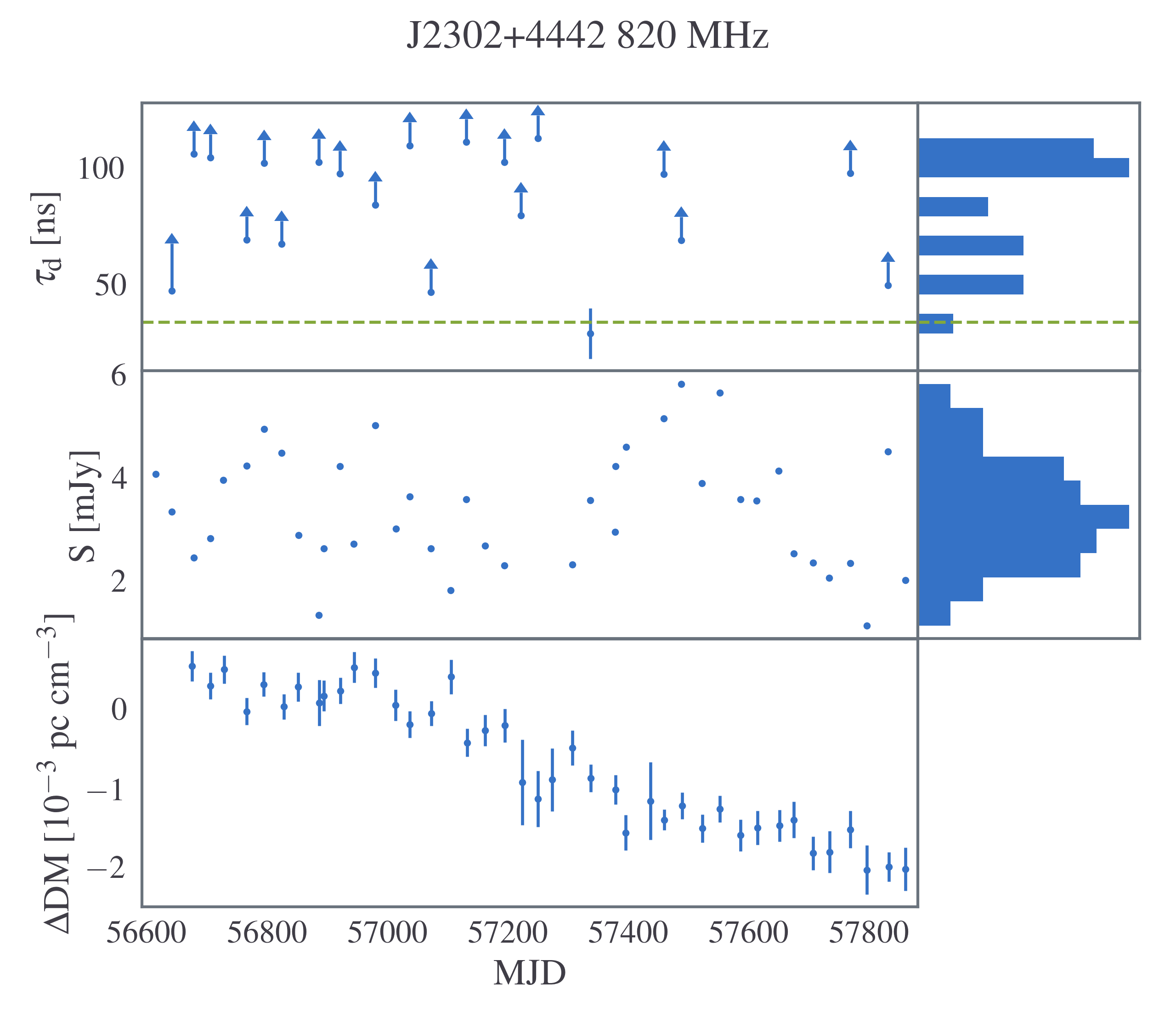}\\
\includegraphics[width=.47\textwidth]{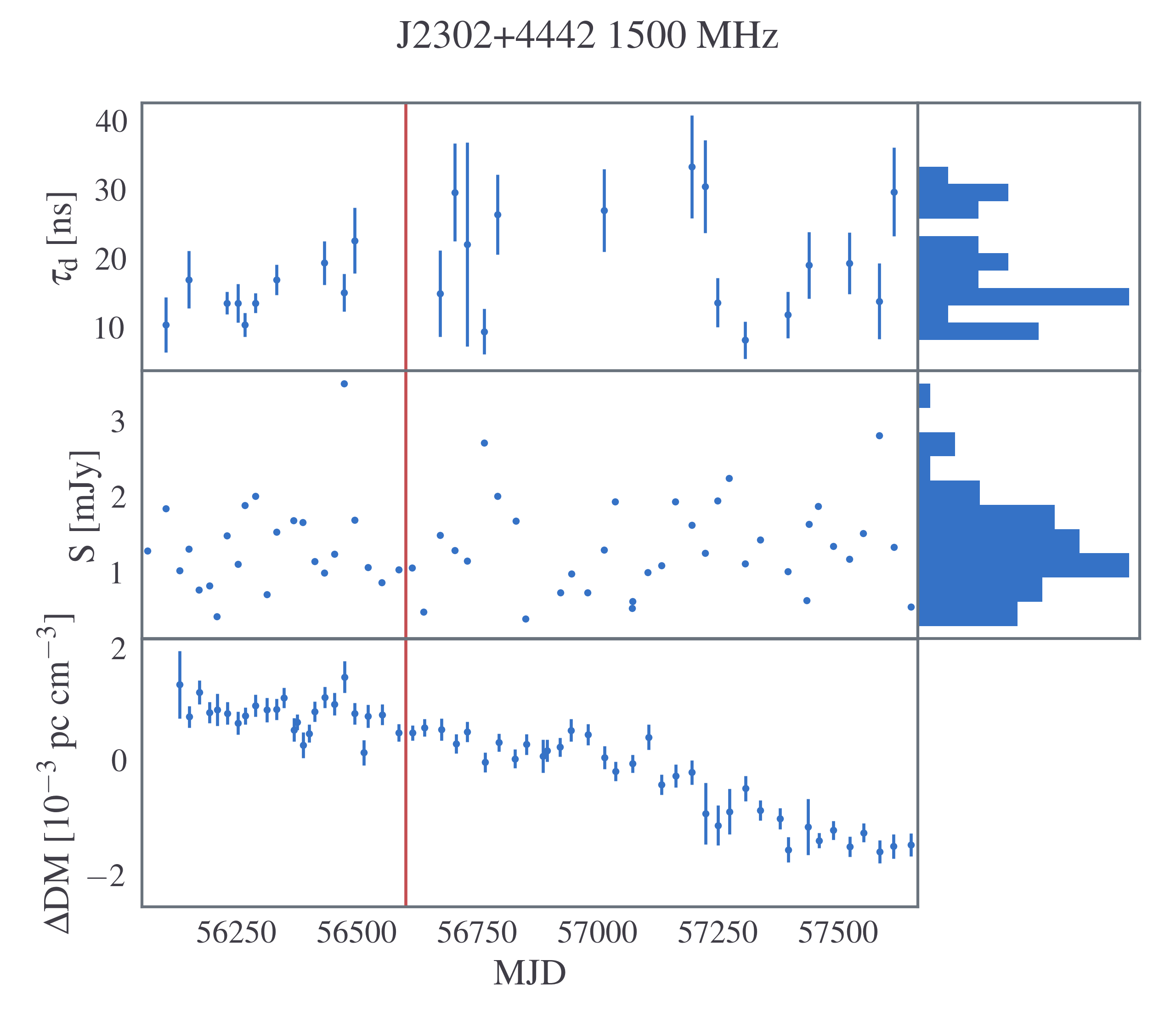}\qquad
\includegraphics[width=.47\textwidth]{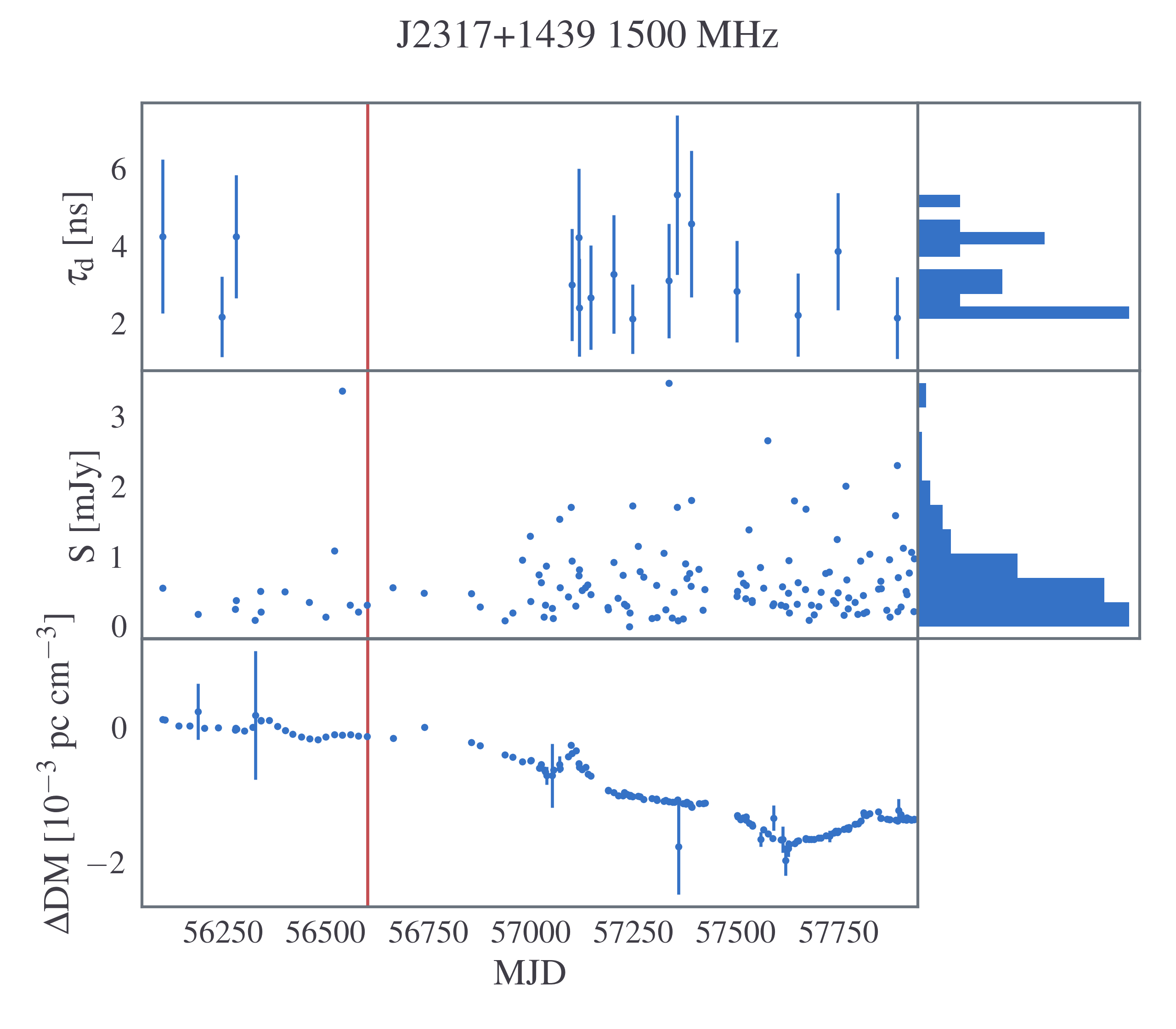}

\caption{Variation of $\tau_{\textrm{d}}$ with time from this paper and \cite{Levin_Scat} at 820 and 1500 MHz along with flux density measurements and $\Delta$DM values that have been mean subtracted and do not account for solar wind effects. Histograms showing the distributions of $\tau_{\textrm{d}}$ and $S$ are shown on the right side of each plot. Errors shown on $\Delta$DM represent the variance. Only pulsars with at least 10 $\tau_{\textrm{d}}$ measurements are shown. Vertical red lines indicate dates separating measurements from \cite{Levin_Scat} and this paper. Horizontal dashed green lines indicate the maximum scattering delay below which we consider measurements lower limits, taken as the scattering delay corresponding to three channel widths (approximately 30 ns).  Fluxes and DMX values were obtained from NANOGrav's wideband timing analysis \citep{wideband}.}
\label{dmx_plots_3}
\vspace{-2.05cm}
\end{figure*}
\clearpage

\noindent velocities if $\sigma_{\textrm{PX}}/\textrm{PX}>0.25$. The values are weighted averages over all measured epochs. Scattering delays over time from this paper and \cite{Levin_Scat} can be found in Figure \ref{dmx_plots_3}. A more detailed discussion of these plots can be found in Section \ref{Dispersion Measure Variations}.

\par In our new observations we were unable to measure scintillation parameters for five pulsars (PSRs J0023$+$4130, J1741$+$1451, B1953$+$29, J2017$+$0603, and J2214$+$3000) that had measureable parameters in \cite{Levin_Scat}. All of these pulsars had 10 or fewer usable observations in that paper and three of them had five or fewer usable measurements. For some of these pulsars, the scintles were too faint, or RFI corrupted too large a portion of each spectrum to obtain scintillation parameters. For example, in quite a few 1.5 GHz observations, the bottom 100--200 MHz of the band was completely corrupted by RFI, and so we either were unable to use many of those epochs or were forced to work with reduced-bandwidth data. In general, the number of measurements obtained on a pulsar-by-pulsar basis in this paper is still largely consistent with \cite{Levin_Scat} over a similar period of time.
\par Some of the pulsars in Figure \ref{dmx_plots_3} show more variability in their scattering delays in our data than in \cite{Levin_Scat} (see PSR J1614--2230 for a good example of this). One reason for this is our ACF calculation method: \cite{Levin_Scat} limited their scintillation bandwidth estimates to integer multiples of the channel bandwidths, whereas our fit interpolated between bins, which means we had more possibilities for quoted scintillation bandwidths and therefore a higher likelihood of variation in our values. The errors in some pulsars are also noticeably larger in our data; this can be attributed largely to RFI, which resulted in  larger finite scintle errors due to the smaller effective observing bands.

\par
\begin{figure}[!ht]
\includegraphics[scale=.55]{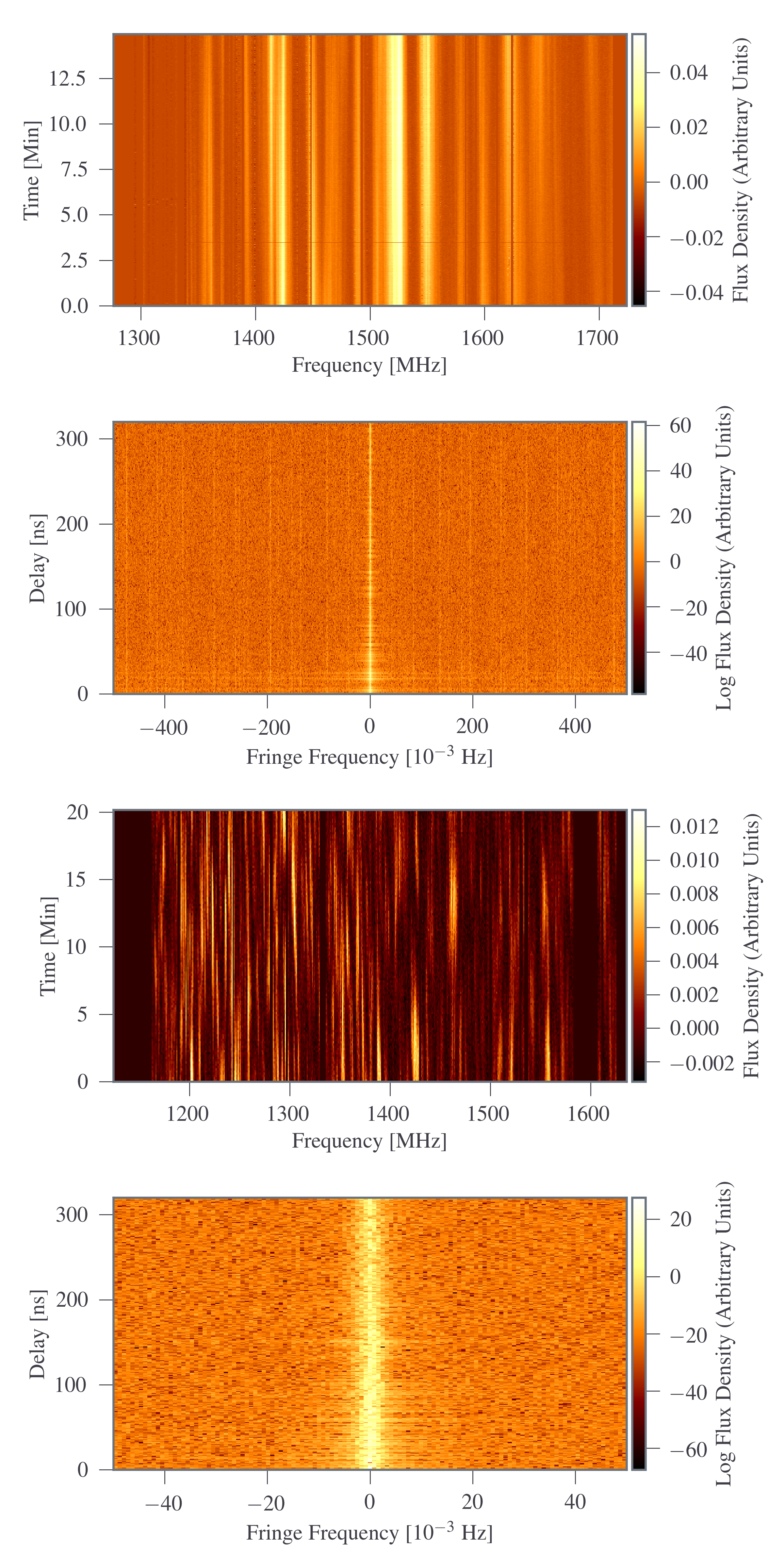}
\centering
\caption{Dynamic (top) and secondary (second from top) spectra of PSR B1855$+$09 on MJD 56640 and dynamic (second from bottom) and secondary (bottom) spectra of PSR B1937$+$21 on MJD 56892. For PSR B1855+09, the intensity in the secondary spectrum drops off at higher delays, indicating we are fully resolving most of its power and properly measuring its scattering delays. For PSR B1937+21, the intensity in the secondary spectrum does not drop off at higher delays, indicating we are not fully resolving its power and are therefore underestimating its scattering delays.}
\label{all_sec}
\end{figure}

 We treat the weighted average scintillation bandwidth measurements for five pulsars at 820 MHz and two pulsars at 1500 MHz as upper limits, as their estimates are less than three times the channel width. For these pulsars, in particular PSRs B1937$+$21 and J1910$+$1256, there are typically many epochs on which the bandwidth was unresolved. We can also demonstrate insufficient frequency resolution for some pulsars through the calculation of secondary spectra, which are the two-dimensional Fourier power spectra of the dynamic spectra. The delay axis in a given secondary spectrum is directly proportional to the relative time delay incurred from scattering \citep{secondary, Hemberger_2008}. The Fourier relation between the two spectra is also especially useful, as small features such as unresolved scintles in a dynamic spectrum will manifest as large, clearly visible features in a secondary spectrum. As an example, in Figure \ref{all_sec}, we show dynamic and secondary spectra of PSRs B1855$+$09 and B1937$+$21.

\par
\begin{figure}[!ht]
\includegraphics[scale=.52]{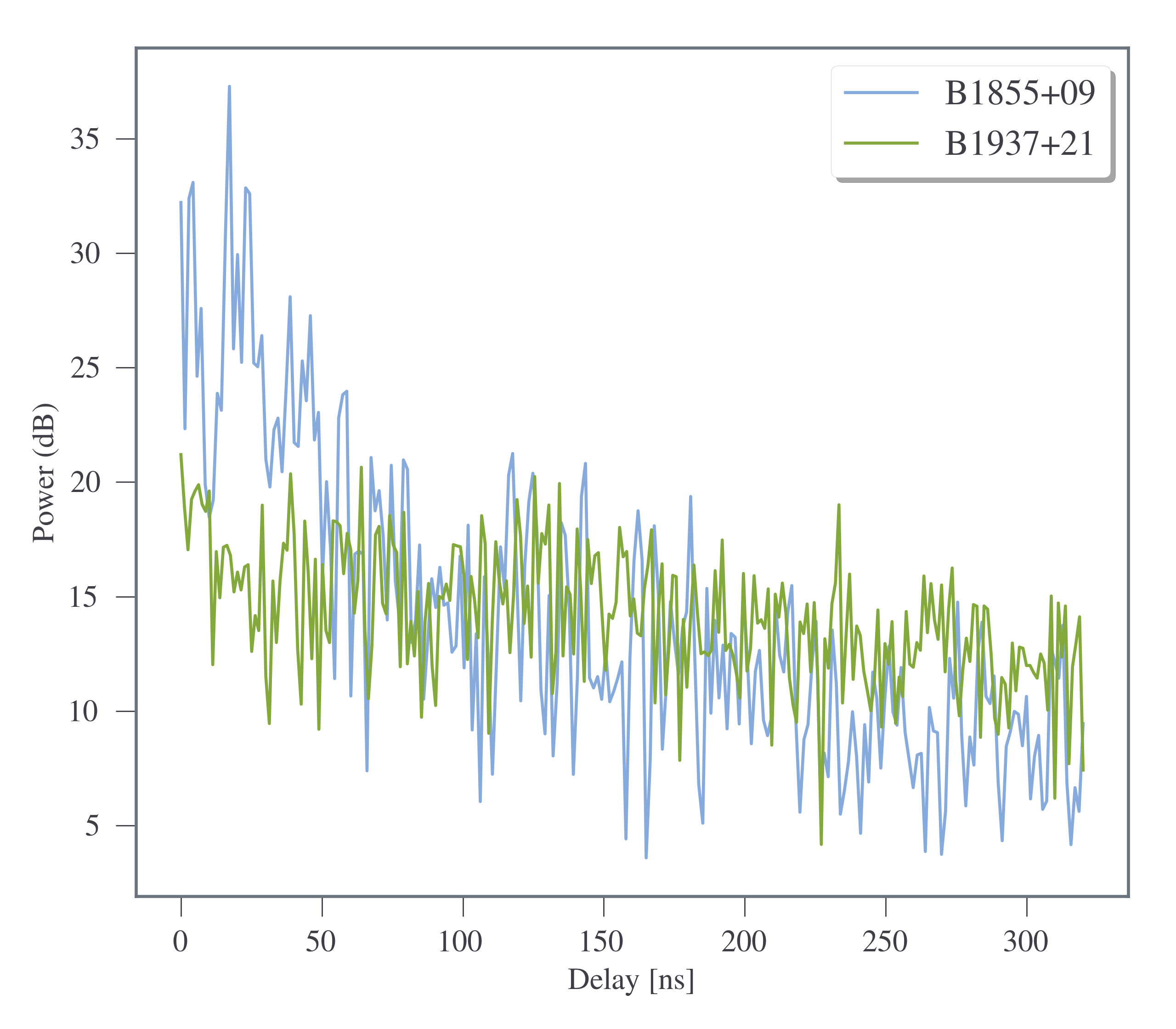}
\centering
\caption{The power in the secondary spectra of PSRs B1855$+$09 and B1937$+$21 as a function of delay, calculated by summing over only the fringe frequencies with visible power. The power of PSR B1937$+$21 does not decrease at higher delays, indicating that we are underestimating scattering delays. Conversely, the power of PSR B1855$+$09 falls off with delay, indicating we are more accurately estimating its scattering delays and resolving a greater fraction of its scintles.}
\label{power_compare}
\end{figure}

 In Figure \ref{power_compare}, we show the average power in the secondary spectrum as a function of delay over the fringe frequency channels in the secondary spectra where power was visible. This corresponds approximately to the middle four channels for PSR B1855$+$09 and the middle seven channels for PSR B1937$+$21. We find little dependence on the exact number of channels used for this analysis. PSR B1937+21 displays no decrease in intensity with delay in its secondary spectrum, indicating its scintles are not fully resolved. For this pulsar, we see this effect in all of its secondary spectra, meaning we should interpret all measured delays as lower limits. The flux density in PSR B1855$+$09's secondary spectra drops off at higher delays, indicating we are resolving more of its scintles, even if narrower ones may not be completely resolved. Some of the remaining power at higher delays could also be due in part to unmitigated RFI. 
\par Overall, we find that power does not dissipate for epochs on which the scintillation bandwidth is less than three channel bandwidths, supporting our decision to treat these measurements as upper limits.

\subsection{Scaling Over Multiple Frequency Bands}
\label{4.2}
\par Since we were unable to resolve scintles at 820 MHz, power-law indices for PSRs B1937$+$21 and J2010--1323 as shown in Table \ref{Scale_Table} were found exclusively by splitting the 1500 MHz passband into 200 MHz subbands, measuring the scattering in each, and fitting via a power law (as in \cite{Levin_Scat}). Indices for PSRs B1855$+$09 and J2302$+$4442 were found using both this method and extended fits that included the 820 MHz passband. The measured indices for all four of these pulsars were shallower than the $-4.4$ expected for a Kolmogorov medium, with only PSR B1937$+$21 yielding an index steeper than $-3$ while the other three pulsars clustered around $-2.5$.   
\par We were also able to obtain first-order estimates of the scaling index for 15 pulsars using the method described in Section \ref{scale_beh}. These results, along with the results described above, are shown in Table \ref{Scale_Table}. As in \cite{Levin_Scat}, all of our measured scaling indices are shallower than the value of $-4.4$ that is expected for a Kolmogorov medium under the simplest assumptions, with only two of these indices being steeper than $-3$. There was also considerable range in the indices measured, with values spanning from $-0.7$ to $-3.5$. We quote upper limits on indices where the majority of 820-MHz scattering delays are lower limits.
\par\cite{Levin_Scat} found noticeably different scaling indices from multiple measurements of various pulsars, indicting that a pulsar's scaling index may vary with time as it moves through the ISM. As mentioned earlier, examining time variability was not possible using the multiband method, since we rarely, if ever, had detectable measurements for two frequencies within about a week of each other. A large part of this was because RFI contamination was much more prominent at the 1500 MHz band than in the 820 MHz band, so it was generally easier to get consistent measurements only at lower frequencies. This RFI contamination also made it difficult to get many epochs that were useable for the subband analysis. Overall, since our multiband method can tell whether a scaling index is shallower or steeper than $-4.4$, even if it is not as precise due to the frequency scaling, and both methods found shallower indices, we can conclude that these two methods agree with each other. 

\begin{figure}[!ht]
\includegraphics[scale=.52]{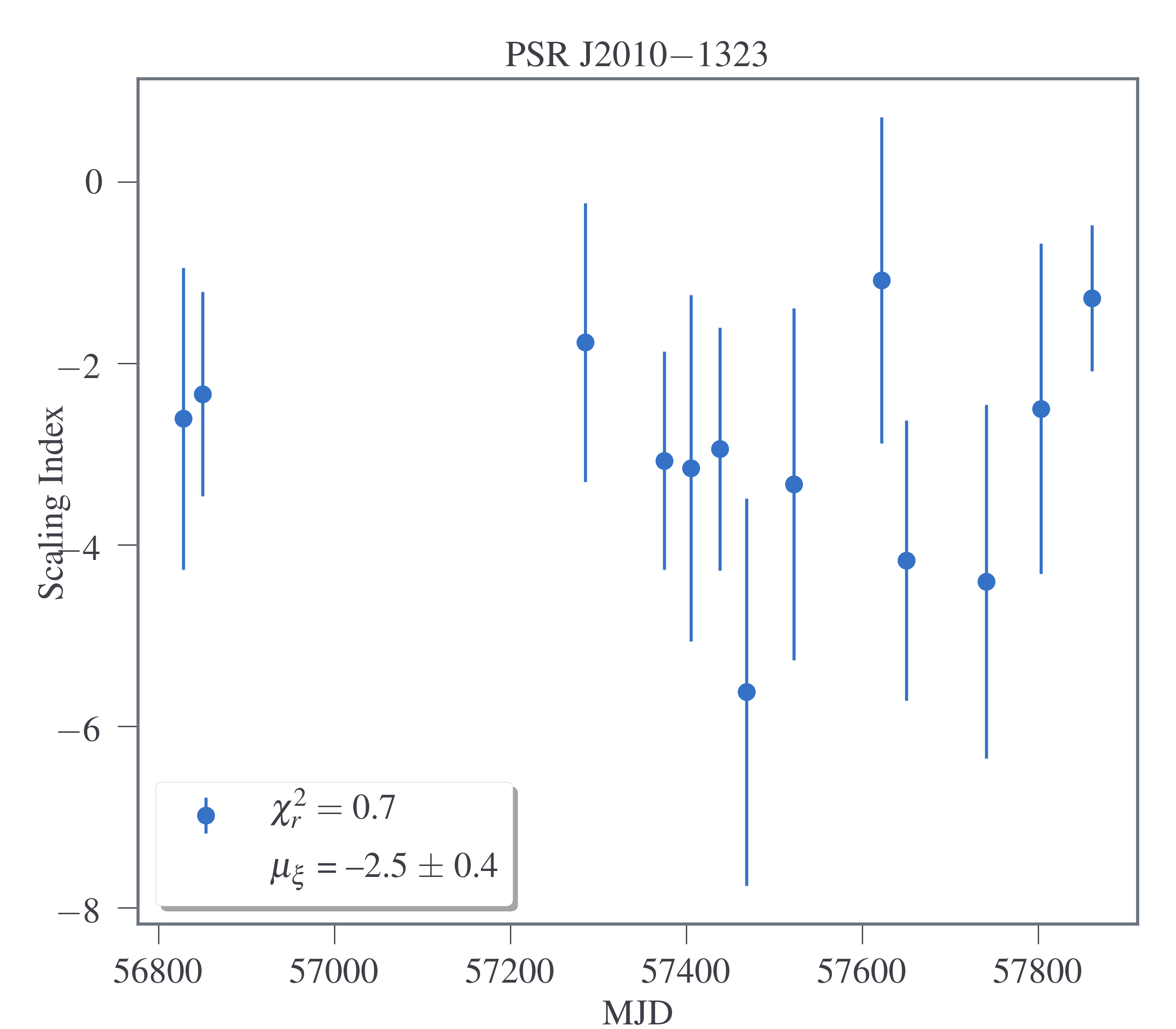}
\centering
\caption{Measured scaling indices from individual epochs of PSR J2010--1323. The level of variation from epoch to epoch is low and the weighted average scaling index is much shallower than $-4.4$.}
\label{20_scale}
\end{figure}

\begin{figure}[!ht]
\includegraphics[scale=.52]{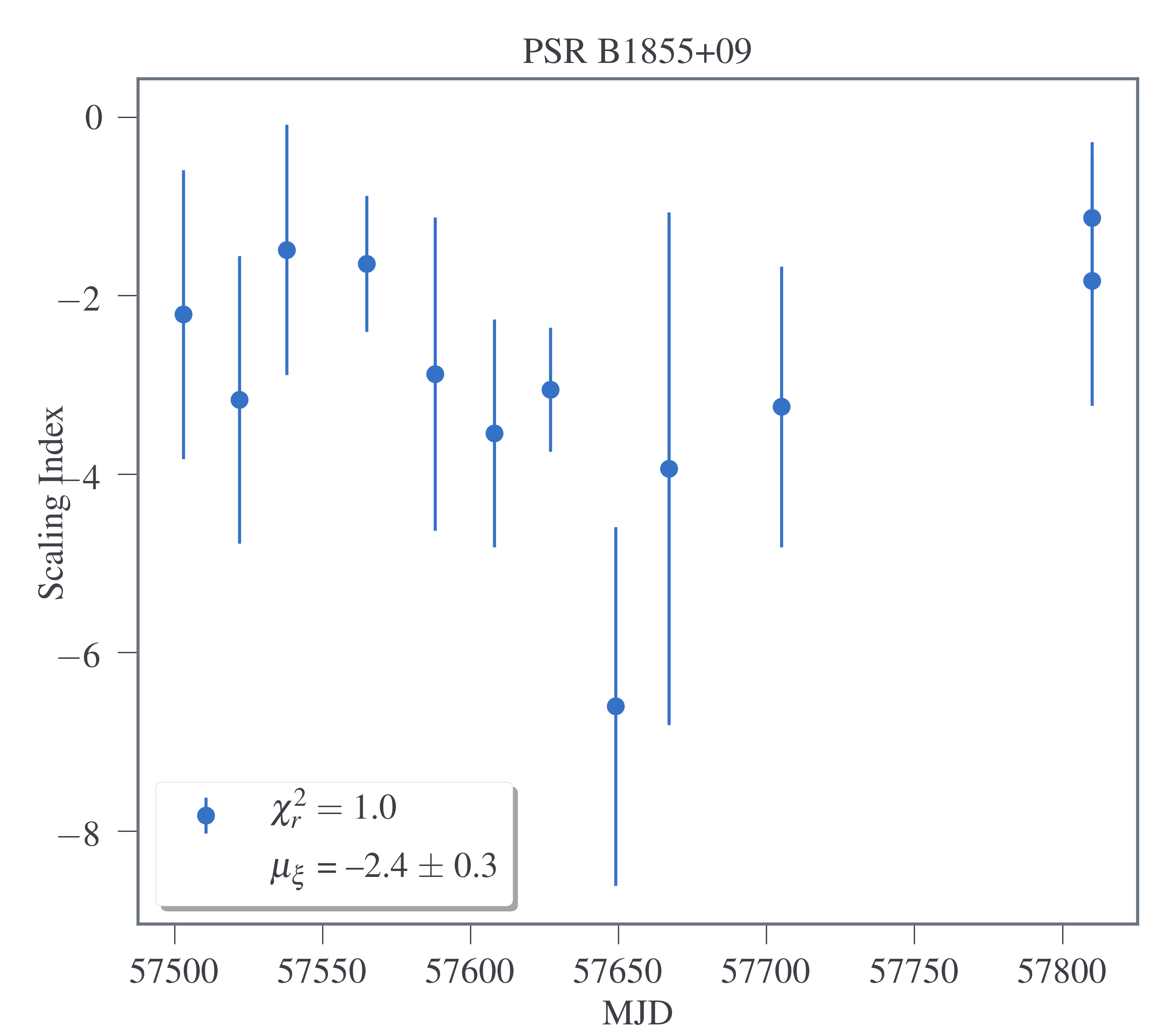}
\centering
\caption{Measured scaling indices from individual epochs of PSR B1855$+$09. The level of variation from epoch to epoch is low and the weighted average scaling index is much shallower than $-4.4$.}
\label{19_scale}
\end{figure}

\subsection{Transverse Velocity Measurements}
\par
\begin{figure}[!ht]
\includegraphics[scale=.52]{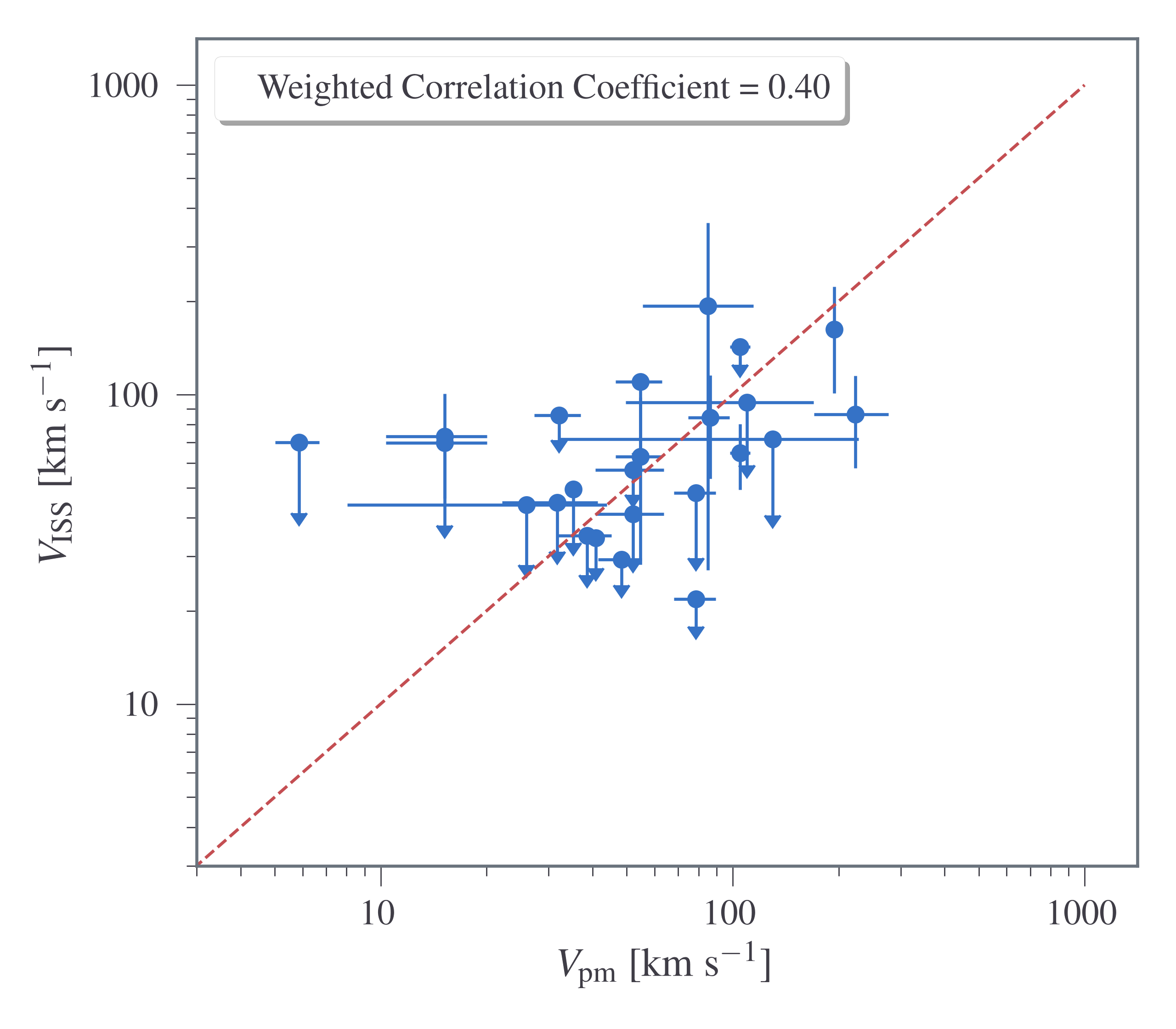}
\centering
\caption{Transverse velocities derived from proper motion vs those determined through scintillation. Downward facing arrows indicate upper limits. The modest correlation suggests that our assumption of a screen at the midpoint between us and the pulsar is roughly correct.}
\label{v_trans_plot}
\end{figure}

 A comparison between transverse velocities derived from scintillation parameters and those derived from proper motions is shown in Table \ref{table_3} and Figure \ref{v_trans_plot}. Because we expect that $V_{\textrm{ISS}}$ and $V_{\textrm{pm}}$ should be equal under the assumptions made in Section \ref{t_velo} and based on surveys such as \cite{nicastro}, we use the Pearson correlation coefficient, 
\begin{equation}
\label{pearson}
r_{p} = \frac{\sigma_{x,y}^2}{\sqrt{\sigma_{x}^2 \sigma_{y}^2}},
\end{equation} 
where $\sigma_{x,y}^2$ is the covariance between some parameters $x$ and $y$ and $\sigma_x$ and $\sigma_y$ are the variances of $x$ and $y$, respectively.
Relative screen distances calculated by assuming that $V_{\textrm{ISS}}$ is equal to $V_{\textrm{pm}}$ are also shown. All $V_{\textrm{pm}}$ values were calculated using proper motions found in \cite{alam2020nanograv}. We are not sensitive to epoch-to-epoch variations in $V_{\textrm{ISS}}$ because our scintles are not always resolved in time for every epoch that they are resolved in frequency. Because of this, all $V_{\textrm{ISS}}$ values were calculated by using the weighted averages of scintillation bandwidth and timescale from Table \ref{table_2} in Equation \ref{iss_velo} for pulsars with at least two epochs for which $\Delta \nu_{\textrm{d}}$ was measured.

\par For pulsars where we could not resolve scintles in time, we have assigned $\Delta t_{\textrm{d}}$ lower limits of 30 minutes, resulting in upper limits on transverse velocity. We used parallax distances for calculating $V_{\textrm{ISS}}$ and $V_{\textrm{pm}}$ if the distance error was $<$ 25\%; otherwise we used distances determined by DM from NE2001 and assumed a 20\% uncertainty \citep{NE2001}.

\begin{deluxetable}{CC}
\tablewidth{\columnwidth}
\tablecolumns{2}
\tablecaption{Estimated Scattering Delay Scaling Indices \label{Scale_Table}}
\tablehead{\colhead{Pulsar} &  \colhead{$\xi$}}
\startdata
\text{J}0613\textrm{$--$}0200 & <$-$1.8 $\pm$ 0.8 \\
\text{J}0636{+}5128 & <$-$2.5 $\pm$ 0.1 \\
\text{J}0740{+}6620 & $-$2.4 $\pm$ 0.6 \\
\text{J}1024\text{$--$}0719 & $-$1.5 $\pm$ 0.6 \\
\text{J}1125{+}7819 & $-$3.5 $\pm$ 0.2 \\
\text{J}1455\textrm{$--$}3330 & $-$3.5 $\pm$ 0.4 \\
\text{J}1614\textrm{$--$}2230 & <$-$1.3 $\pm$ 0.9 \\
\text{J}1744\textrm{$--$}1134 & $-$1.8 $\pm$ 0.3 \\
\text{B}1855{+}09 & $-$0.7 $\pm$ 0.5 \\
\text{B}1855{+}09 & $-$2.4 $\pm$ 0.3$^{\dagger}$ \\
\text{J}1909\textrm{$--$}3744 & <$-$2.9 $\pm$ 0.3 \\
\text{J}1910{+}1256 & <$-$1.0 $\pm$ 0.3 \\
\text{B}1937{+}21 & $-$3.6 $\pm$ 0.1$^{\dagger}$ \\
\text{J}1944{+}0907 & $-$1.0 $\pm$ 0.3 \\
\text{J}2010\text{$--$}1323 & $-$2.5 $\pm$ 0.4$^{\dagger}$ \\
\text{J}2145\text{$--$}0750 & $-$2.1 $\pm$ 0.4 \\
\text{J}2302{+}4442 & <$-$1.3 $\pm$ 0.4 \\
\text{J}2302{+}4442 & $-$2.6 $\pm$ 1.1$^{\dagger}$ \\
\text{J}2317{+}1439 & $-$2.3 $\pm$ 0.8
\enddata
\tablecomments{Measurements with a dagger were calculated using non-stretched subbands, and others used measurements at two frequencies based on stretched spectra. Uncertainties on values with daggers represent the weighted error of all measured indices, while uncertainties on values without daggers represent the 1$\sigma$ errors on $\xi$ in the model fits. We quote upper limits on indices where the majority of 820-MHz scattering delays are lower limits.}
\end{deluxetable}

\begin{deluxetable}{CCCCC}[h!]
\tablewidth{0pt}
\tablecolumns{5}
\tablecaption{Pulsar Transverse Velocities Inferred from Interstellar Scattering and Proper Motions \label{table_3}}
\tablehead{\colhead{Pulsar} &  \colhead{Frequency} &       \colhead{$V_{\textrm{ISS}}$} &     \colhead{$V_{\textrm{pm}}$} & \colhead{$D_{\textrm{o}}/D_{\textrm{p}}$}\\ \colhead{} & \colhead{(MHz)} & \colhead{(km s$^{-1}$)} & \colhead{(km s$^{-1}$)} & \colhead{}}
\startdata
\text{J}0613\text{$--$}0200 &        820 &   110 $\pm$ 45 &    55 $\pm$ 8 &  0.3 $\pm$ 0.2\\
\text{J}0613\text{$--$}0200 &       1500 &   63 $\pm$ 35 &    55 $\pm$ 8 & 0.8 $\pm$ 0.8 \\
\text{J}0636{+}5128 &        820 &    $<$75 $\pm$ 43 &    15 $\pm$ 5 & 0.04 $\pm$ 0.05 \\
\text{J}0636{+}5128 &       1500 &    70 $\pm$ 32 &    15 $\pm$ 5 & 0.05 $\pm$ 0.04\\
\text{J}0931\text{$--$}1902 &       1500 &    $<$44 $\pm$ 16 &   26 $\pm$ 18 & $>$0.4 $\pm$ 0.3\\
\text{J}1024\text{$--$}0719 &        820 &  86$\pm$ 12 &  220 $\pm$ 90 & 6.7 $\pm$ 4.4\\ 
\text{J}1125{+}7819 &        820 &    84 $\pm$ 21 &   86 $\pm$ 12 & 1.1 $\pm$ 0.8\\
\text{J}1614\text{$--$}2230 &        820 &   140 $\pm$ 20 &  110  $\pm$ 10 & 0.5 $\pm$ 0.1 \\
\text{J}1614\text{$--$}2230 &       1500 &    65 $\pm$ 16 &  110 $\pm$ 10 & 2.6 $\pm$ 1.3 \\
\text{J}1640{+}2224 &       1500 &    $<$94 $\pm$ 35 &   110 $\pm$ 60 & $>$1.4 $\pm$ 1.0\\
\text{J}1713{+}0747 &       1500 &     $<$44 $\pm$ 12 &    36 $\pm$ 1 & $>$0.7 $\pm$ 0.3\\
\text{J}1738{+}0333  &    1500 & $<$72 $\pm$ 31 &   130 $\pm$ 100 & $>$3.3 $\pm$ 2.9 \\  
\text{J}1744\text{$--$}1134 &       1500 &  <34 $\pm$ 7 &   41 $\pm$ 1 & 1.4 $\pm$ 0.6\\
\text{J}1853{+}1303 &       1500 &     $<$45 $\pm$ 14 &     32 $\pm$ 10 & $>$0.5 $\pm$ 0.3\\
\text{B}1855{+}09 &       1500 &     $<$35 $\pm$ 9 &    39 $\pm$ 7 & $>$1.2 $\pm$ 0.6\\
\text{J}1909\text{$--$}3744 &        820 &   <160 $\pm$ 60 &   190 $\pm$ 4 & >1.4 $\pm$ 1.1\\
\text{J}1910{+}1256 &        820 &     $<$48 $\pm$ 19 &    79 $\pm$ 11 &  $>$2.7 $\pm$ 2.1\\
\text{J}1910{+}1256 &       1500 &     $<$22 $\pm$ 4 &    79 $\pm$ 11 & $>$ 12.9 $\pm$ 4.9\\
\text{J}1918\text{$--$}0642 &       1500 &     $<$29 $\pm$ 5 &    48 $\pm$ 7 & $>$2.7 $\pm$ 0.9\\
\text{B}1937{+}21 & 1500 & < 70 $\pm$ 29 & 6 $\pm$ 1 & > 0.01 $\pm$ 0.01  \\
\text{J}2010\text{$--$}1323 &       1500 &    190 $\pm$ 170 &    85 $\pm$ 29 & 0.2 $\pm$ 0.3\\
\text{J}2145\text{$--$}0750 &        820 &    $<$41 $\pm$ 12 &   52 $\pm$ 11 & $>$1.6 $\pm$ 0.9\\
\text{J}2145\text{$--$}0750 &       1500 &    $<$57 $\pm$ 9 &   52 $\pm$ 11 & $>$0.8 $\pm$ 0.3\\
\text{J}2317{+}1439 &       1500 &     $<$86 $\pm$ 14 &    32 $\pm$ 5 & $>$0.1 $\pm$ 0.1
\enddata
\tablecomments{ $V_{\textrm{ISS}}$ values were calculated using the weighted averages of $\Delta \nu_{\textrm{d}}$ and $\Delta t_{\textrm{d}}$ found in Table \ref{table_2} and the assumption that the scattering screen is equidistant from the pulsar and Earth. Uncertainties are calculated by propagating the weighted errors on the scintillation measurements with the uncertainties on pulsar distance and proper motion. Many of the $V_{\textrm{ISS}}$ estimations are upper limits, since scintillation timescale lower limits were used. We calculated $D_{\textrm{o}}/D_{\textrm{p}}$ by assuming that $V_{\textrm{pm}}$ is correct, setting it equal to $V_{\textrm{ISS}}$, and solving for $D_{\textrm{o}}/D_{\textrm{p}}$. Some measurements are also upper limits due to resolution limits on scintillation bandwidths. Due to the large uncertainty in both PSRs J1125$+$7819 and J1910$+$1256's parallax measurements, their distance was determined by DM via NE2001. All measurements and errors have been rounded to the last significant digit shown.}
\end{deluxetable}

\section{Discussion}\label{ne compare}

\subsection{Scattering Variability and Correlations with Dispersion Measure \& Flux Density Variations}\label{Dispersion Measure Variations}
 \par We have searched for correlations between scattering delays and DM variations and scattering delay and flux density for pulsars with at least 10 scattering measurements. These coefficients were determined using only epochs where both scattering delay and $\Delta$DM or flux data were available. The data and their corresponding correlations with $\Delta$DM and flux density are shown in Figure \ref{dmx_plots_3} and Table \ref{table_chi}, respectively. Here, we only examine the linear Pearson correlation (Equation \ref{pearson}) since theoretical predictions in the literature show support for this type of correlation. DMs and flux densities were obtained from NANOGrav's wideband timing analysis \citep{wideband}. DMX determines DM variations by treating DM($t$) as a piecewise constant and fitting for a new DM at up to six-day intervals along with the rest of the parameters in our timing model. 
\par In order to examine the variability of the flux density in each pulsar as a function of time, we performed a reduced $\chi^2$ analysis using a model consisting of the weighted average of the combined flux densities. For a given time series with measurements of a parameter, $x$, we define our reduced $\chi^2$ as 

\begin{equation}
\label{red_chi}
\chi_r^2 = \frac{1}{N -1}\sum \frac{(x(t)-\overline{x})^2}{\sigma^2(t)},
\end{equation}
where $N$ is the number of measurements, $\overline{x}$ is the weighted average of the measurements, $x(t)$ is the measurement at time $t$, and $\sigma^2(t)$ is the measurement variance at time $t$. In the case of the fluxes, $x(t)$ and $\overline{x}$ in Equation \ref{red_chi} represent the flux density as a function of time and the weighted average of the flux density, respectively. We list these values in Table \ref{table_chi}.

\par While there may visually be correlations between scattering delay and DM or flux, these correlations do not appear linear. For this reason, in addition to examining linear correlations using the Pearson correlation coefficient (Equation \ref{pearson}), we examine general correlations using the Spearman correlation coefficient,

\begin{equation}
\label{spearman}
r_s = 1- \frac{6\sum_{i=1}^N[\textrm{rg}(y_i)-\textrm{rg}(x_i)]^2}{N(N^2-1)},
\end{equation}
where rg($y_i$) and rg($x_i$) are the ranks of the $i^{\textrm{th}}$ values of $y$ and $x$, respectively, and $N$ is the number of data points being used. The rank of a value is defined by its size relative to other quantities in a shared data set, with the smallest value having a rank of one, the second smallest value having a rank of two, and so on. 

\par We also examined the variability of scattering delays as a function of time by performing a reduced $\chi^2$ analysis using the combined scattering delays from this paper and \cite{Levin_Scat}, with $x(t)$ and $\overline{x}$ in Equation \ref{red_chi} representing the scattering delays as a function of time and the weighted average of the scattering delays, respectively. The results are shown in Table \ref{table_chi}.

\par Finally, we explored the variability of scattering delays by examining the scintillation bandwidth modulation index, defined as

\begin{equation}
\label{mod_index}
m_\textrm{b} = \frac{1}{\langle \Delta\nu_{\textrm{d}}\rangle}\Bigg(\frac{1}{N_{\textrm{obs}} -1} \sum_{i = 1}^{N_{\textrm{obs}}}\big(\Delta\nu_{\textrm{d,i}}-\langle \Delta\nu_{\textrm{d}} \rangle \big)^2 \Bigg)^{1/2},
\end{equation}
where $\langle \Delta\nu_{\textrm{d}} \rangle$ is the average scintillation bandwidth, $N_{\textrm{obs}}$ is the number of observations, and $\Delta\nu_{\textrm{d},i}$ is the scintillation bandwidth at the $i^{\textrm{th}}$ epoch \citep{Bhat_1999}. We correct these indices for the estimation error following \cite{Bhat_1999},
\begin{equation}
\label{mod_corrected}
m_{\textrm{b;corrected}}^2 = m_{\textrm{b;measured}}^2 - m_{\textrm{b;error}}^2,
\end{equation}
where $m_{\textrm{b;measured}}$ and $m_{\textrm{b;error}}$ are the modulation indices found from using the scintillation bandwidth measurements and errors, respectively, in Equation \ref{mod_index}. The estimation error-corrected modulation indices for pulsars with more than 10 measurements can be found in Table \ref{table_chi}. Some low-DM pulsars had negative $m_{\textrm{b;corrected}}$ values due to the finite scintle effect; we do not list modulation indices in these cases. There are also likely some instances where the scintles are not fully resolved, such as PSR B1937$+$21. In cases like this, the $m_{\textrm{b;error}}^2$ may be overestimated.

\par We can also compare these results with the theoretic prediction, assuming a Kolmogorov medium with a thin screen halfway between us and the pulsar, as in \cite{Romani}:

\begin{equation}
\label{m_theoretic}
m_{\textrm{b;Kolmogorov}} \approx 0.202 (C_{\textrm{n}}^2)^{-1/5}\nu_{\textrm{obs}}^{3/5}D^{-2/5},
\end{equation}
 where $\nu_{\textrm{obs}}$ is the observing frequency in GHz, $D$ is the distance to the pulsar in kpc, and $C_{\textrm{n}}^2$ describes the strength of scattering effects in units of 10$^{-4}$ m$^{-20/3}$ \citep{Bhat_1999}, and is given by 
 \begin{equation}
\label{strength_scatter}
C_{\textrm{n}}^2 = 0.002\Delta \nu_{\textrm{d}}^{-5/6}D^{-11/6}\nu_{\textrm{obs}}^{11/3} \textrm{m}^{-20/3}
 \end{equation}
 for a Kolmogorov medium, with $\Delta \nu_{\textrm{d}}$ in MHz \citep{Cordes1986}. We calculated $m_{\textrm{b;Kolmogorov}}$ for each pulsar with at least 10 measurements. The results are shown in Table \ref{table_chi}.

\par As mentioned in Section \ref{intro}, dispersion is the largest source of delay from the ISM and is usually the only ISM effect that is corrected for by PTAs. Since both dispersion and scattering are ISM effects originating from the same structures along the LOS, it would be reasonable to expect a correlation between the two quantities on an epoch-to-epoch basis. \cite{rankin_crab} examined correlations between dispersion and scattering for the Crab pulsar during a period of activity from late 1969 to late 1970. It appears that there may be an approximately one-month lag between changes in dispersion and scattering, although it is difficult to say these events are actually correlated. \cite{mckee} looked at around six years of observations of the Crab pulsar and claimed evidence of correlations between $\tau_{\textrm{d}}$ and DM. However, the strength of these correlations is mild, with a correlation coefficient of only 0.56$\pm$0.01. \cite{event} also made observations of the Crab pulsar over a 200 day period coinciding with a large ISM event due to an ionized cloud or filament crossing the LOS, over which time both $\tau_{\textrm{d}}$ and DM followed very similar time signatures. Correlations between these two parameters have also been explored in millisecond pulsars in many contexts. \cite{Coles_2015} examined case of extreme scattering events and found sharp increases in DM are seen to clearly mirror sharp increases in scattering delay in by-eye examinations of the data. In simulated data, \cite{lentati} found scattering delays to be correlated in a non-linear way with both the pulse TOA and DM. \cite{mckee_2019} looked at giant pulses from PSR B1937$+$21 and found no correlation between scattering and DM despite earlier studies finding such correlations using giant pulses in the Crab pulsar \citep{mckee}. \cite{Main} examined the scintillation arcs of PSR J0613--0200 and found that the arc curvature followed the annual variation seen in DM. It is possible that in our data the scattering delays are partially absorbed into DMX fits, decreasing any measured correlation, as suggested by \cite{pss}.

\par It is well known that RISS affects flux densities. \cite{Stinebring_2000} found that pulsars with larger DMs had more stable flux densities, suggesting that  flux density variations in nearby pulsars were due to propagation effects such as RISS.
 \cite{Romani} found that scintillation bandwidth and flux should be strongly anticorrelated, given a thin-screen ISM model. In addition, both \cite{Stine_corr} and \cite{Bhat_1999b} observed these same correlations, though weaker than those predicted by \cite{Romani}, in different samples of pulsars. RISS and DISS are related through flux density, and therefore we might also expect DISS properties to be correlated with flux. 

\par Despite these predictions and earlier work, we do not find any meaningful correlations between $\tau_{\textrm{d}}$ and flux or $\tau_{\textrm{d}}$ and DM in our data. Some also show anti-correlations, although this is likely just due to the small sample of delay measurements relative to $\Delta$DM measurements (e.g., PSRs J2317$+$1439 and J1614$-$2230).

\par
\begin{figure}[!ht]
\includegraphics[scale=.505]{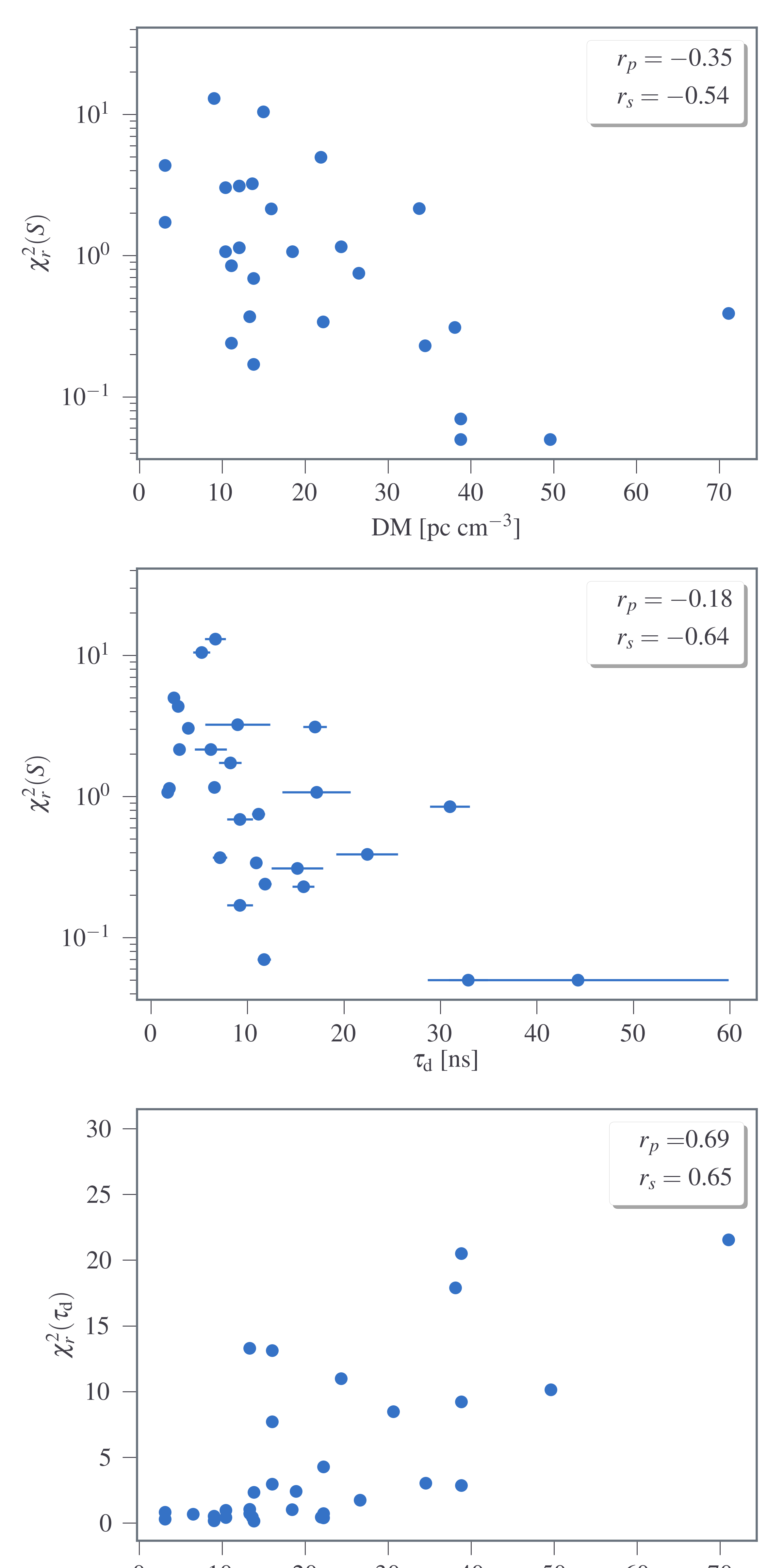}
\caption{Top: A semi$-$log comparison of a pulsar's average scattering delay and the $\chi_r^2$ variability in its flux density. The Spearman correlation coefficient shows moderate evidence of an inverse correlation, indicating flux densities are less variable for more distant pulsars. Middle: A semi$-$log comparison of a pulsar's DM and the $\chi_r^2$ variability in its flux density. The Spearman correlation coefficient shows moderate evidence an inverse correlation, indicating flux densities are less variable for more highly scattered pulsars. We are able to examine DM variations at scales of $10^{-4}$ pc cm$^{-3}$ using DMX, and so the errors on DM in this plot are too small to see. Bottom: The DM vs the $\chi_r^2$ variability in the scattering delay. The Pearson correlation coefficient shows moderate evidence for linear correlations, indicating that pulsars at higher DMs ($\gtrsim$ 20 pc cm$^{-3}$) experience greater variability in their scattering delays, and the Spearman correlation coefficient also shows moderate evidence of general increasing correlations. We are able to examine DM variations at scales of $10^{-4}$ pc cm$^{-3}$ using DMX, and so the errors on DM in this plot are too small to see.}
\label{all_cor}
\end{figure}

 While the evidence for linear correlations between the flux density variability and both $\tau_{\textrm{d}}$ and DM is rather weak, with Pearson coefficients of $-0.18$ and $-$0.35, respectively, there was moderate evidence for general correlations, with Spearman coefficients of $-$0.64 and $-$0.54, respectively. These results are shown in the top and middle panels of Figure \ref{all_cor}. This indicates that flux density variability decreases as $\tau_{\textrm{d}}$ and DM increase, likely due to the higher number of scintles at larger DMs. 

\par We expect that the flux density distributions in Figure \ref{dmx_plots_3} should be exponential at low DMs and small scattering delays and Gaussian at higher DMs and scattering delays \citep{scheuer, hesse} as we transition from small to large numbers of scintles. Indeed, most of the pulsars analyzed have low DMs and show exponential flux distributions. Pulsars with high DM and high scattering, such as PSRs J0340$+$4130, J0613--0200, and B1937$+$21, all exhibit Gaussian flux density distributions.

\begin{deluxetable*}{CCCCCCCCC}[ht]
\tablecaption{Scattering Delay Trends \& Correlations \label{table_chi}}
\centering
\tablecolumns{9}
\tablehead{\colhead{\textrm{Pulsar}} & \colhead{\textrm{Freq}} & \colhead{$\chi_{r}^2(\tau_\textrm{d})$} & \colhead{$\chi_{r}^2(S)$} & \colhead{r($\tau_{\textrm{d}}$, S)} & \colhead{$\overline{S}(\textrm{mJy})$} & \colhead{r($\tau_{\textrm{d}}$, DM)} & \colhead{$m_{\textrm{b;corrected}}$} & \colhead{$m_{\textrm{b;Kolmogorov}}$}} 
\startdata
\textrm{J}0340{+}4130 & 1500 & 10.1 & 0.1 & -0.2 $\pm$ 0.2 & 0.5 $\pm$ 0.00 & 0.7 $\pm$ 0.1 & \text{---} & 0.13 $\pm$ 0.01 \\
\text{J}0613\text{$--$}0200 &             820 &                          21 &            0.1 &                    -0.5 $\pm$ 0.2 &        6.7 $\pm$ 0.00 &                       -0.1 $\pm$ 0.2 & \text{---} & 0.12 $\pm$ 0.00\\
\text{J}0613\text{$--$}0200 &            1500 &                           9.2 &            0.1 &                      0.2 $\pm$ 0.1 &        1.9 $\pm$ 0.00 &                     -0.4 $\pm$ 0.1 & \text{---} & 0.13 $\pm$ 0.01\\
\text{J}0636{+}5128 & 820 &                          13.1 &            0.9 &                      -0.2 $\pm$ 0.2 &        1.9 $\pm$ 0.00 &                       0.1 $\pm$ 0.2 & \text{---} & 0.12 $\pm$ 0.01 \\
\text{J}0636{+}5128 &            1500 &                           7.7 &            0.2 &                     -0.4 $\pm$ 0.2 &        0.7 $\pm$ 0.00 &                       0.3 $\pm$ 0.2 & \text{---} & 0.15 $\pm$ 0.01 \\
\text{J}0740{+}6620 &             820 &                           0.7 &           10.5 &                     -0.1 $\pm$ 0.2 &        2.7 $\pm$ 0.00 &                      0.2 $\pm$ 0.2 & \text{---} & 0.20 $\pm$ 0.02\\
\text{J}1024\text{$--$}0719 &            1500 &                           0.7 &            3.8 &                    -0.1 $\pm$ 0.4 &        1.7 $\pm$ 0.00 &                     -0.5 $\pm$ 0.3 & 0.30 & 0.19 $\pm$ 0.00\\
\text{J}1125{+}7819 &             820 &                           1.0 &            3.1 &                     -0.4 $\pm$ 0.2 &        4.1 $\pm$ 0.00 &                       0.2 $\pm$ 0.2 & \text{---} & 0.20 $\pm$ 0.02\\
\text{J}1125{+}7819 &            1500 &                           0.7 &            1.1 &                     -0.3 $\pm$ 0.2 &        0.9 $\pm$ 0.00 &                       0.2 $\pm$ 0.2 & \text{---} & 0.16 $\pm$ 0.01\\
\text{J}1455\text{$--$}3330 &             820 &                           0.5 &            3.2 &                    -0.1 $\pm$ 0.3 &        2.9 $\pm$ 0.00 &                     0.1 $\pm$ 0.3 & \text{---} & 0.15 $\pm$ 0.01\\
\text{J}1614\text{$--$}2230 &            1500 &                           3.0 &            0.2 &                      0.3 $\pm$ 0.1 &        1.1 $\pm$ 0.00 &                     -0.5 $\pm$ 0.2  & \text{---}& 0.14 $\pm$ 0.01 \\
\text{J}1640{+}2224 &            1500 &                           1.0 &            1.1 &                    -0.4 $\pm$ 0.2 &        0.1 $\pm$ 0.00 &                       -0.03 $\pm$ 0.2 & \text{---}& 0.20 $\pm$ 0.02  \\
\text{J}1713{+}0747 &            1500 &                           3.0 &            2.2 &                    -0.4 $\pm$ 0.1&        4.7 $\pm$ 0.00 &                       0.5 $\pm$ 0.1 & \text{---}& 0.18 $\pm$ 0.02\\
\text{J}1738{+}0333 &            1500 &                           2.8 &            2.2 &                    -0.3 $\pm$ 0.2 &        0.70 $\pm$ 0.00 &                     0.1 $\pm$ 0.2 & 0.44 & 0.16 $\pm$ 0.01\\
\text{J}1744\text{$--$}1134 &             820 &                           0.3 &            1.7 &                      0.1 $\pm$ 0.2 &        7.2 $\pm$ 0.00 &                       0.1 $\pm$ 0.2 & 0.13 & 0.17 $\pm$ 0.01  \\
\text{J}1744\text{$--$}1134 &            1500 &                           0.8 &            4.4 &                    -0.4 $\pm$ 0.2 &        2.6 $\pm$ 0.00 &                     -0.04 $\pm$ 0.19& \text{---} & 0.19 $\pm$ 0.01\\
\text{J}1853{+}1303 &            1500 &                          8.5 &            0.5 &                    -0.4 $\pm$ 0.3 &        0.3 $\pm$ 0.00 &                     -0.3 $\pm$ 0.3 & 0.24 & 0.15 $\pm$ 0.01 \\
\text{B}1855{+}09 &            1500 &       13 &                     0.4 &                       -0.3 $\pm$ 0.1 & 4.6 $\pm$ 0.00 & -0.06 $\pm$ 0.15 & \text{---} & 0.15$\pm$0.01 \\
\text{J}1909\text{$--$}3744 &             820 &                           0.4 &            1.1 &                    -0.2 $\pm$ 0.2 &        4.5 $\pm$ 0.00 &                     -0.7 $\pm$ 0.2 & 0.18 & 0.14 $\pm$ 0.01 \\
\text{J}1909\text{$--$}3744 &            1500 &                           1.0 &            3.0 &                    -0.1 $\pm$ 0.1 &        1.60 $\pm$ 0.00 &                     -0.4 $\pm$ 0.2 & \text{---} & 0.18 $\pm$ 0.01 \\
\text{J}1910{+}1256 &            1500 &                          18 &            0.3 &                      0.3 $\pm$ 0.2 &        0.6 $\pm$ 0.00 &                     -0.1 $\pm$ 0.2 & \text{---} & 0.12 $\pm$ 0.01 \\
\text{J}1918\text{$--$}0642 &            1500 &                           1.7 &            0.8 &                      0.02 $\pm$ 0.13 &        1.5 $\pm$ 0.00 &                     -0.2 $\pm$ 0.1 & \text{---} & 0.15 $\pm$ 0.01 \\
\text{J}1923{+}2515 &            1500 &                           2.4 &            1.4 &                    -0.4 $\pm$ 0.3 &        0.4 $\pm$ 0.00 &                       0.3 $\pm$ 0.3 & \text{---}& 0.17 $\pm$ 0.02 \\
\text{B}1937{+}21 &            1500 &       22 &                      0.4 &                       0.1 $\pm$ 0.1  & 12.3 $\pm$ 0.0 & -0.2 $\pm$ 0.2 & \text{---} & 0.11 $\pm$ 0.01\\
\text{J}1944{+}0907 &1500 &  11 & 1.2 & -0.2 $\pm$ 0.2 & 2.6 $\pm$ 0.00 & -0.1 $\pm$ 0.2 & \text{---} & 0.14 $\pm$ 0.02\\
\text{J}2010\text{$--$}1323 &            1500 &                           4.3 &            0.3  &                    -0.2 $\pm$ 0.1 &        0.7 $\pm$ 0.00 &                       0.3 $\pm$ 0.1 & \text{---} & 0.14 $\pm$ 0.01 \\
\text{J}2145\text{$--$}0750 &             820 &                           0.5 &           13.0 &                      0.03 $\pm$ 0.24 &       23.2 $\pm$ 0.00 &                     -0.2 $\pm$ 0.3 & 0.16 & 0.16 $\pm$ 0.01 \\
\text{J}2145\text{$--$}0750 &   1500 &  0.2 & 5.0 & 0.3 $\pm$ 0.2 & 6.4 $\pm$ 0.00 & 0.2 $\pm$ 0.3 & \text{---}& 0.20 $\pm$ 0.01 \\
\text{J}2302{+}4442 & 820 &   0.1 &  0.2 & -0.2 $\pm$ 0.2 &  3.30 $\pm$ 0.00 &  0.3 $\pm$ 0.2 & \text{---} & 0.12 $\pm$ 0.00 \\
\text{J}2302{+}4442 & 1500 &   2.3 &  0.7 &  -0.3 $\pm$ 0.2 & 1.3 $\pm$ 0.0 &   -0.2 $\pm$ 0.2 & \text{---}& 0.14 $\pm$ 0.01\\
\text{J}2317{+}1439 &            1500 &    0.4 &  5.0 &  -0.1 $\pm$ 0.3 &  0.04  $\pm$ 0.00 &  -0.9 $\pm$ 0.1 & 0.23 & 0.20 $\pm$ 0.01\\
\enddata
\tablecomments{Reduced $\chi^2$ measurements, flux density and $\Delta$DM correlations, and measured and predicted modulation indices for all pulsars with at least 10 scattering delay measurements. We find that no strong correlations between scattering delays and flux density and scattering delays and $\Delta$DM. The unusually strong correlation coefficient seen in PSR J2317$+$1439 is likely not physical, as the scattering delay data is very sparsely sampled relative to $\Delta$DM estimates. A similar argument can be made for PSR J1614$-$2230 for correlations between flux and scattering delay. For most, if not all, of the pulsars shown above, the difference in sample rates between scattering delay and flux density and $\Delta$DM are too different to draw any meaningful conclusions on correlations. All correlations in this table use the Pearson correlation coefficient. We have not reported modulation indices in cases of negative $m_{\textrm{b;corrected}}^2$.}
\end{deluxetable*}

Of the 24 pulsars we analyzed for scattering delay variability at 1500 MHz, seven showed no variation ($\chi_r^2\leq1$), 11 had moderate variations ($1<\chi_r^2\leq10$), and six had significant variations ($\chi_r^2>10$), indicating that scattering delays can be variable among MSPs and are highly dependent on the LOS to each pulsar. We also performed the same analysis on the nine pulsars with more than 10 $\tau_{\textrm{d}}$ measurements at 820 MHz. Seven of the nine pulsars showed no variation, while the other two showed significant variations. Pulsars with at least 10 measurements at both frequencies generally had a similar degrees of variation at both frequencies, although it is unclear if this independence would hold if the observing frequencies were much farther apart. The exception to this was PSR J0613--0200, although the level of variability was still high at both frequencies.  

\par Both the Pearson and Spearman coefficients also indicate greater scattering delay variability for pulsars with higher DMs (see bottom of Figure \ref{all_cor}). Note that all of the $\chi_r^2(\tau_{\textrm{d}})$ values are slightly underestimated due to the finite scintle approximation we made in Equation \ref{finite_scintle} in Section \ref{analysis}. 

 \par Predicted modulation indices range from around 0.1$\lesssim m_{\textrm{b;Kolmogorov}} \lesssim 0.2$, which is in agreement with most of our $m_{\textrm{b;corrected}}$ values. As expected, those that disagree with theoretical predictions tend to be biased high, likely either due to excess refraction or an overly simplistic thin screen model. 

\subsection{Measuring Scaling Indices}
\label{scale_dis}
\par We have determined the scaling of scattering delays with frequency by splitting each frequency band into subbands for four pulsars and by using an average delay measurement at each frequency for 15 pulsars.  

\par Every scaling index we found using our multiband method was significantly shallower than $-4.4$, with a weighted average of $-2.6\pm0.1$. We find that our results agree with \citep{Levin_Scat}, who found an index weighted average of $-3.1 \pm 0.1$ for 10 pulsars over 26 epochs, with the vast majority of measured scaling indices being shallower than $-4.4$. This also agrees with the $-3.4 \pm 0.1$ weighted average we found for scaling indices determined using the subband method. In the two pulsars for which we were able to use both scaling analyses, the index measured using the multiband analysis was at least twice as shallow ($-2.4 \pm 0.3$ compared with $-0.7\pm 0.5$ for PSR B1855$+$09 and $-2.6 \pm 1.1$ compared with $-1.3 \pm 0.4$ for PSR J2302$+$4442). This could indicate that the index of $-4.4$ used for stretching may not be properly scaling the scintles in each band, with the stretching index being too shallow in these two cases. However, this could partially be the result of the large discrepancy in the number of measurements used at 820 and 1500 MHz for the multiband analysis, as is discussed below.
\par There are several reasons why these indices may not agree with the $-$4.4 expected for a Kolmogorov medium. The thin screen approximation which is commonly used assumes an infinite scattering screen. However, many of the pulsars have lower DMs, which means they could be subject to finite or truncated scattering screens \citep{Cordes_2001}. Scaling indices shallower than $-4.4$ (as low as $-4.0$) have been found to be better fits to the data for assuming the existence of these finite screens for a given inner scale cutoff \citep{Rickett_2009}.

\par While using a wider frequency range for this analysis is an improvement over \cite{Levin_Scat}, they were able to measure scaling indices for individual days, whereas, as mentioned earlier, we rarely, if ever, had days in which we had detectable measurements for two frequencies in a given epoch.

\par We may also be biased by the ratio of our measurable observations from both frequencies. For five of the 15 pulsars we analyzed, only 1$-$3 measurements at 820 MHz were obtained, and so these fits are much more constrained at higher frequencies. We were able to make at least six measurements for the other pulsars at both frequencies, although for two of them there are more than twice as many 820 MHz measurements. As mentioned earlier, for PSR B1855$+$09, for which we have many more measurements at 1500 MHz, the subband method returns a much steeper scaling index at each frequency than the multiband method. This implies that similar effects may impact the measurements for other pulsars, for which we were unable to apply the subband measurements.  

\par Finally, as we discussed in Section \ref{scint_param} and the beginning of Section \ref{ne compare}, our limited bandwidths and frequency resolution may cause underestimations on high scintillation bandwidths and overestimations on low ones. Because scintillation bandwidths are smaller at lower frequencies, there will be more underestimations of scattering delay at lower frequencies and more overestimations at higher frequencies. With wider bandwidths and better resolution, our scaling indices would likely be closer to $-4.4$. 

\par The trend of shallow scaling indices has been found in multiple studies in addition to this paper and \cite{Levin_Scat}. \cite{Bansal_2019} performed observations on seven pulsars and found five of them to have shallower indices than $-4.4$. While the other two were close to $-4.4$ when considering their weighted averages, there were deviations on an epoch-to-epoch basis. \cite{Bhat_2004} observed several pulsars at at least two frequencies and determined scaling indices with a pulse broadening function that assumed a thin screen between the Earth and the pulsars. While a few of their pulsars were consistent with a $-4.4$ scaling index, the average index for their sample was $-$3.12 $\pm$ 0.13. Using a pulse broadening function that assumed scattering material uniformly distributed along the LOS, they found an average index of $3.83 \pm 0.19$. The latter was in better agreement with the global fit to their data, which resulted in an index of $-3.86 \pm 0.16$, which they found via a parabolic fit of $\tau_{\textrm{d}}$ vs DM using a variation of the model from \cite{ne2001_b}. They determined that such trends could still be expected for a Kolmogorov medium if the spectrum of turbulence had an inner cutoff between around 300--800 km. Other studies show higher DM pulsars seem to exhibit indices that are shallower than expected for a Kolmogorov medium \citep{L_hmer_2002} (although this can be explained by a truncation of the scattering region), while lower DM pulsars tend to have indices much more in line with a Kolmogorov medium \citep{1985_cordes}.  New techniques such as cyclic spectroscopy will allow for more accurate single-epoch scaling index measurements than are currently obtainable by ACF analyses \citep{Demorest_2012}.

\par A benefit of the subband method is that we can look for variability in the scaling index of a given pulsar over time, which was not possible with our multiband method due to limited epochs having frequency-resolvable, same-day measurements. As briefly mentioned in Section \ref{4.2}, and clearly visible by eye in both Figures \ref{20_scale} and \ref{19_scale}, for pulsars with fully resolved scintles there appears to be a low degree of variation in the scaling index from epoch to epoch, which is further evidenced by both pulsars having $\chi_r^2 \leq 1.0$. This consistency implies that scaling indices are intrinsically stable, as expected. Conversely, for a pulsar like PSR B1937$+$21, which likely has many epochs with unresolved scintles, we found a much larger degree of variation, with $\chi_r^2=7.7$. However, it is likely this variation would decrease significantly once sufficient resolution was achieved. 

\par We have also computed Pearson correlation coefficients between average scaling index and $\overline{\Delta\nu_{\textrm{d}}}$ and average scaling index and DM for both the subband and multiband methods. We find no current evidence of correlations for any of these quantities for either approach.
\subsection{Transverse Velocity Measurements}
 Transverse velocity measurements listed in Table \ref{table_3} are shown in Figure \ref{v_trans_plot}. We find, under the assumption of an equidistant scattering screen between us and a given pulsar, poor agreement between velocities derived from both methods, as indicated by the low correlation coefficient. As mentioned earlier, we are likely biased low on most of our average scintillation timescales due to our short observation lengths, and as a result more of our $V_{\rm ISS}$ values may be upper limits than our averages would indicate. The exception to this is PSR B1937$+$21, for we are confident with our measurement of its scintillation timescale, as all of its epochs had scintles that were clearly resolved in time. However, as mentioned before, we are likely overestimating its scintillation bandwidth, which will also lead to an overestimation of its $V_{\rm ISS}$. Additionally, the discrepancy between $V_{\rm ISS}$ and $V_{\rm pm}$ also demonstrates that knowledge of the scattering screen distance is crucial to accurately determine transverse velocities in this manner, provided other assumptions about the geometry and electron density of the ISM are correct. Many of the $V_{\textrm{ISS}}$ upper limits are consistent with their corresponding $V_{\textrm{pm}}$ values. 

\par We used weighted averages of the scintillation parameters to estimate $V_{\textrm{ISS}}$. However, even though scintillation variability seen on shorter timescales in many pulsars is comparatively small (see Figure \ref{dmx_plots_3} for many examples of this), these changes can have drastic effects on the calculated transverse velocity. For example, \cite{McLaughlin_2002} measured scintillation parameters for the pulsar PSR J1740$+$1000 at seven epochs that spanned over 700 days and found the changes in scintillation behavior lead in the most extreme cases to factor-of-two variations in transverse velocity estimations. While some of this fluctuation was due to measurement uncertainties, they also partially attributed it to ISM effects, particularly modulations in RISS. There are also quite a few pulsars in Figure \ref{dmx_plots_3} where we can see at least factor-of-two variations in the scattering delay, which would either imply significant changes in $V_{\textrm{ISS}}$ or significant changes in the screen location from epoch to epoch. As a result, the average of $V_{\textrm{ISS}}$ is a better measure of velocity than the measurement at a single epoch. We do not expect $V_{\textrm{ISS}}$ and $V_{\textrm{pm}}$ to fully agree without accounting for the screen distance. However, \cite{Reardon_2020} were able to use 16 years of scintillation measurements for PSR J0437--4715 to determine orbital parameters with  higher precision than through timing, indicating that similar levels of precision and accuracy may be obtainable for scintillation-derived transverse velocities. Among pulsars for which both $V_{\textrm{pm}}$ and $V_{\textrm{ISS}}$ measurements were possible, proper motion provided higher precision for the majority of the pulsars. However, there are still benefits to using $V_{\textrm{ISS}}$, as discrepancies between $V_{\textrm{ISS}}$ and $V_{\textrm{pm}}$ could imply a significant motion of the ISM along a LOS or whether a uniform medium or thin screen structure is more accurate for a LOS \citep{Reardon}. 

\par Calculated scattering screen fractional distances are also shown in Table \ref{table_3}, with values greater than one indicating a screen closer to the Earth, and less than one indicating a screen closer to the pulsar. Of the pulsars with measurements that are not upper limits, two pulsars at 820 MHz and three at 1500 MHz require screens that are closer to Earth, three pulsars at 820 MHz require screens that are equidistant, and one pulsar at 1500 MHz and one at 820 MHz had a screen closer to the pulsar. If we look at pulsars with upper limits, two pulsars at 820 MHz and three pulsars at 1500 MHz require a screen closer to the pulsar, while it could be argued that four of the pulsars at 1500 MHz and one pulsar at 820 MHz likely have screens that are equidistant.

\par We assume that velocities from the ISM provide negligible contributions to a given pulsar's transverse velocity. Additionally, contributions to the ISM velocity from the transverse component of differential Galactic rotation (DGR), even if the velocity of DGR is large, can be ignored for nearby pulsars for which the ISM will co-rotate with the LOS. As most of the pulsars we analyzed are no more than 1.5 kpc away, and only one is more than two kpc away, we are unable to probe the regime where contributions from Galactic rotation become significant and whether our assumptions about a uniform Kolmogorov medium break down at these distances.

\begin{figure}[!ht]
\includegraphics[scale=.52]{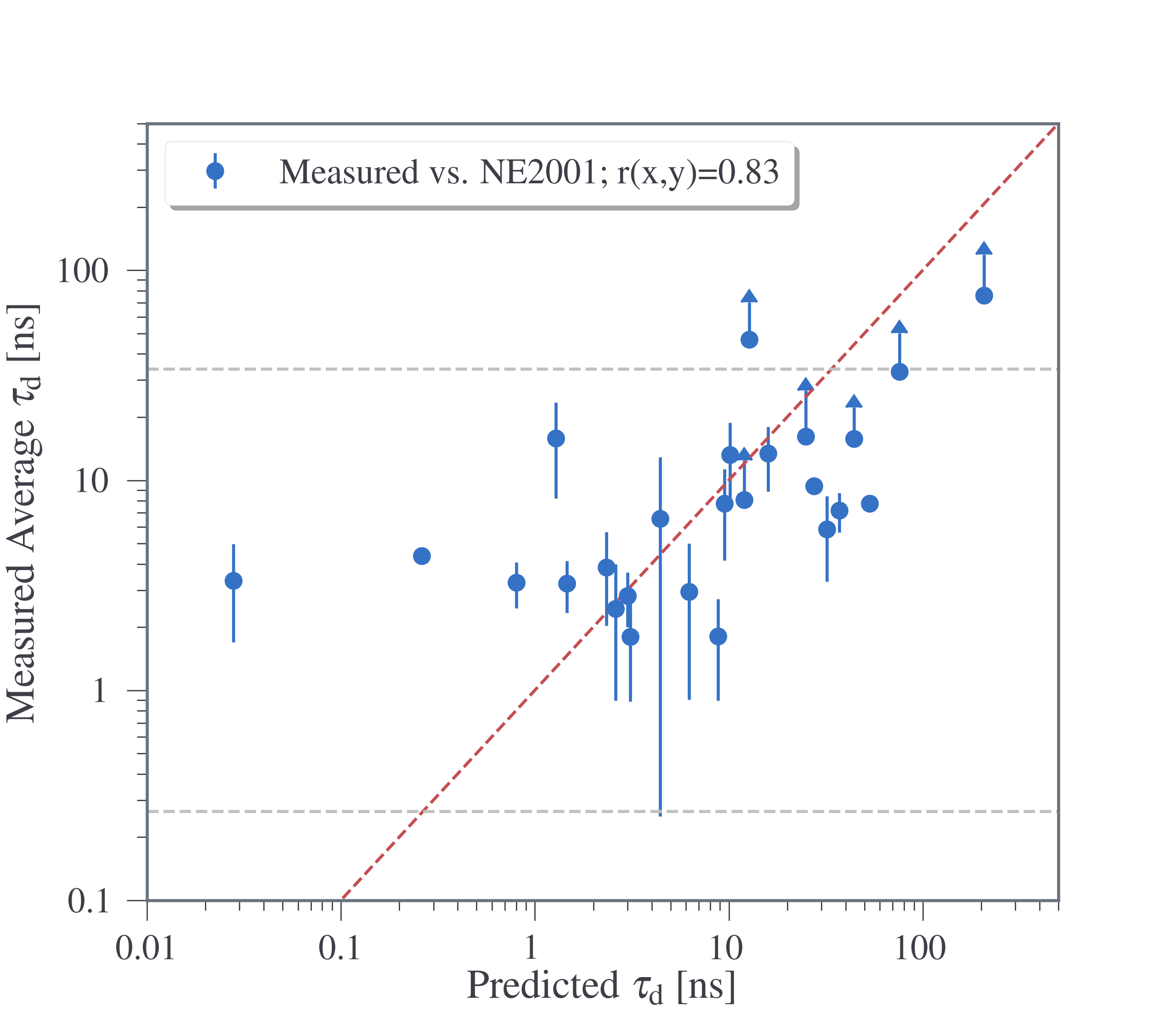}
\caption{The average scattering delay measured at 1500 MHz from this paper compared with the predicted delays by the NE2001 model. The dotted red line indicates a trend with a Pearson correlation coefficient of one, the grey dotted lines indicate the largest and smallest scattering delays we can resolve, corresponding with three channel widths and our effective bandwidth at 1500 MHz, respectively, and the points with arrows indicate delay averages that are lower limits. Generally, values above 0.7$-$0.8 indicate a fairly strong correlation, depending on how precisely it is expected a given model will agree with data.}
\label{ne2001_delay}
\end{figure}

\subsection{Scaling of Scattering Delay with DM}\label{scale_dm}
\par The most commonly used model for mapping electron densities in the Milky Way is NE2001 \citep{NE2001}. This model uses a pulsar's DM and position in the Galaxy to estimate its distance, as well as scintillation parameters such as scintillation bandwidth and timescale. We have plotted the predicted scattering delay vs. the measured weighted means for all pulsars in Figure \ref{ne2001_delay}. As indicated by the high correlation coefficient, we find reasonably strong agreement between our measured delays and those predicted by NE2001. Improved frequency resolution and/or wider observing bandwidths are necessary in order to probe the relationship between NE2001 predictions and our measurements at high and low delays. Other models, such as \cite{Bhat_2004}, \cite{YMW16}, and \cite{Krishnakumar} use an empirical fit to scattering delays vs DM for prediction.
\par
\begin{figure}[!ht]
\includegraphics[scale=.52]{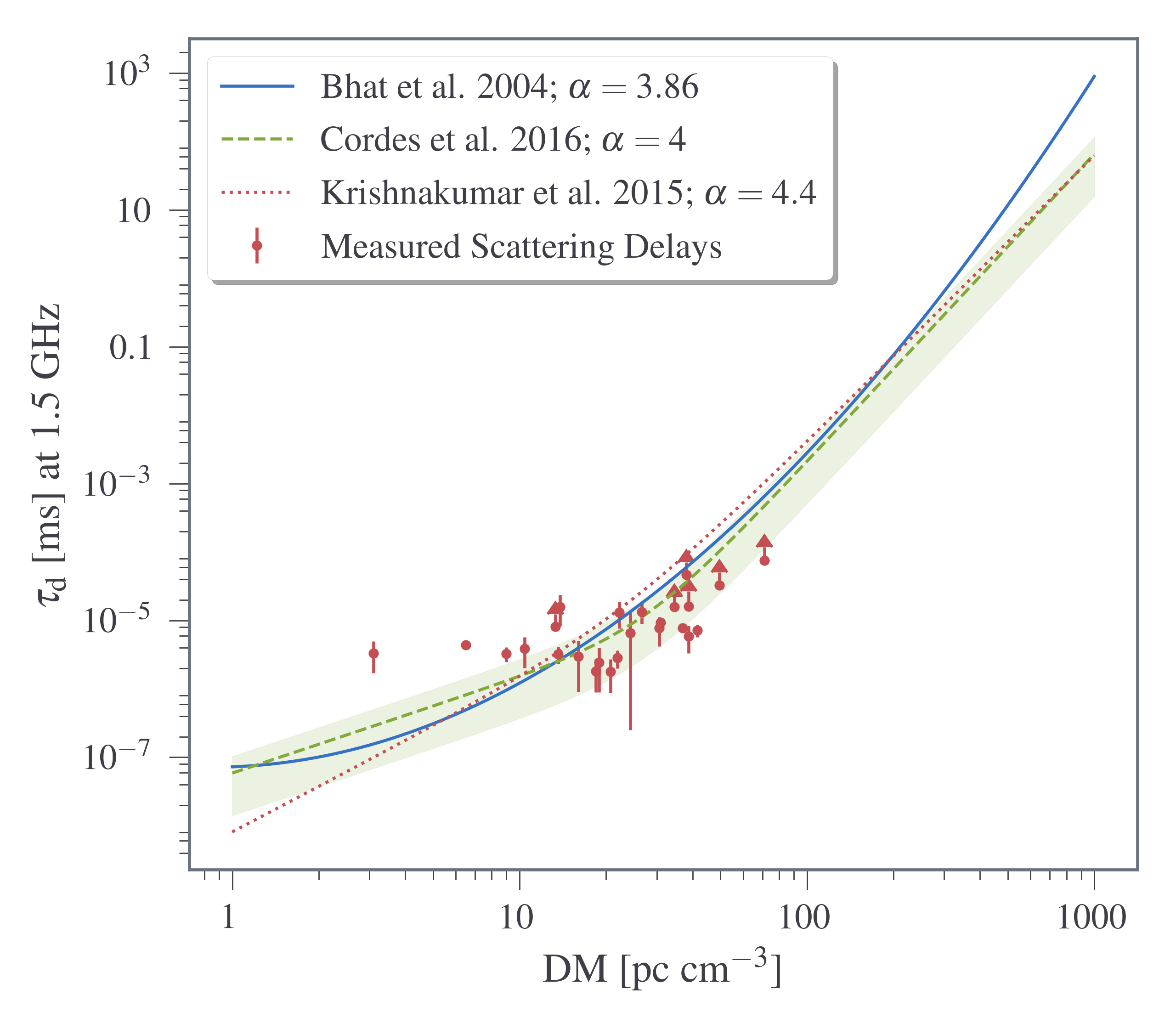}
\caption{A comparison between measured scattering delays at 1500 MHz and fits made by \cite{Bhat_2004}, \cite{Cordes_2016}, and \cite{Krishnakumar}, along with the 1$\sigma$ errors from \cite{Cordes_2016}, shaded in green. The scales shown in the plot were chosen based on the spread of data over which the fits were initially determined.}
\label{scat_models}
\end{figure}

 In Figure \ref{scat_models}, we plot average scattering delay as a function of DM. \cite{Bhat_2004} surveyed over 100 pulsars and fit a parabolic relation of the form
\begin{equation}
\label{DM_fit}
\log \tau_{\textrm{d,$\mu$s}}= a+b(\log \textrm{DM})+c(\log \textrm{DM})^{2}-\alpha \log \nu_{\textrm{GHz}},
\end{equation}
where $a$, $b$, and $c$ are dimensionless scaling coefficients and $\alpha$ is the scaling index of the medium. In their fit, they assumed $\alpha=4.4$ and found $a$, $b$, and $c$ to be 6.46, 0.154, and 1.07, respectively, with a resulting scaling index of $\alpha=3.86 \pm 0.16$. While this index is slightly shallower than the fiducial Kolmogorov index of 4.4, they provide a number of detailed explanations for this discrepancy, including a finite wavenumber cutoff based on the inner scale for a Kolmogorov medium and abrupt changes in the medium transverse to the LOS.

\par \cite{Krishnakumar} used the relation from \cite{Ramachandran}, fitting an exponential equation of the form

\begin{equation}
\label{rama}
\tau_{\textrm{d,s}} = a\textrm{DM}^{\gamma}(1+b\textrm{DM}^{\zeta})\nu^{-\alpha},
\end{equation}
where $a$, $b$, $\gamma$, and $\zeta$ are dimensionless coefficients and $\alpha$ is again the scaling index of the medium. They set $\gamma=2.2$, as expected for a Kolmogorov medium.  

\par \cite{Krishnakumar} then set $\alpha = 4.4$ and fit for $a$, $b$, and $\zeta$ using scattering data from 358 pulsars, finding values of 4.11$\times 10^{-11}$, 1.94$\times 10^{-3}$, and 2.0, respectively. However, unlike our approach, in which we set $C_1=1$, scattering delays used in this fit were calculated with $C_1=1.16$. It should be noted that since these three fits are not direction dependent, they can only serve as first order approximations for how DM correlates with $\tau_{\textrm{d}}$ within our galaxy.
\par \cite{Cordes_2016} set $\alpha = 4$ and fit for all remaining parameters using 531 lines of sight from pulsars, magnetars, and FRBs and found $a=2.98\times 10^{-7}$, $b=3.55\times 10^{-5}$, $\gamma = 1.4$, and $\zeta = 3.1$.

\par We compared these three fits with our measured scattering delays in Figure \ref{scat_models}, with the 1$\sigma$ errors from \cite{Cordes_2016}. The scales shown were chosen based on the scales over which the initial models were fit. The delays we measure are comparable to those predicted by all of these models, but we do not have data over a wide enough DM range to discriminate among them. Also note that the delays we measure for the lowest DM pulsars are much higher than model predictions, indicating the LOS-dependence of scattering at these low DMs.
\section{Conclusions}
We used dynamic spectra made from observations with the GUPPI and PUPPI spectrometers to obtain scintillation parameters of pulsars in the NANOGrav 12.5-year data set. 
\par We looked for correlations between scattering delays and both DM and flux density as a function of time. We do not find any significant correlations, and any instances of high correlation could be attributed to the scale of our scattering delays and their limited sample size.  Additional contributions to flux density may also be masking existing correlations with DISS, and a lack of change in the electron density structure of the ISM along our LOSs may be further limiting correlations with DMX. 
\par We then examined the variability of our scattering delay measurements via a reduced $\chi^2$ analysis on 24 of the pulsars. We also found that, for most pulsars where at least 10 measurements of $\tau_{\textrm{d}}$ were available at both 820 and 1500 MHz, the degree of variation was virtually the same at both frequencies, meaning that scattering variation might be independent of observing frequency. 
\par We measured scaling indices for 17 pulsars and found that all of the pulsars exhibited a shallower than $\nu^{-4.4}$ scaling. We concluded that although the ISM along these LOSs might follow shallower scaling laws than expected, biases introduced by uneven sampling of our two frequencies and resolution issues provide plausible explanations for this.
\par We were able to use scintillation parameters to estimate transverse velocities. We also calculated the location of the scattering screen, assuming that $V_{\textrm{ISS}}$ and $V_{\textrm{pm}}$ are equal. Much of the disagreement is likely the result of our scintillation timescale averages being biased low as the result of our short observation length. 
\par We were also able to determine scattering screen fractional distances using our measured scintillation parameters and $V_{\textrm{pm}}$ values. 
\par Finally, we examined how scattering delays compare with electron density models as well as scale with DM and plotted our results against empirical fits of scattering delay vs DM \citep{Cordes_2016, Bhat_2004, Krishnakumar}. We find that, on the DM scales these fits consider, both the spread of and trends in our data agree with all three fits above a DM of around 10 pc cm $^{-3}$, below which the models begin to follow a steeper trend than our measurements. We also found our results largely agree with predictions made by  NE2001 for pulsars where scintles were resolvable.
\par As we continue to observe PTAs with higher precision and get closer to gravitational wave detection additional sources of TOA residual uncertainty will become significant enough that they cannot be ignored by our timing models. We have already reached that stage with scattering delays in some pulsars, as we have shown that they exhibit average delays comparable in magnitude to the 10 ns precision believed to be necessary for gravitational wave detection.  Additionally, pulsars such as PSR B1937$+$21 already have scattering delays comparable to or greater than their median TOA uncertainty at certain frequencies \citep{11yr_Obs}. Many more pulsars are on track to reach these levels of precision in TOA uncertainty within the next few years, at which point it will be increasingly detrimental to ignore effects from scattering. It is crucial to incorporate methods to mitigate these delays in our timing pipelines as soon as possible.

\par Our analysis illustrates the need for finer frequency resolution in our standard timing observations. New techniques like cyclic spectroscopy allow for the determination of ISM-related delays and unscattered pulse profiles from single observations, making it much more efficient to mitigate these delays than the current method of ACF fitting \citep{cyc_spec, Palliyaguru_2015}. This technique will allow us to obtain much better scattering estimations for highly scattered and high S/N pulsars \citep{dolch} has already been used with fine frequency resolutions \citep{Archibald_2014} Ongoing efforts are taking place to implement real-time cyclic spectroscopy pipelines into NANOGrav's existing observing pipelines, with the goal of removing scattering effects before any further timing analysis has taken place.

\par \textit{Acknowledgements}: The NANOGrav project receives support from National Science Foundation (NSF) Physics Frontier Center award number 1430284. TD and MTL acknowledge NSF AAG award number 2009468. NANOGrav research at UBC is supported by an NSERC Discovery Grant and Discovery Accelerator Supplement and by the Canadian Institute for Advanced Research. Data for this project were collected using the facilities of the Green Bank Observatory and the Arecibo Observatory. The Green Bank Observatory is a facility of the National Science Foundation operated under cooperative agreement by Associated Universities, Inc. The Arecibo Observatory is a facility of the National Science Foundation operated under cooperative agreement by the University of Central Florida in alliance with Yang Enterprises, Inc. and Universidad Metropolitana. The National Radio Astronomy Observatory is a facility of the National Science Foundation operated under cooperative agreement by Associated Universities, Inc.
\par Part of this research was carried out at the Jet Propulsion Laboratory, California Institute of Technology, under a contract with the National Aeronautics and Space Administration.
\par The majority of data processing for this work took place on the Bowser computing cluster at West Virginia University.
\par We would like to thank Dick Manchester for a useful discussion on the relations between scattering delay and DM. We would also like to thank Lina Levin for providing most of the initial groundwork for this paper and for the scripts we used to process and analyze dynamic spectra.

\par \textit{Author contributions}: JET undertook all of the data analysis, wrote most of the pipelines and supplementary scripts, and wrote the paper. MAM provided mentorship and suggestions regarding the analysis and writing of the paper. JMC provided valuable suggestions on data interpretation and analyses to run, as well as discussions on ISM structure and behavior. BJS provided useful discussion on transverse velocities, scintillation behavior and analysis techniques, and software development. MTL provided useful discussion on software development and suggesting the inclusion of certain analyses. DRS provided important discussion regarding underestimations of scattering delays in our data. SC and TJWL provided useful discussion on the structure of the ISM and the nature of scattering delays. ZA, HB, PRB, HTC, PBD, MED, TD, JAE, RDF, ECF, EF, NGD, PAG, DCG, MLJ, MTL, DRL, RSL, MAM, CN, DJN, TTP, SMR, RS, IHS, KS, JKS, and WWZ developed the 12.5-year data set.
\par \textit{Software}: \textsc{psrchive} \cite{2004PASA}, \textsc{tempo} \cite{tempo}, \textsc{scipy} \cite{scipy}, \textsc{numpy} \cite{numpy}, and \textsc{matplotlib} \cite{matplotlib}.
\bibliography{12_5_yr_Scattering_Paper_Draft.bib}{}
\bibliographystyle{aasjournal}
\end{document}

%% file: authors.tex
\author[0000-0002-2451-7288]{Jacob E. Turner}
\affiliation{Department of Physics and Astronomy, West Virginia University, P.O.~Box 6315, Morgantown, WV 26506, USA}
\affiliation{Center for Gravitational Waves and Cosmology, West Virginia University, Chestnut Ridge Research Building, Morgantown, WV 26505, USA}

\author[0000-0001-7697-7422]{Maura A. McLaughlin}
\affiliation{Department of Physics and Astronomy, West Virginia University, P.O. Box 6315, Morgantown, WV 26506, USA}
\affiliation{Center for Gravitational Waves and Cosmology, West Virginia University, Chestnut Ridge Research Building, Morgantown, WV 26505, USA}

\author[0000-0002-4049-1882]{James M. Cordes}
\affiliation{Cornell Center for Astrophysics and Planetary Science and Department of Astronomy, Cornell University, Ithaca, NY 14853, USA}

\author[0000-0003-0721-651X]{Michael T. Lam}
\affiliation{School of Physics and Astronomy, Rochester Institute of Technology, Rochester, NY 14623, USA}
\affiliation{Laboratory for Multiwavelength Astrophysics, Rochester Institute of Technology, Rochester, NY 14623, USA} 

\author[0000-0002-7283-1124]{Brent J. Shapiro-Albert}
\affiliation{Department of Physics and Astronomy, West Virginia University, P.O. Box 6315, Morgantown, WV 26506, USA}
\affiliation{Center for Gravitational Waves and Cosmology, West Virginia University, Chestnut Ridge Research Building, Morgantown, WV 26505, USA}

\author[0000-0002-1797-3277]{Daniel R. Stinebring}
\affiliation{Department of Physics and Astronomy, Oberlin College, Oberlin, OH 44074, USA}

\author{Zaven Arzoumanian}
\affiliation{X-Ray Astrophysics Laboratory, NASA Goddard Space Flight Center, Code 662, Greenbelt, MD 20771, USA}

\author[0000-0003-4046-884X]{Harsha Blumer}
\affiliation{Department of Physics and Astronomy, West Virginia University, P.O. Box 6315, Morgantown, WV 26506, USA}
\affiliation{Center for Gravitational Waves and Cosmology, West Virginia University, Chestnut Ridge Research Building, Morgantown, WV 26505, USA}

\author[0000-0003-3053-6538]{Paul R. Brook}
\affiliation{Department of Physics and Astronomy, West Virginia University, P.O. Box 6315, Morgantown, WV 26506, USA}
\affiliation{Center for Gravitational Waves and Cosmology, West Virginia University, Chestnut Ridge Research Building, Morgantown, WV 26505, USA}

\author[0000-0002-2878-1502]{Shami Chatterjee}
\affiliation{Cornell Center for Astrophysics and Planetary Science and Department of Astronomy, Cornell University, Ithaca, NY 14853, USA}

\author[0000-0002-6039-692X]{H. Thankful Cromartie}
\affiliation{University of Virginia, Department of Astronomy, P.O. Box 400325, Charlottesville, VA 22904, USA}
\author[0000-0002-2185-1790]{Megan E. DeCesar}
\affiliation{Department of Physics, Lafayette College, Easton, PA 18042, USA}
\affiliation{AAAS, STPF, ORISE Fellow hosted by the U.S. Department of Energy}
\author[0000-0002-6664-965X]{Paul B. Demorest}
\affiliation{National Radio Astronomy Observatory, 1003 Lopezville Rd., Socorro, NM 87801, USA}
\author[0000-0001-8885-6388]{Timothy Dolch}
\affiliation{Department of Physics, Hillsdale College, 33 E. College Street, Hillsdale, MI 49242, USA}
\affiliation{Eureka Scientific, Inc.  2452 Delmer Street, Suite 100, Oakland, CA 94602-3017}

\author{Justin A. Ellis}
\affiliation{Infinia ML, 202 Rigsbee Avenue, Durham NC, 27701}
\author[0000-0002-2223-1235]{Robert D. Ferdman}
\affiliation{School of Chemistry, University of East Anglia, Norwich, NR4 7TJ, United Kingdom}

\author{Elizabeth C. Ferrara}
\affiliation{NASA Goddard Space Flight Center, Greenbelt, MD 20771, USA}

\author[0000-0001-8384-5049]{Emmanuel Fonseca}
\affiliation{Department of Physics, McGill University, 3600  University St., Montreal, QC H3A 2T8, Canada}

\author[0000-0001-6166-9646]{Nathan Garver-Daniels}
\affiliation{Department of Physics and Astronomy, West Virginia University, P.O. Box 6315, Morgantown, WV 26506, USA}
\affiliation{Center for Gravitational Waves and Cosmology, West Virginia University, Chestnut Ridge Research Building, Morgantown, WV 26505, USA}
\author[0000-0001-8158-638X]{Peter A. Gentile}
\affiliation{Department of Physics and Astronomy, West Virginia University, P.O. Box 6315, Morgantown, WV 26506, USA}
\affiliation{Center for Gravitational Waves and Cosmology, West Virginia University, Chestnut Ridge Research Building, Morgantown, WV 26505, USA}
\author[0000-0003-1884-348X]{Deborah C. Good}
\affiliation{Department of Physics and Astronomy, University of British Columbia, 6224 Agricultural Road, Vancouver, BC V6T 1Z1, Canada}
\author[0000-0001-6607-3710]{Megan L. Jones}
\affiliation{Center for Gravitation, Cosmology and Astrophysics, Department of Physics, University of Wisconsin-Milwaukee,\\ P.O. Box 413, Milwaukee, WI 53201, USA}

\author{T. Joseph W. Lazio}
\affiliation{Jet Propulsion Laboratory, California Institute of Technology, 4800 Oak Grove Drive, Pasadena, CA 91109, USA}
\author[0000-0003-1301-966X]{Duncan R. Lorimer}
\affiliation{Department of Physics and Astronomy, West Virginia University, P.O. Box 6315, Morgantown, WV 26506, USA}
\affiliation{Center for Gravitational Waves and Cosmology, West Virginia University, Chestnut Ridge Research Building, Morgantown, WV 26505, USA}
\author{Jing Luo}
\affiliation{Department of Astronomy \& Astrophysics, University of Toronto, 50 Saint George Street, Toronto, ON M5S 3H4, Canada}
\author[0000-0001-5229-7430]{Ryan S. Lynch}
\affiliation{Green Bank Observatory, P.O. Box 2, Green Bank, WV 24944, USA}

\author[0000-0002-3616-5160]{Cherry Ng}
\affiliation{Dunlap Institute for Astronomy and Astrophysics, University of Toronto, 50 St. George St., Toronto, ON M5S 3H4, Canada}

\author[0000-0002-6709-2566]{David J. Nice}
\affiliation{Department of Physics, Lafayette College, Easton, PA 18042, USA}
\author[0000-0001-5465-2889]{Timothy T. Pennucci}
\altaffiliation{NANOGrav Physics Frontiers Center Postdoctoral Fellow}
\affiliation{National Radio Astronomy Observatory, 520 Edgemont Road, Charlottesville, VA 22903, USA}
\affiliation{Institute of Physics, E\"{o}tv\"{o}s Lor\'{a}nd University, P\'{a}zm\'{a}ny P. s. 1/A, 1117 Budapest, Hungary}
\author[0000-0002-8826-1285]{Nihan S. Pol}
\affiliation{Department of Physics and Astronomy, West Virginia University, P.O. Box 6315, Morgantown, WV 26506, USA}
\affiliation{Center for Gravitational Waves and Cosmology, West Virginia University, Chestnut Ridge Research Building, Morgantown, WV 26505, USA}
\author[0000-0001-5799-9714]{Scott M. Ransom}
\affiliation{National Radio Astronomy Observatory, 520 Edgemont Road, Charlottesville, VA 22903, USA}

\author[0000-0002-6730-3298]{Ren\'{e}e Spiewak}
\affiliation{Centre for Astrophysics and Supercomputing, Swinburne University of Technology, P.O. Box 218, Hawthorn, Victoria 3122, Australia}
\author[0000-0001-9784-8670]{Ingrid H. Stairs}
\affiliation{Department of Physics and Astronomy, University of British Columbia, 6224 Agricultural Road, Vancouver, BC V6T 1Z1, Canada}

\author[0000-0002-7261-594X]{Kevin Stovall}
\affiliation{National Radio Astronomy Observatory, 1003 Lopezville Rd., Socorro, NM 87801, USA}
\author[0000-0002-1075-3837]{Joseph K. Swiggum}
\altaffiliation{NANOGrav Physics Frontiers Center Postdoctoral Fellow}
\affiliation{Department of Physics, Lafayette College, Easton, PA 18042, USA}

\author[0000-0003-4700-9072]{Sarah J. Vigeland}
\affiliation{Center for Gravitation, Cosmology and Astrophysics, Department of Physics, University of Wisconsin-Milwaukee,\\ P.O. Box 413, Milwaukee, WI 53201, USA}

%% file: abstract.tex
\begin{abstract}
We extract interstellar scintillation parameters for pulsars observed by  the NANOGrav radio pulsar timing program. Dynamic spectra for the observing epochs of each pulsar were used to obtain estimates of scintillation timescales, scintillation bandwidths, and the corresponding scattering delays using a stretching algorithm to account for frequency-dependent scaling. We were able to measure scintillation bandwidths for 28 pulsars at 1500 MHz and 15 pulsars at 820 MHz. We examine  scaling behavior for 17 pulsars and find power-law indices ranging from $-0.7$ to $-3.6$, though these may be biased shallow due to insufficient frequency resolution at lower frequencies.
We were also able to measure scintillation timescales for six pulsars at 1500 MHz and seven pulsars at 820 MHz. There is fair agreement between our scattering delay measurements and electron-density model predictions for most pulsars. We derive interstellar scattering-based transverse velocities assuming isotropic scattering and a scattering screen halfway between the pulsar and earth. We also estimate the location of the scattering screens assuming proper motion and interstellar scattering-derived transverse velocities are equal. We find no correlations between variations in scattering delay and either variations in dispersion measure or flux density. For most pulsars for which scattering delays were measurable, we find that time of arrival uncertainties for a given epoch are larger than our scattering delay measurements, indicating that variable scattering delays are currently subdominant in our overall noise budget but are important for achieving precisions of tens of ns or less. 
\end{abstract}